\documentclass[12pt]{article}

\usepackage[font=small,format=plain,labelfont=bf,textfont=it]{caption}
\usepackage{setspace}
\usepackage{apptools}
\usepackage{derivative}

\usepackage{fullpage, verbatim, amsfonts, amsmath, amssymb, amsthm, latexsym, setspace, graphicx}
\allowdisplaybreaks

\usepackage[T1]{fontenc}
\usepackage[latin9]{inputenc}
\usepackage{geometry}
\usepackage{xcolor}
\usepackage{natbib}
\usepackage{overpic}

\usepackage[french, english]{babel}
\usepackage{hyperref,bookmark}

\hypersetup{linkcolor=MyDarkBlue, citecolor=MyDarkBlue, colorlinks=true} 
	\definecolor{MyDarkBlue}{rgb}{0,0.08,0.45}
	\definecolor{MyDarkGreen}{rgb}{0,0.55,0.08}

\theoremstyle{plain}
\newtheorem{thm}{Theorem}
\newtheorem{lem}{Lemma}
\newtheorem{cor}{Corollary}
\newtheorem{prop}{Proposition}

\newtheorem{claim}{Claim}

\newtheorem{assume}{Assumption}

\newtheorem{ex}{Example}

\theoremstyle{definition}
\newtheorem{rem}{Remark}

\AtAppendix{\counterwithin{lem}{section}}
\AtAppendix{\counterwithin{prop}{section}}
\AtAppendix{\counterwithin{thm}{section}}
\AtAppendix{\counterwithin{cor}{section}}
\AtAppendix{\counterwithin{ex}{section}}
\AtAppendix{\counterwithin{assume}{section}}

\newcommand{\argmax}{\mathop{\mathrm{argmax}}}

\newcommand\cites[1]{\citeauthor{#1}'s \citeyearpar{#1}}

\newcommand{\finex}{\leavevmode\unskip\penalty9999 \hbox{}\nobreak\hfill\quad\hbox{$\blacktriangle$}}

\begin{document}

	\title{Multidimensional Screening with Precise Seller Information\thanks{Previous versions of this paper were titled ``Multidimensional Screening with Rich Consumer Data.'' Frick: Princeton University (mfrick@princeton.edu); Iijima: Princeton University (riijima@princeton.edu); Ishii: Pennsylvania State University (yxi5014@psu.edu). For helpful comments that substantially improved the paper, we thank the co-editor, three anonymous referees, numerous seminar and conference audiences, as well as Dirk Bergemann, Tilman B\"orgers, Yang Cai, Piotr Dworczak, Soheil Ghili, Yannai Gonczarowski, Nima Haghpanah, Nicole Immorlica, Zi Yang Kang, Andreas Kleiner, Yingkai Li, Teddy Mekonnen, Axel Niemeyer, Alessandro Pavan, Ron Siegel, Philipp Strack, Juuso Toikka, Frank Yang, Kai Hao Yang, and Jidong Zhou. We thank Michelle Hyun for excellent research assistance. We also gratefully acknowledge the financial support from Sloan Research Fellowships.}}
	\author{Mira Frick \and Ryota Iijima \and Yuhta Ishii}
	\date{First posted version: February 12, 2024 \\ \vspace{3mm} This version: \today}
	
	\maketitle
	
	\begin{abstract} A multi-product monopolist faces a buyer who is privately informed about his valuations for the goods. As is well-known, optimal mechanisms are in general complicated, while simple mechanisms---such as pure bundling or separate sales---can be far from optimal and do not admit clear-cut comparisons. We show that this changes if the monopolist has sufficiently precise information about the buyer's valuations: Now, pure bundling always outperforms separate sales; moreover, there is a sense in which pure bundling performs essentially as well as the optimal mechanism. To formalize this, we characterize how fast the corresponding revenues converge to the first-best revenue as the monopolist's information grows precise: Pure bundling achieves the same convergence rate to the first-best as optimal mechanisms; in contrast, the convergence rate under separate sales is suboptimal.

		\end{abstract}
	
	\newpage
	\onehalfspacing
\section{Introduction}

A classic problem in economic theory is how a multi-product monopolist should sell its goods to a buyer whose valuations for the goods are unknown to the monopolist.  While of clear economic relevance from both a positive and normative perspective, this problem---and related multi-dimensional screening problems---is well-known to be quite intractable. Even in simple subcases (e.g., two goods for which the buyer has independently distributed, additive values), the seller's optimal mechanism can be difficult to characterize and look complicated: For instance, it may require the buyer to choose between a continuum of differently priced lotteries over product bundles \citep[e.g.,][]{daskalakis2017}.\footnote{Finding the optimal mechanism can also be computationally intractable \citep[e.g.,][]{daskalakis2014}, and it can exhibit counterintuitive features, e.g., nonmonotonic revenues with respect to first-order stochastic dominance increases in buyer values \citep[]{hart2015}.}  At the same time, the multi-product selling mechanisms used in practice are often quite simple: For example, many firms only present buyers with some limited number of deterministic product bundles, or even engage in \emph{pure bundling}, i.e., offer only the grand bundle of all products at a take-it-or-leave-it price.\footnote{See Section~\ref{sec:conclusion}.}

This paper provides a novel perspective on the use of such simple multi-product selling mechanisms. We consider a revenue-maximizing monopolistic seller endowed with a finite set $G$ of indivisible goods. There is one potential buyer, whose valuations for the goods are summarized by a type vector $\theta \in \mathbb{R}_{++} ^{|G|}$ (drawn from an arbitrary prior distribution) and whose payoffs are additive across goods. (Section~\ref{sec:extensions} extends the analysis to nonadditive buyer utilities, negative valuations, and seller production costs). The realization of $\theta$ is only known to the buyer. However, as in the literature on price discrimination, we assume that the seller observes some information about the buyer's type that she can use in designing a selling mechanism. For example, the seller might observe various buyer characteristics (or noisy signals thereof) that are correlated with $\theta$ within the population from which the buyer is drawn.


The question we ask is: Are there simple mechanisms that perform well at exploiting such information, and, if so, which ones? If the seller's information is arbitrary, the comparison between different simple mechanisms is not clear-cut: For example, it is well-known that, depending on the seller's posterior, pure bundling may yield higher or lower revenue than \emph{separate sales} (i.e., setting a separate price for each individual good). Moreover, both these simple mechanisms are in general far from optimal. However, the key insight of our paper is that there is a sharp answer if the seller's information is sufficiently \emph{precise}: In this case, we show that the seller is always better off using pure bundling than separate sales; what is more, we formalize a sense in which pure bundling performs essentially as well as the optimal mechanism.

The assumption of precise seller information may, for instance, reflect the proliferation of consumer data to which retailers have access in digital marketplaces.\footnote{Remark~\ref{rem:data} provides a stylized example in the context of past purchasing data. More broadly, retailers have begun to draw inferences about consumers' preferences from data such as their browsing history, geolocation, or operating system, and to personalize selling mechanisms based on such data, not only via outright price discrimination but also more subtle channels such as personalized discounts \citep[e.g.,][]{OECD2018}. \label{fn:data}} To tractably formalize such information, our baseline model assumes that the seller observes $n$ independent signals from some distribution $P_\theta$ that depends on $\theta$. (Section~\ref{sec:extensions} discusses more general settings). Thus, the seller's information is more precise the richer the amount of data $n$, and we are interested in the case where $n$ is large.

%

Crucially, to evaluate and compare the performance of different classes of mechanisms, we do not explicitly derive the seller's revenues at any particular $n$. Instead, we take a \emph{convergence rate} approach: We analyze how fast the gap between the seller's expected revenue and the first-best (i.e., known type) revenue vanishes as $n$ grows large. This provides a parsimonious way of comparing revenues across different classes of mechanisms under rich consumer data: Mechanisms with a faster convergence rate to the first-best yield higher expected revenues than mechanisms with a slower convergence rate at \emph{all} large enough $n$. As a one-dimensional statistic, convergence rates also succinctly characterize how revenues at large $n$ depend on features of the environment, such as the seller's signal distribution $P_\theta$. Most importantly, while optimal mechanisms by definition approximate the first-best revenue at the fastest rate, this does not rule out that some simple (but suboptimal) mechanisms may achieve this same optimal convergence rate. If this is the case, this suggests that a seller who observes rich consumer data loses very little from using such simple mechanisms rather than the optimal mechanism.

In particular, our main result (Theorem~\ref{thm:main}) shows that, regardless of whether the seller optimizes over general selling mechanisms or is restricted to pure bundling, her expected revenue converges to the first-best equally fast as $n$ grows large: In both cases, the revenue gap relative to the first-best vanishes as fast as $\lambda^G \sqrt{\frac{\ln n}{n}}$. Here, the coefficient $\lambda^G$ captures how fast the seller's posterior standard deviation of the value $\sum_{g \in G} \theta_g$ of the grand bundle decays; this coefficient depends on the signal distribution $P_\theta$ only through a standard statistical informativeness measure, Fisher information. In contrast, if the seller optimizes over separate sales mechanisms, Theorem~\ref{thm:main} shows that her expected revenue converges to the first-best more slowly: Now, the revenue gap relative to the first-best vanishes as fast as $\sum_{g \in G} \lambda^g \sqrt{\frac{\ln n}{n}}$, where the coefficients $\lambda^g$ capture how fast the seller's posterior standard deviation of the values $\theta_g$ of each individual good decays. The coefficient $\sum_{g \in G} \lambda^g$ is greater than the optimal coefficient $\lambda^G$: By subadditivity of standard deviations, the sum of the standard deviations of all $\theta_g$ always exceeds the standard deviation of the sum $\sum_{g \in G} \theta_g$.

To interpret, note that in the limit as $n \to \infty$, both pure bundling and separate sales approximate the first-best revenue; however, at any finite $n$, both mechanisms in general yield strictly lower revenues than the optimal mechanism, which may screen the buyer via (possibly elaborate) menus of bundles or lotteries over bundles. This leaves open the possibility that, even at large $n$, the revenue gap relative to the first-best under these simple mechanisms may be many times bigger than under the more complicated optimal mechanism. The key implication of Theorem~\ref{thm:main} is that under pure bundling this is not the case: Since pure bundling and the optimal mechanism approximate the first-best revenue equally fast, using the optimal mechanism rather than pure bundling reduces the revenue gap to the first-best by only a negligible amount at all large enough $n$. For example, as we discuss, the revenue gain from using the optimal mechanism rather than pure bundling is smaller than the gain from having access to even an arbitrarily small fraction of additional signals. In contrast, under separate sales, Theorem~\ref{thm:main} implies that the revenue gap to the first-best at all large enough $n$ is $\sum_{g \in G} \lambda^g /\lambda^G$ times as big as under the optimal mechanism and pure bundling, where depending on parameters this ratio can be arbitrarily large. Thus, even though at small $n$, the seller's revenue can be higher under separate sales or pure bundling, pure bundling always outperforms separate sales when $n$ is large enough.

A potential concern in assessing the economic relevance of Theorem~\ref{thm:main} is whether it requires such unrealistically rich data $n$ that the seller can almost perfectly discriminate between different buyer types. This concern can be mitigated by analyzing numerical examples. For instance, in a Gaussian environment, Section~\ref{sec:numerical} finds that pure bundling starts to approximate the optimal mechanism and to significantly outperform separate sales at $n$ where the corresponding revenues are still only a moderate fraction of the first-best.

The proof of Theorem~\ref{thm:main} first exploits the Bernstein-von Mises theorem to reduce the analysis to a setting in which the seller's posterior is deterministic and Gaussian. The analysis is then based on three main insights that we illustrate in Section~\ref{sec:gaussian analysis}:

First, focusing on the single-good case, we shed light on how rich data affects the tradeoff between two forms of revenue loss that the seller incurs away from the first-best benchmark: Losses on the \emph{intensive margin} (due to having to shade the price relative to the known type case) and on the \emph{extensive margin} (due to the probability that the buyer is unwilling to pay the seller's chosen price). We show that, as the seller's posterior grows more and more precise, these two margins become strikingly imbalanced: The seller
optimally prices the good in such a way that her revenue losses are driven almost entirely by the intensive margin; in contrast, extensive-margin losses become negligible at large $n$. This dominance of the intensive margin lies at the heart of why, in the multi-good setting, the convergence to the first-best under pure bundling cannot be improved upon by general deterministic mechanisms. Under the latter, the seller can engage in \emph{mixed bundling}, i.e., screen the buyer by offering a menu of product bundles. Relative to pure bundling, this benefits the seller by extracting revenue from buyers who are unwilling to purchase the grand bundle. However, since this benefit only affects the extensive margin, it has a negligible impact on the revenue gap relative to the first-best at large $n$.\footnote{Relative to pure bundling, mixed bundling in principle also allows the seller to reduce the magnitude of intensive-margin losses by raising the grand bundle's price. However, we show this increases extensive-margin losses by an order of magnitude that outweighs intensive-margin gains.}


Second, we show that intensive-margin losses under single-good monopoly vanish as fast as the (scaled) \emph{standard deviation} of the seller's posterior about the buyer's type. In the multi-good case, this implies that 
comparing the revenue gap to the first-best at large $n$ under pure bundling vs.\ separate sales boils down to comparing the seller's posterior standard deviation about the value $\sum_{g \in G} \theta_g$ of the grand bundle vs.\ the sum of her posterior standard deviations about each $\theta_g$. As noted, the former is smaller than the latter, as standard deviation is subadditive. At a high level, this relates to the classic intuition \citep[e.g.,][]{adams1976, armstrong1999} that bundling reduces the seller's uncertainty relative to separate sales, as overestimating the valuations of some goods but underestimating those of others can cancel out when estimating the value of the grand bundle. However, for this intuition to be valid in our setting, it is crucial to show that the relevant measure of the seller's uncertainty is standard deviation, as other measures of uncertainty (e.g., variance) are not subadditive.

Finally, we show that pure bundling achieves the same convergence rate as the optimal mechanism, which may additionally involve randomization. To get around the aforementioned challenge that optimal mechanisms are difficult to characterize, we introduce a more tractable \emph{relaxed problem} that yields an upper bound on the seller's optimal revenue. We then prove that even this upper bound converges to the first-best no faster than the pure bundling revenue. The relaxed problem partitions the type space into line segments and imposes incentive compatibility only within each segment. As Section~\ref{sec:1-dim type} explains, optimal mechanisms in this problem are tractable to characterize and take an ``almost deterministic'' form. Moreover, we show that by carefully choosing the line segments, we can ensure that the pure bundling revenue in the relaxed problem is the same as in the original problem. Based on this, similar arguments to the comparison of pure vs.\ mixed bundling yield that pure bundling achieves the optimal convergence rate.

\subsection{Related Literature}\label{sec:literature}

We relate to the large classical literature on multi-dimensional screening \citep[][among many others]{adams1976, mcafee1989, armstrong1996, rochet1998, manelli2006, manelli2007}.
Several of these papers develop general methodologies to characterize optimal mechanisms. However, as discussed in the Introduction, optimal mechanisms are complicated unless specific conditions are imposed on type distributions. Some papers derive conditions under which pure bundling is optimal \citep[e.g.,][]{pavlov2011, daskalakis2017, haghpanah2021, ghili2023}.\footnote{Other recent work \citep[e.g.,][]{yang2023, bergemann2021} provides conditions under which more general menus of bundles (e.g., nested bundling) are optimal.} We allow for general type distributions, and pure bundling is in general suboptimal at any $n$. Instead, departing from the criterion of exact optimality, we provide a rationale for pure bundling based on the idea that this simple mechanism allows a seller who has precise information about buyers' types to approximate the first-best revenue at the optimal rate. 

%

Our approach of studying a seller with precise information can be contrasted with a recent strand of the literature that considers settings with minimal seller information: the seller perceives a large \emph{set} of possible type distributions and maximizes her worst-case revenue across this set \citep[e.g.,][]{carroll2017, carroll2019, deb2021, che2021}. Optimal mechanisms in these settings can be simple, but the form (e.g., separate sales in \citet{carroll2017} vs.\ pure bundling in \citet{deb2021}) depends on the structure of the set of type distributions.

A different notion of approximate optimality is often studied in computer science: worst-case guarantees, i.e., lower bounds on the performance ratio of simple vs.\ optimal mechanisms that are uniform with respect to type distributions \citep[or other features of the environment; see the survey by][]{roughgarden2019}.\footnote{Our question also differs from work that studies how fast single- or multi-good sellers who observe samples from an unknown \emph{distribution} of buyer types can approximate the second-best (i.e., known distribution) revenue \citep[e.g.,][]{cole2014, gonczarowski2021}.} Neither pure bundling, separate sales, nor more generally the class of all deterministic mechanisms admit a non-zero worst-case guarantee \citep{hart2019selling}.\footnote{However, positive worst-case guarantees can be derived under more restrictive assumptions on the environment \citep{hart2017, babaioff2020}.}

While we fix a set of goods $G$, \cite{armstrong1999} and \cite{bakos1999} study the many-good limit. Under independent and additive valuations, they show that pure bundling approximates the first-best as $|G| \to \infty$, because the value of the grand bundle becomes deterministic by a law of large numbers; in contrast, separate sales does not, because the value of individual goods is random.\footnote{\cite{fang2006} provide joint non-asymptotic conditions on the number of goods and type distribution under which pure bundling outperforms separate sales and vice versa.} In our setting, bundling also reduces uncertainty relative to separate sales, but in a different sense: Regardless of the type distribution, both bundling and separate sales approximate the first-best as the amount of consumer data grows large, because this data allows the seller to learn the valuations for all goods; however, the rate of convergence under bundling is always faster than under separate sales. More importantly, we show that the convergence rate under bundling is the same as under the optimal mechanism.

As noted, a key step of our proof is to upper-bound the optimal revenue via a relaxed problem that partitions buyer types into line segments. Exploiting the linear structure of this problem, Proposition~\ref{prop:simplemech} shows that optimal mechanisms involve at most one random allocation (that is consumed only by types with binding IR). This almost deterministic structure makes optimal mechanisms tractable to analyze and is notable, as even under one-dimensional types, optimal mechanisms with multi-dimensional allocations in general rely on more extensive randomization. As we discuss, this result may be useful in economic settings beyond multi-good monopoly. \cite{haghpanah2021} impose a stochastic monotonicity assumption on type distributions and use this to construct a different relaxed problem: There, types are decomposed into (not necessarily linear) one-dimensional paths and the optimal mechanism is pure bundling.\footnote{For related decomposition approaches, see also \cite{armstrong1996} and \cite{wilson1993}.}

More broadly, we contribute to a resurgent literature on price discrimination, which studies the implications of a seller's ability to condition selling mechanisms on additional information about buyers \citep[e.g.,][]{BBM, haghpanah2023}. 
Motivated by the availability of rich consumer data in digital marketplaces, some papers focus on sellers with very precise information \citep[e.g.,][]{rhodes2024}. In the context of multi-good monopoly, we highlight that sufficiently precise information can serve as a rationale for particular classes of simple selling mechanisms, such as pure bundling. An important step in our analysis is the insight that, under single-good monopoly, such information leads to a stark imbalance between intensive and extensive-margin considerations.

Finally, convergence rates to a perfect information benchmark have been used as a performance measure in other contexts: In single-agent decision problems, \cite{moscarini2002} study how fast an agent who observes many i.i.d.\ signals about the state and follows an optimal strategy can approximate the perfect information payoff; they use this to define a ranking over signal structures. Subsequent work conducts related exercises in other learning settings.\footnote{This includes multi-agent learning and misspecified learning \citep[e.g.,][]{harel2021, frick2022, FII21}.} Aside from our different economic environment and research question, a technical departure from these papers is that our setting features continuous actions and types. This necessitates different mathematical techniques from the large-deviation theory tools that apply in the finite state and action settings of these papers. This technical difference also applies relative to our previous work, \cite{FIImonitoring}, which uses convergence rates to evaluate the performance of simple but suboptimal contracts in a moral hazard problem where a principal observes rich monitoring data about an agent's action. The ``hidden action'' nature of that problem makes the analysis quite different from the current ``hidden type'' setting in other respects as well.\footnote{More broadly, convergence rates to full efficiency have been used to study the performance of simple mechanisms or strategies in the context of trade in large markets \citep[e.g.,][]{rustichini1994} and repeated games with patient players \citep{sugaya2023}.}

\section{Setting}\label{sec:model}

A monopolistic seller (``she'') is endowed with a finite set $G$ of indivisible goods. The seller is risk-neutral, attaching zero value to the goods herself and seeking to maximize her expected revenue from selling them. There is one potential buyer (``he''), described by a type vector $\theta \in\mathbb{R}^{|G|}_{++}$ whose $g$th entry $\theta_g$ represents his valuation of good $g$. The type $\theta$ is drawn from a prior density $h$ whose support is some compact set $\Theta\subseteq \mathbb R_{++}^{|G|}$ with non-empty interior. The buyer's utility from receiving the bundle $B \subseteq G$ and paying a monetary transfer $t$ to the seller takes the additively separable form
\[
\sum_{g\in B}\theta_g-t = \mathbf{1}^B \cdot \theta - t,
\]
where $\mathbf{1}^B \in \mathbb{R}^{|G|}$ denotes the indicator vector on bundle $B$ (i.e., $\mathbf{1}^B_g := \mathbf{1}_{\{g \in B\}}$ for each $g \in G$). Section~\ref{sec:extensions} incorporates nonadditive buyer utilities, negative valuations, and seller production costs into the analysis.


While the realization of $\theta$ is only known to the buyer, the seller has access to fairly precise (but imperfect) information about the buyer's type. As discussed in the Introduction, such information may represent various observable consumer data that is correlated with $\theta$ within the population from which the buyer is drawn (e.g., past purchasing data in Remark~\ref{rem:data} below). For tractability, our main focus is on the following formulation of seller information (Section~\ref{sec:extensions} discusses more general formulations): Let $X$ be a measurable space of signals that is endowed with some $\sigma$-finite measure. The seller observes $n$ signals, $x^n = (x_1, \ldots, x_n)$, that are drawn i.i.d.\ conditional on $\theta$ from a distribution $P_\theta \in \Delta(X)$ with density $f(\cdot, \theta)$. Thus, the seller's information is more precise the richer her amount of data $n$, and we will be interested in settings where $n$ is large.

Upon observing the signals $x^n$, the seller updates her prior $h$ about the buyer's type. She then chooses a direct mechanism, which asks the buyer to report his type and, as a function of this report, specifies allocation probabilities $q:\Theta\to \Delta(2^G)$ for each bundle along with an expected transfer $t:\Theta\to \mathbb R$.\footnote{This formulation rules out asking the buyer to place ``bets'' \citep[\`a la][]{cremer1988} on the realization of $x^n$. This is in line with the above interpretation of $x^n$ as a collection of personal data that is known to the buyer.} Denote by $q (B; \theta)$ the allocation probability of bundle $B \subseteq G$ to type $\theta$.

 In the second-best problem, the seller chooses $q$ and $t$ conditional on each signal sequence $x^n$ to maximize her expected revenue
 \begin{equation}\label{eq:SB}
R^{\rm SB}_{x^n}:=\sup_{q, t}\mathbb E[t(\theta) \mid x^n]
\end{equation}
subject to the incentive compatibility (IC) and individual rationality (IR) constraints 
\begin{equation}\tag{IC}\label{eq:IC}
\sum_{B\subseteq G}q(B; \theta) \left(\mathbf{1}^B \cdot \theta \right) -t(\theta) \geq \sum_{B\subseteq G}q(B; \theta')\left(\mathbf{1}^B \cdot \theta \right) -t(\theta'), \;\; \forall\theta,\theta' \in \Theta,
\end{equation}
\begin{equation}\tag{IR}\label{eq:IR}
\sum_{B\subseteq G}q(B; \theta)\left(\mathbf{1}^B \cdot \theta \right) -t(\theta) \geq 0, \;\; \forall \theta \in \Theta.
\end{equation}
The seller's \textbf{\textit{second-best revenue}} is then the ex-ante expectation $R_n^{\rm SB}:={\mathbb E}[R^{\rm SB}_{x^n}]$.

As discussed in the Introduction, the solution to the second-best problem (i.e., the optimal mechanism) is in general complicated. Thus, we are interested in understanding how well the seller can perform at large $n$ by using simpler mechanisms. To evaluate the performance of different classes of mechanisms, we do not explicitly derive the corresponding revenues at any given $n$. Instead, we analyze the convergence rates of these revenues to the first-best as $n$ grows large. Formally, denote by $R^{\rm FB}:=\mathbb{E}\left[ \sum_{g \in G} \theta_g \right]$ the \textbf{\textit{first-best revenue}}, i.e., the expected revenue the seller can achieve when she directly observes the buyer's type and thus can fully extract his valuation $\theta_g$ for each good $g$. Note that, as $n$ grows large, the second-best revenue converges to the first-best, i.e., $\lim_{n \to \infty} R_{n}^{\rm SB} = R^{\rm FB}$, as observing infinitely many signals perfectly reveals the buyer's type to the seller given Assumption~\ref{asp}.4 below. Moreover, by definition, the rate at which $R_{n}^{\rm SB}$ converges to $R^{\rm FB}$ represents the optimal (i.e., fastest) convergence rate of the seller's revenue to the first-best across all IC-IR mechanisms. 

Convergence rates to the first-best provide a parsimonious way to compare the performance of simple but suboptimal mechanisms: Whenever $n$ is large enough, mechanisms whose revenues converge to the first-best faster yield higher expected revenues than mechanisms with a slower convergence rate. At the same time, achieving the optimal convergence rate to the first-best is a less demanding criterion than maximizing revenue at each $n$. Thus, there may be simpler classes of mechanisms than the second-best that nevertheless attain the optimal convergence rate.

We will be particularly interested in the following two classes of simple mechanisms:
Under \textbf{\textit{pure bundling}} (which we often refer to as \textbf{\textit{bundling}} for short), the seller posts a single price $p (G)$ for the entire grand bundle $G$ of goods and makes a sale only if the buyer is willing to purchase all goods at this price. Under \textbf{\textit{separate sales}}, the seller posts a separate price $p (g)$ for each good $g$ and sells $g$ to any buyer willing to pay this price.\footnote{Under bundling, $q (G;\theta) = \mathbf{1}_{ \{ \sum_{g \in G} \theta_g \geq p (G) \} }$,  $q (\emptyset;\theta) = \mathbf{1}_{ \{ \sum_{g \in G} \theta_g < p (G) \} }$, $t (\theta) = \mathbf{1}_ { \{ \sum_{g \in G} \theta_g \geq p (G) \} } p (G)$ for all $\theta$. Under separate sales, $q (B; \theta) = \mathbf{1}_{ \{\theta_g \geq p (g) \forall g\in B \text{ and }  \theta_{g'} < p (g') \forall g'\not\in B\} }$, $t(\theta) = \sum_{g \in G} \mathbf{1}_{ \{ \theta_g \geq p (g) \}} p (g)$ for all $\theta$.} Let $R_{x^n}^{\text{bd}}$ (resp.\ $R_{x^n}^{\text{sep}}$) denote the seller's expected revenue when in (\ref{eq:SB}) she is restricted to optimizing over bundling (resp.\ separate sales) mechanisms. Let $R_n^{\text{bd}} := \mathbb{E} [R_{x^n}^{\text{bd}}]$ and $R_n ^{\text{sep}} := \mathbb{E} [R_{x^n}^{\text{sep}}]$.

  Throughout, we impose the following regularity conditions. For technical convenience, we extend the signal distribution $P_\theta$ and signal density $f(\cdot,\theta)$ to all types in some compact neighborhood $\hat \Theta \supseteq \Theta$ with $\hat\Theta \subseteq \mathbb{R}^{|G|}_{++}$:

\begin{assume}\label{asp} \ 
\begin{enumerate}
\item  The prior density $h$ is strictly positive and locally Lipschitz continuous for all $\theta \in \Theta$.

\item The signal densities $f(x,\theta)$ are strictly positive and bounded in $(x, \theta)$, and $C^2$ in $\theta\in {\rm int} \, \hat\Theta$. Moreover, for all $g,g'\in G$,  $\frac{\partial^2 \ln f(x, \theta)}{\partial\theta_g\partial\theta_{g'}}$ is bounded in $(x, \theta)$ and Lipschitz continuous in $\theta$ uniformly across $x$.

\item We have $\sup_{\theta\in\Theta} \int \left(\inf_{\theta'\in\Theta}\ln f(x,\theta')\right)^2 \, dP_\theta(x)<\infty$.\footnote{A simple sufficient condition for Assumption~\ref{asp}.3 is if $f(x, \theta)$ is bounded away from 0. When $X$ is finite (e.g., in Remark~\ref{rem:data}), the latter condition is implied by Assumption~\ref{asp}.2.} 

\item The \textbf{\textit{Fisher information}} matrix $I(\theta)\in{\mathbb R}^{|G|\times |G|}$, given by
 \begin{equation}\label{eq:Fisher}
I(\theta) := \left(-\int \frac{\partial^2}{\partial \theta_g \partial \theta_{g'}}\ln f(x,\theta) \, dP_\theta(x)\right)_{g,g' \in G},
\end{equation}
 is well-defined and positive definite for each $\theta\in\hat\Theta$. 
 \end{enumerate}
 \end{assume}

The main content of Assumption~\ref{asp} is that the log-signal densities $\ln f(x, \theta)$ are sufficiently well-behaved. Most importantly, Assumption~\ref{asp}.4 requires that, conditional on $\theta$, their expected curvatures $\frac{\partial^2}{\partial \theta_g \partial \theta_{g'}} \ln f(x,\theta)$ with respect to changes in $\theta$ are well-defined. The Fisher information matrix $I(\theta)$, which summarizes these curvatures, is a standard statistical measure of how informative signals are about type $\theta$ and will feature prominently in our analysis. Fisher information provides a local approximation of the Kullback-Leibler divergence between the signal distributions at type $\theta$ vs.\ nearby types $\theta'$, as ${\rm KL}(P_\theta, P_{\theta'}):=\int \ln \frac{f(x,\theta)}{f(x,\theta')} \, dP_\theta(x) =(\theta-\theta')\cdot I(\theta)(\theta-\theta')+o(\|\theta-\theta'\|^2)$. Imposing positive definiteness on each $I(\theta)$ implies that ${\rm KL}(P_\theta, P_{\theta'})>0$ for all $\theta\not=\theta'$, so signal distributions differ across different types.
For example, if signals conditional on each $\theta$ are distributed Gaussian with mean $\theta$ and a fixed positive definite covariance matrix $\Sigma$, then $I(\theta) = \Sigma^{-1}$ for all $\theta$, and positive definiteness rules out perfect (positive or negative) correlation across signal coordinates.

\begin{rem}[{\bf Past purchase data}]\label{rem:data} As a stylized example of the kind of data that signals $x_i$ might represent, suppose that for each good $g \in G$, the seller has access to $n$ observations of the buyer's past purchase decisions for similar goods (``$g$-variants''). Formally, signals $x_i$ ($i = 1, 2, \ldots, n$) are vectors drawn from $X = \{0, 1\}^{|G|}$, where $x_{ig} =1$ (resp.\ $x_{ig} = 0$) means that the buyer purchased (resp.\ did not purchase) the $i$th $g$-variant. Assume that $x_{ig} = 1$ if and only if $\theta_g + \varepsilon_{ig} > p_g$. Here, $\varepsilon_{ig}$ is an unobserved mean-zero error term that is i.i.d.\ across $i$ and $g$ and captures how similar the buyer's valuation of the $i$th $g$-variant is to his valuation $\theta_g$ of good $g$; and $p_g > 0$ is an observed price, which we assume to be fixed across observations $i$.\footnote{For simplicity, we treat $g$-variants and their prices as exogenous. It is straightforward to allow for prices $p_{ig}$ that are i.i.d.\ across $i$ by incorporating price observations into the signal space} Assuming that $\varepsilon_{ig}$ follows a logistic distribution with scale parameter $1/\beta$ ($\beta > 0$), the signal density (with respect to the counting measure on $X$) is given by $f(x, \theta)=\prod_{g\in G} \frac{x_g\exp[\beta(\theta_g-p_g)]+1-x_g}{\exp[\beta(\theta_g-p_g)]+1}$ for all $x \in X$. Thus, the Fisher information matrix is a diagonal matrix with $(I(\theta))_{gg}=
\beta^2 \frac{\exp[\beta(\theta_g-p_g)]}{(\exp[\beta(\theta_g-p_g)]+1)^2}$ for all $g \in G$.
Observe that Assumption~\ref{asp} is satisfied. \finex
\end{rem}

\section{Main Result}

\subsection{Optimal Convergence Rate}\label{sec:convergence}

Our main result, Theorem~\ref{thm:main} below, shows that bundling allows the seller to achieve the optimal convergence rate to the first-best, whereas the convergence under separate sales is slower. Naturally, how fast the seller can approximate the perfect information benchmark depends on how informative her signals are about the buyer's type. By explicitly characterizing these convergence rates in terms of the underlying signal distributions, Theorem~\ref{thm:main} also sheds light on the relevant measure of informativeness in this setting.

To formalize this, the Fisher information matrix $I(\theta)$ introduced in Assumption~\ref{asp}.4 plays a key role. By the Bernstein-von Mises theorem, this measure is closely tied to the covariance of the seller's posterior at large $n$: Conditional on (almost) any true type $\theta^*$, the seller's posterior as $n \to \infty$ is approximated by a normal distribution $\mathcal{N} \left(\theta^*, \frac{1}{n} I(\theta^*)^{-1} \right)$ whose covariance matrix is the scaled down inverse Fisher information $\frac{1}{n} I(\theta^*)^{-1}$. Consequently, for any bundle $B$, the standard deviation of the seller's posterior about the buyer's valuation $\mathbf{1}^B \cdot \theta$ of $B$ is approximated by 
$\frac{\lambda^B (\theta^*)}{ \sqrt{n}}$, where 
$$\lambda^B(\theta^*) :=\sqrt{{\mathbf 1}^B\cdot I(\theta^*)^{-1}  {\mathbf 1}^B}. $$
An important property of standard deviation is subadditivity. In particular, if $|G| \geq 2$, then letting $\lambda^g (\theta^*) := \lambda^{\{g\}} (\theta^*)$ for each $g \in G$, we have
\begin{equation}\label{eq:subadd}
\lambda^G (\theta^*) < \sum_{g \in G} \lambda^g (\theta^*),
\end{equation}
where the inequality is strict by the positive definiteness of $I (\theta^*)$.\footnote{Since $I(\theta^*)$ is symmetric and positive definite, $\|y\|_{\theta^*}:=\sqrt{y\cdot I(\theta^*)^{-1}y}$ yields a norm on $\mathbb{R}^{|G|}$ induced by an inner product. Thus, for any non-empty and disjoint $B, B'\subseteq G$,  the triangle/Minkowski inequality implies $\lambda^{B\cup B'}(\theta^*)=\|{\bf 1}^{B\cup B'}\|_{\theta^*}<\|{\bf 1}^B\|_{\theta^*}+\|{\bf 1}^{B'}\|_{\theta^*}=\lambda^{B}(\theta^*)+\lambda^{B'}(\theta^*)$.}

Let $\lambda^B := \mathbb{E} [\lambda^B (\theta)]$ for each bundle $B$.  For any sequences of real numbers $x_n$, $y_n$ with $\lim_{n \to \infty} x_n = \lim_{n \to \infty} y_n = 0$, write $x_n\sim y_n$ if $\lim_{n\to\infty}\frac{x_n}{y_n}=1$, i.e., if $x_n$ and $y_n$ vanish equally fast as $n$ grows large.

\begin{thm}\label{thm:main}Under the second-best mechanism and optimal bundling mechanism, the revenue gap to the first-best vanishes equally fast:
\begin{equation}\label{eq:SB-conv}
R^{\rm FB}-R^{\rm SB}_n  \sim   R^{\rm FB}-R_n^{\text{\rm bd}} \sim \lambda^G \sqrt{\frac{\ln n}{n}}. 
\end{equation}
Under separate sales, the revenue gap to the first-best vanishes more slowly:
\begin{equation}\label{eq:sep-conv}
R^{\rm FB}-R_n^{\text{\rm sep}} \sim \sum_{g\in G}\lambda^g \sqrt{\frac{\ln n}{n}}.
\end{equation}
\end{thm}

By (\ref{eq:SB-conv}), when the seller optimizes over all mechanisms, the gap between her revenue $R^{\rm SB}_n$ and the first-best revenue $R^{\rm FB}$ vanishes as fast as $\lambda^G \sqrt{\frac{\ln n}{n}}$ as the amount of data $n$ grows large. Crucially, (\ref{eq:SB-conv}) also shows that the seller can achieve this same optimal convergence rate to the first-best under pure bundling, because $R^{\rm FB}-R_n^{\text{\rm bd}}$ likewise vanishes as fast as $\lambda^G \sqrt{\frac{\ln n}{n}}$. In both cases, how fast the seller can approximate the first-best depends on the signal distribution only through the coefficient $\lambda^G$, which, as noted above, captures how fast the seller's posterior standard deviation about the value $\mathbf{1}^G \cdot \theta$ of the grand bundle vanishes.

To interpret this result, note that at each $n$, pure bundling is in general suboptimal, as optimal mechanisms may involve menus of bundles and/or lotteries over bundles. However, the result shows that bundling is an effective way for the seller to exploit rich consumer data, providing a rationale for this simple class of mechanisms. As $n$ grows large, any revenue gain from optimizing over general mechanisms vs.\ pure bundling is second-order: By (\ref{eq:SB-conv}), $R^{\rm SB}_n$ exceeds $R_n^{\text{\rm bd}}$ by a term that is $o\left(\sqrt{\frac{\ln n}{n}} \right)$, but such a term reduces the revenue gap to the first-best, $\lambda^G \sqrt{\frac{\ln n}{n}}$, by only a negligible amount at large $n$.\footnote{For any $f, g : \mathbb{N} \to \mathbb{R}_{++}$, write $f (n) = o(g(n))$ if $\lim_{n \to \infty} \frac{f(n)}{g(n)} = 0$.} One way to quantify this is by saying that at large $n$, the benefit of optimizing over general mechanisms vs.\ pure bundling is smaller than the benefit of having access to even an arbitrarily small fraction of additional signals: For any $\varepsilon > 0$, no matter how small, we have
$$R^{\rm bd}_{\lceil(1+\varepsilon)n\rceil} > R^{\rm SB}_n \, \text{ for all large enough } n,$$
i.e., the seller is better off under pure bundling with $(1 + \varepsilon)n$ signals than under the optimal mechanism with $n$ signals.\footnote{This has a similar flavor to \cites{bulow1996} seminal (non-asymptotic) result that, in independent private value auctions, an auctioneer's benefit of using optimal vs.\ second-price auctions is smaller than the benefit of having just one additional bidder.}

In contrast, (\ref{eq:sep-conv}) shows that under separate sales the seller's revenue approximates the first-best more slowly: $R^{\rm FB}-R_n^{\text{\rm sep}}$ only vanishes as fast as $\sum_{g\in G}\lambda^g \sqrt{\frac{\ln n}{n}}$, and by (\ref{eq:subadd}), $\sum_{g\in G}\lambda^g> \lambda^G$ for $|G|\geq 2$. Thus, whereas for small $n$, the seller's revenue may be higher under separate sales or bundling depending on parameters, Theorem~\ref{thm:main} implies that bundling always outperforms separate sales when the seller has access to rich enough consumer data. Indeed, at all large enough $n$, the revenue gap to the first-best is $\frac{\sum_{g\in G}\lambda^g}{\lambda^G}$ times bigger under separate sales than under bundling (or the optimal mechanism). Thus, to attain the same revenue as under bundling with $n$ signals, a seller who uses separate sales would need to have access to at least $\left(\frac{\sum_{g\in G}\lambda^g}{\lambda^G}\right)^2 n$ signals. Depending on parameters, the ratio $\frac{\sum_{g\in G}\lambda^g}{\lambda^G}$ can be arbitrarily large:

\begin{ex}\label{ex:gauss}
Suppose $G = \{1,2\}$ and signals $x_i \in \mathbb{R}^2$ are distributed Gaussian with mean $\theta$ and covariance matrix $\Sigma = \left(\begin{array}{cc}\sigma^2 &  \rho \sigma^2 \\  \rho \sigma^2 & \sigma^2\end{array} \right)$. 
Then $\frac{\lambda^1(\theta)+\lambda^2(\theta)}{\lambda^G(\theta)} = \frac{2\sigma}{\sigma\sqrt{2(1 + \rho)}} > 1$ for all $\theta$. This ratio, and hence the performance gap of bundling vs.\ separate sales, is decreasing in the correlation $\rho$ of signals across goods and approaches $\infty$ as $\rho \to -1$ (while it approaches 1 in the perfect correlation limit $\rho \to 1$).\footnote{This is broadly reminiscent of classical findings \citep[e.g.,][]{adams1976, schmalensee1984} that negatively (resp.\ positively) correlated valuations across goods favor bundling (resp.\ separate sales) under some conditions. However, note that in this example $\rho$ parametrizes \emph{signal} correlations and that bundling outperforms separate sales for \emph{all} $\rho$ when $n$ is large enough.}  \finex
\end{ex}

Another way to understand the contrast between (\ref{eq:SB-conv}) and (\ref{eq:sep-conv}) is by considering the performance ratios relative to the \emph{second-best} under bundling and separate sales (i.e., $\frac{R_n^{\text{\rm bd}}}{R^{\rm SB}_n}$ and $\frac{R_n^{\text{\rm sep}}}{R^{\rm SB}_n}$), which are commonly studied measures of
efficiency. Theorem~\ref{thm:main} shows that $R^{\text{\rm FB}} - R_n^{\text{\rm bd}}$ vanishes at a faster \emph{rate} than $R^{\text{\rm FB}} - R_n^{\text{\rm bd}}$, but the \emph{order} of convergence of these revenue gaps is the same, as they differ only in the constants $\lambda^G$ and $\sum_{g \in G} \lambda^g$ multiplying $\sqrt{\frac{\ln n}{n}}$. However, the fact that $\lambda^G$ also describes the rate of convergence of $R_n^{\text{\rm SB}}$ implies that the performance ratios $\frac{R_n^{\text{\rm bd}}}{R^{\rm SB}_n}$ and $\frac{R_n^{\text{\rm sep}}}{R^{\rm SB}_n}$ have different orders of convergence to $1$: the latter is of order $\sqrt{\frac{\ln n}{n}}$, while the former has a smaller order of magnitude. Thus, if the comparison between
bundling and separate sales is based on these performance ratios (or, equivalently, the revenue gaps $R^{\rm SB}_n - R_n^{{\rm bd}}$ and $R^{\rm SB}_n - R_n^{{\rm sep}}$ to the second-best), then bundling outperforms separate sales not just in terms of the
rate but in terms of the order of convergence:

\begin{cor}\label{cor:performance-ratio}
The performance ratio to the second-best under bundling has a faster order of convergence than that under separate sales:
$$1 - \frac{R_n^{\text{\rm bd}}}{R^{\rm SB}_n}=o\left(\sqrt{\frac{\ln n}{n}} \right), \quad \text{ while } \quad 1 - \frac{R_n^{\text{\rm sep}}}{R^{\rm SB}_n}=\sqrt{\frac{\ln n}{n}}\frac{\left(\sum_g\lambda^g-\lambda^G\right)}{R^{\rm FB}}  + o\left(\sqrt{\frac{\ln n}{n}} \right).$$
\end{cor}

We prove Theorem~\ref{thm:main} in Appendix~\ref{app:main-pf}. Section~\ref{sec:gaussian analysis} illustrates the main ideas. Notably, the proof implies that under bundling the seller can achieve the optimal convergence rate to the first-best using a 
pricing algorithm that does not require her to know the prior type distribution: At each realized signal sequence $x^n$, compute the maximum-likelihood estimate $\hat\theta_{x^n}$ of the buyer's type; then price the grand bundle at
\begin{equation}\label{eq:price-bd}
p_{x^n}  ={\bf 1}^G\cdot \hat\theta_{x^n}-\sqrt{\frac{\ln n}{n}} \sqrt{ \mathbf{1}^G \cdot I_{x^n}(\hat\theta_{x^n})^{-1} \mathbf{1}^G}.
\end{equation}
Here $I_{x^n}(\cdot)$ denotes the empirical Fisher information, which replaces the expectation with respect to $P_\theta$ in (\ref{eq:Fisher}) with the sample average (see Appendix~\ref{app:approximation}). Thus, under (\ref{eq:price-bd}), the price depends only on two statistics: The seller's best estimate ${\bf 1}^G\cdot \hat\theta_{x^n}$ of the buyer's valuation of the grand bundle, shaded by (an amount that approximates) the standard deviation of this valuation scaled by $\sqrt{\ln n}$. That the price (\ref{eq:price-bd}) can be set independently of the prior reflects the fact that the effect of the prior on the seller's (approximately Gaussian) posterior at large $n$ is negligible relative to $\sqrt{\frac{\ln n}{n}}$ (see Lemma~\ref{lem:gauss-approx} below). While (\ref{eq:price-bd}) need not coincide with the optimal pure bundling price (which does not typically admit a closed-form solution, even when the seller's posterior is exactly Gaussian), we show that the induced expected revenue gap $R^{\rm FB}-{\mathbb E}[\mathbf{1}_{\{p_{x^n}\leq {\bf 1}^G\cdot \theta\}}p_{x^n}]$ again vanishes at the optimal rate $\lambda^G   \sqrt{\frac{\ln n}{n}}$ (see Section~\ref{sec:1dim}).

\begin{rem}\label{rem:thm}
{\bf Convergence rates of other simple mechanisms.} Theorem~\ref{thm:main} highlights that the optimal convergence rate is achieved by a particularly simple, albeit stark, class of mechanisms---pure bundling. However, a fortiori, this implies that the optimal convergence rate is also attained by optimizing over other simple and less extreme classes of mechanisms that generalize pure bundling. This includes commonly observed mechanisms such as (i) nested bundling (the seller offers a menu of bundles that are nested by set inclusion) and (ii) two-part tariffs (the seller charges an entry fee along with a usage fee for each good).\footnote{Pure bundling corresponds to the special case of (ii) where the entry fee is the price of $G$ and the usage fee is $0$ for all $g$.} Like pure bundling, these mechanisms involve no randomization and only limited menus of bundles and are in general suboptimal at any fixed $n$. At the same time, the suboptimal convergence of separate sales extends to mechanisms where the seller chooses a more general partition $B_1, \ldots, B_k$ ($1 < k < |G|$) of $G$ and prices $p(B_\ell)$ for each $\ell$ \citep[e.g.,][]{palfrey1983}: The convergence rate is $\sum_{\ell =1}^k  \lambda^{B_\ell}\sqrt{\frac{\ln n}{n}}$, which is in between that under separate sales and bundling and faster the coarser the partition.

{\bf Countervailing forces to pure bundling.} In various natural settings, Theorem~\ref{thm:main} breaks down as even the first-best may not involve all goods being sold. Section~\ref{sec:extensions} considers two such cases---seller production costs and negative buyer valuations. A generalization of Theorem~\ref{thm:main} remains valid: The optimal convergence rate to the first-best is achieved by mechanisms that, similar to pure bundling, offer just a single bundle (albeit generally a strict subset of $G$) at a take-it-or-leave price. This preserves the message that a seller with rich enough consumer data derives only a negligible benefit from using more complicated menus or lotteries over bundles. \finex
\end{rem}

\subsection{Gaussian Environment and Illustration of Convergence}\label{sec:numerical}

As Theorem~\ref{thm:main} highlights, an advantage of analyzing convergence rates to the first-best is that this allows for sharp comparisons across different mechanisms that do not depend on all the specifics of the environment (e.g., the details of the prior and signal distributions). On the flipside, convergence rate results are silent about revenues at fixed $n$, which would depend on the details of the environment. However, this does not mean that the predictions of Theorem~\ref{thm:main} only apply once $n$ is so large that revenues are very close to the first-best. This can be illustrated by considering numerical examples.

Specifically, we introduce the following deterministic Gaussian setting that will also play a central role in the proof of Theorem~\ref{thm:main}. For any type $\theta^*$, denote by $R^i _{{\rm gauss}, n} (\theta^*)$ ($i \in  \{ {\rm SB}, {\rm bd}, {\rm sep} \}$) the seller's optimal expected revenue (under general mechanisms, bundling, and separate sales) when her belief about the buyer's type is $\mathcal{N} (\theta^*, \frac{1}{n} I(\theta^*)^{-1})$.\footnote{That is, $R^{\rm SB} _{{\rm gauss}, n} (\theta^*)= \sup_{q, t} \mathbb E_{\mathcal{N} (\theta^*, \frac{1}{n} I(\theta^*)^{-1} )} [t(\theta)]$ subject to (\ref{eq:IC})--(\ref{eq:IR}),
and $R^{\rm bd}_{{\rm gauss}, n} (\theta^*)$ ($R^{\rm sep}_{{\rm gauss}, n} (\theta^*)$) is the seller's value when the $\sup$ is over bundling (separate sales) mechanisms.} Thus, whereas in our original model the seller's posterior at each $n$ is stochastic and depends on the realized signal sequence $x^n$, here the seller observes no signals and $n$ parametrizes a deterministic sequence of Gaussian beliefs that concentrate on $\theta^*$ as $n$ grows large. Nevertheless, by the aforementioned Bernstein-von Mises theorem, this deterministic Gaussian setting approximates the seller's expected revenues conditional on type $\theta^*$ in our original model; importantly, this approximation has a faster order of convergence than the order of convergence to the first-best ($\sqrt{\frac{\ln n}{n}}$) in Theorem~\ref{thm:main}:\footnote{Lemma~\ref{lem:gauss-approx} is immediate from Propositions~\ref{prop:gaussian'} and \ref{prop:conditional} in Appendix~\ref{app:main-proofs}.}

\begin{lem}\label{lem:gauss-approx} For almost all $\theta^* \in  \Theta$ and all $i \in \{{\rm SB, bd, sep}\}$, there exists $C_{\theta^*}>0$ such that 
\[
\left| {\mathbb E}[R^i_{x^n}| \theta^*] - R^i _{{\rm gauss}, n} (\theta^*) \right|\leq  C_{\theta^*} \sqrt{\frac{1}{n}} + o\left(\sqrt{\frac{1}{n}} \right).
\]
\end{lem}

\begin{figure}[t]
\begin{center}
\begin{overpic}[unit = 1mm, scale=0.13]{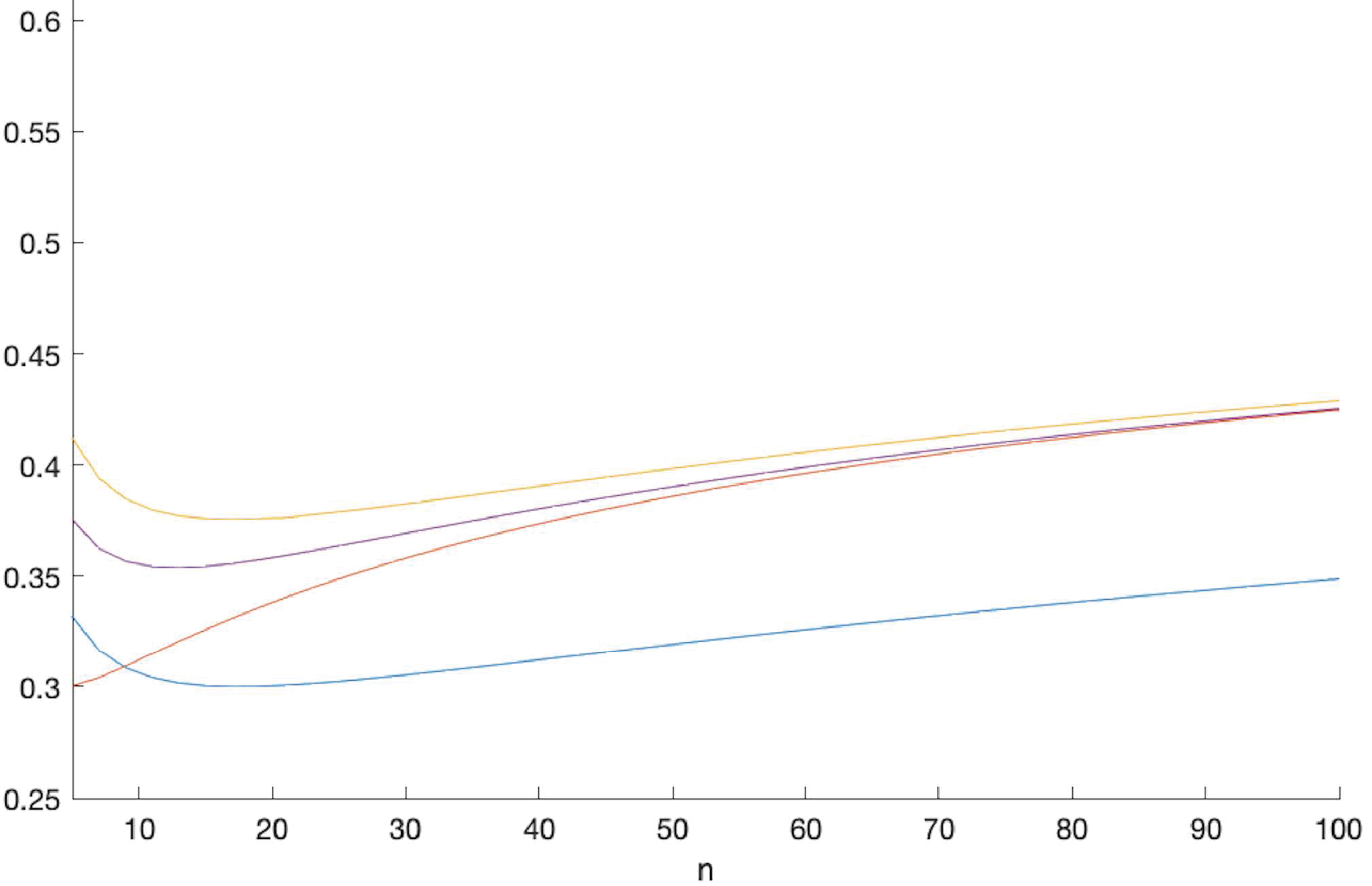}
\put(5.3, 63) {\color{gray} \line(1,0){92}}
\put(46,68){{\tiny $\rho = -0.5$}}
\end{overpic}
\quad
\begin{overpic}[unit = 1mm, scale=0.13]{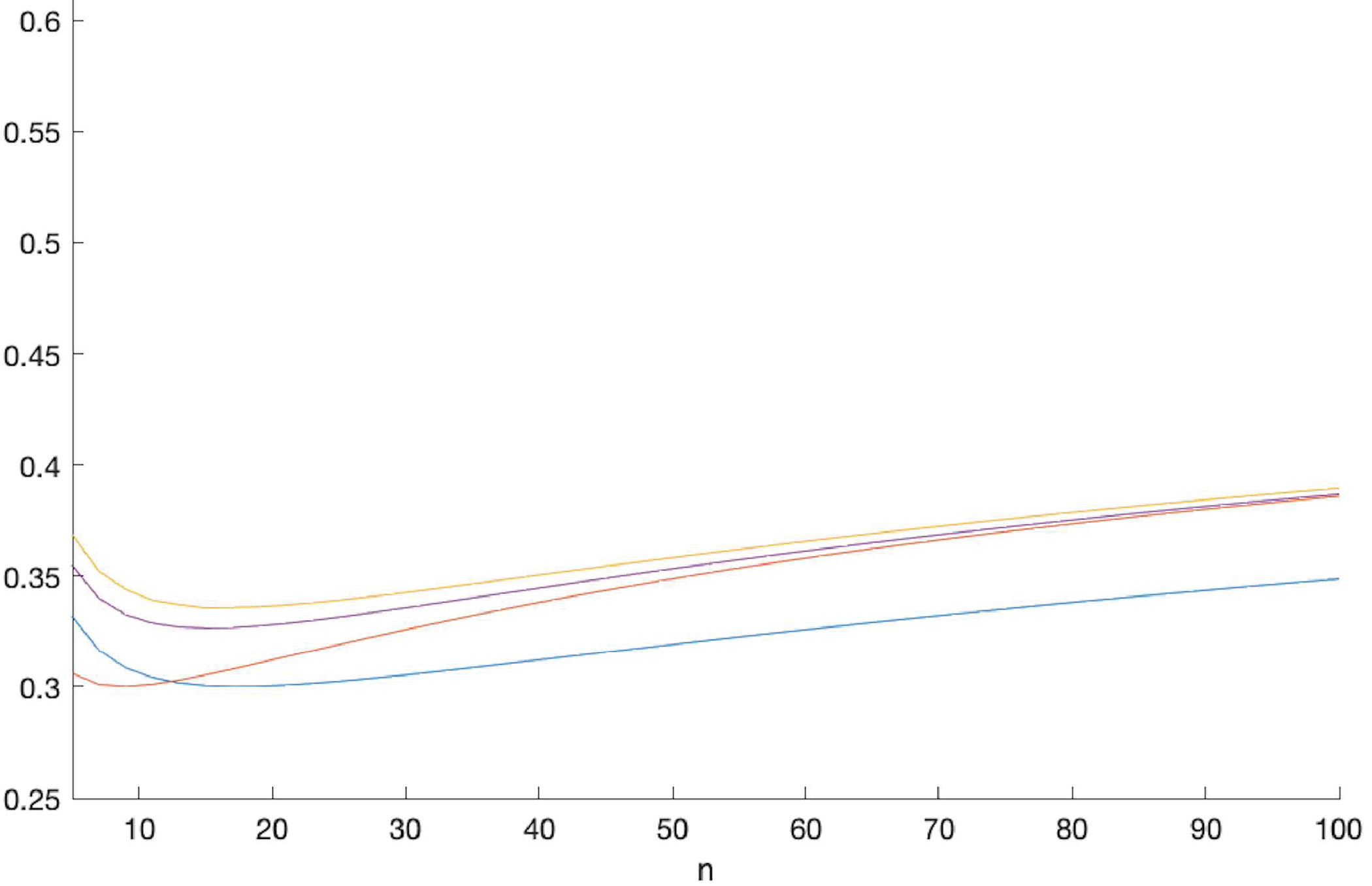} 
\put(5.3, 63) {\color{gray} \line(1,0){92}}
\put(46,68){{\tiny $\rho = 0$}}
\end{overpic}
\quad
\begin{overpic}[unit = 1mm, scale=0.13]{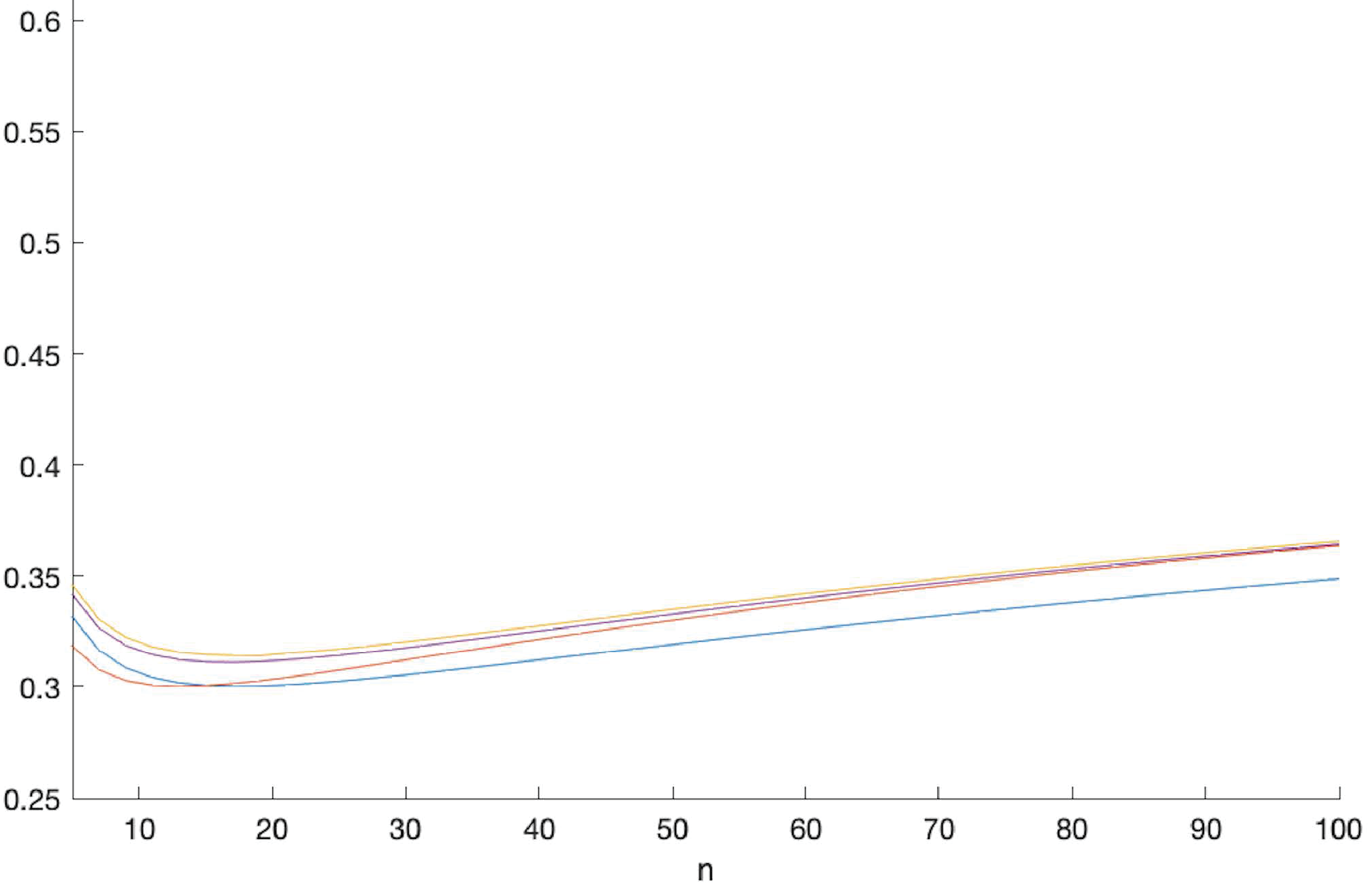}
\put(5.3, 63) {\color{gray} \line(1,0){92}}
\put(46,68){{\tiny $\rho = 0.5$}}
\end{overpic}
\end{center}
\begin{caption}
{\label{fig:simulation} \footnotesize Gaussian setting with $\theta^* = \left(\begin{array}{c}  0.3 \\ 0.3 \end{array} \right)$, $I(\theta^*)^{-1} =  \left(\begin{array}{cc} 1 &  \rho  \\  \rho  & 1 \end{array} \right)$, and $\rho = -0.5$ (left), $\rho = 0$ (middle), or $\rho = 0.5$ (right). We plot revenues $R^{i} _{{\rm gauss}, n} (\theta^*)$ under pure bundling (orange), mixed bundling (purple), separate sales (blue), and the upper bound $\overline R^{\rm SB}_{{\rm gauss}, n} (\theta^*)$ on the second-best revenue (yellow) as a function of $n$. All revenues converge to the first-best $R^{\rm FB} (\theta^*) = 0.6$ (gray) as $n \to \infty$.}
\end{caption}
\end{figure}

The revenues in this Gaussian setting can be studied numerically as a function of $n$. Figure~\ref{fig:simulation} plots the revenues under pure bundling, separate sales, and mixed bundling (i.e., the optimal deterministic mechanism, see Section~\ref{sec:mixed bundling}); it also plots an upper bound on the second-best revenue that we will derive in Section~\ref{sec:vs optimal}.

Figure~\ref{fig:simulation} illustrates all aforementioned implications of Theorem~\ref{thm:main}. First, in line with the slow convergence of revenues to $R^{\rm FB}$ of order $\sqrt{\frac{\ln n}{n}}$, all plotted revenues are well below the first-best of 0.6 throughout the domain of the figure.\footnote{Note that (unlike in our original model) revenues in this deterministic setting are initially decreasing in $n$, as very low $n$ correspond to very high posterior variance, which benefits the seller by allowing her to charge very high prices and focus on extracting surplus from right-tail types.} 
Second, reflecting that $R^{\rm bd}_n$ has a faster order of convergence to $R^{\rm SB}_n$ than $\sqrt{\frac{\ln n}{n}}$ (Corollary~\ref{cor:performance-ratio}), the revenue gap of pure bundling relative to both mixed bundling and the second-best bound all but disappears within the domain of the figure, at a point where these revenues are still only a moderate fraction (less than 70\%) of the first-best. In contrast, since the order of convergence of $R^{\rm sep}_n$ to $R^{\rm SB}_n$ is only $\sqrt{\frac{\ln n}{n}}$,  bundling starts to outperform separate sales very quickly, and the revenue gap between the two remains significant throughout the figure. Finally, reflecting the impact of correlation on the relative performance of separate sales vs.\ bundling (Example~\ref{ex:gauss}), the revenue gap between separate sales and bundling is more pronounced in the left-hand panel of Figure~\ref{fig:simulation} ($\rho = -0.5$) than in the middle panel ($\rho = 0$) and right-hand panel ($\rho = 0.5$).

\section{Main Ideas behind Theorem~\ref{thm:main}}\label{sec:gaussian analysis}

This section illustrates the main ideas behind Theorem~\ref{thm:main}. Section~\ref{sec:gauss} reduces the analysis to the Gaussian setting from Section~\ref{sec:numerical}. 
Section~\ref{sec:1dim} analyzes single-good monopoly, establishing a key result about revenue losses on the intensive vs.\ extensive margins. 
Based on this, we derive the convergence rates under bundling vs.\ separate sales (Section~\ref{sec:bd-sep}), and show that offering menus of bundles does not improve the convergence rate relative to pure bundling
(Section~\ref{sec:mixed bundling}). Finally, Section~\ref{sec:vs optimal} introduces a relaxed problem that yields an upper bound on the second-best revenue. Using this, we show that bundling achieves the same convergence rate as the optimal mechanism.

\subsection{Reduction to Gaussian Setting}\label{sec:gauss}

To prove Theorem~\ref{thm:main}, it suffices to establish the following analog of this result in the deterministic Gaussian setting we introduced in Section~\ref{sec:numerical}. Let $R^{\rm FB}(\theta^*):=\sum_{g\in G} \theta^*_g$.

\begin{prop}\label{prop:gaussian} For all $\theta^* \in \Theta$, we have
\begin{align*}
R^{\rm FB}(\theta^*) - R^{\rm SB}_{{\rm gauss}, n} (\theta^*) \sim  R^{\rm FB}(\theta^*)- R^{\rm bd} _{{\rm gauss}, n} (\theta^*)  &\sim \lambda^G({\theta^*}) \sqrt{\frac{\ln n}{n}},\\
\text{ and } \quad R^{\rm FB}(\theta^*)- R^{\rm sep}_{{\rm gauss}, n} (\theta^*) &\sim \sum_{g\in G}\lambda^g({\theta^*})  \sqrt{\frac{\ln n}{n}}.
\end{align*}
\end{prop}

That Proposition~\ref{prop:gaussian} implies Theorem~\ref{thm:main} follows from the Gaussian approximation in Lemma~\ref{lem:gauss-approx}, along with additional arguments that allow us to go from the conditional revenues in Proposition~\ref{prop:gaussian} to the ex-ante expected revenues in Theorem~\ref{thm:main}.\footnote{Such arguments are needed because the approximation in Lemma~\ref{lem:gauss-approx} is not uniform in $\theta$. } Henceforth, we focus on the Gaussian setting and illustrate the ideas behind Proposition~\ref{prop:gaussian}.

\subsection{Single-Good Monopoly: Intensive vs.\ Extensive Margin}\label{sec:1dim}

First, we analyze the convergence rate to the first-best under single-good monopoly. Suppose $|G| = 1$ and the seller's belief is distributed $ \mathcal{N} (\theta^*, \sigma^2/n)$ for some $\theta^*>0$; the corresponding cdf is $F_n (x) := \Phi \left( \frac{\left(x-\theta^* \right) \sqrt{n}}{\sigma} \right)$, where $\Phi$ is the standard normal cdf.  Under single-good monopoly, the optimal mechanism is a posted price: The seller chooses a price $p_n$ and sells the good to buyers of type $\theta \geq p_n$. Let $R^*_n = \max_p p (1-F_n(p))$ denote the optimal revenue. Let $p^*_n$ denote an optimal price.

%

To understand how fast the revenue gap to the first-best (i.e., $R^{\rm FB} (\theta^*) - R^*_n = \theta^* - R^*_n$) vanishes as $n$ grows large, it is helpful to decompose this gap as follows:
  \[
\theta^* - R^*_n=\theta^* - p^*_n (1-F_n(p^*_n))  = \underbrace{\theta^*-p^*_n}_{\text{intensive margin}}+\underbrace{\theta^* F_n(p^*_n)}_{\text{extensive margin}}
 - \underbrace{(\theta^*-p^*_n)F_n(p^*_n)}_{\text{smaller order terms}}.
 \]
That is, at each $n$, the optimal price $p^*_n$ must trade-off revenue losses on the \textbf{\textit{intensive margin}} (i.e., due to price discounts relative to the first-best) and the \textbf{\textit{extensive margin}} (i.e., due to some buyer types refusing to buy). 
To reduce intensive-margin losses, the seller wants to choose a price sequence $p_n$ that approaches $\theta^*$ as fast as possible. But if the rate at which $p_n \to \theta^*$ is too fast, then extensive-margin losses $F_n (p_n) = \Phi \left( \frac{ \left(p_n - \theta^* \right) \sqrt{n}}{\sigma} \right)$ may vanish too slowly or fail to vanish at all. 

The key observation is that this tradeoff dictates that $p^*_n$ is set in such a way that revenue losses at large $n$ are driven almost entirely by the intensive margin:

\begin{prop}\label{lem:1-dim} 
In the Gaussian environment with $|G| = 1$ and $\theta$ distributed $\mathcal{N} (\theta^*, \sigma^2/n)$ for some $\theta^*>0$, any optimal price sequence $p^*_n$ satisfies
\begin{equation}\label{eq:1-dim R}
R^{\rm FB} (\theta^*) - R^*_n  \sim \theta^*- p^*_n \sim \sqrt{\ln n}\frac{\sigma}{\sqrt{n}} \quad \text{ and } \quad \theta^*F_n(p^*_n)=o\left(\sqrt{\frac{\ln n}{n}}\right).
\end{equation}
Moreover, for any price sequence $p_n$ with $\theta^*-p_n \sim \delta (\theta^*-p^*_n)$ for some $\delta\in [0,1)$, we have $\lim_n \frac{\theta^* F_n (p_n)}{\sqrt{\frac{\ln n}{n}}}=\infty$. 
\end{prop}

Proposition~\ref{lem:1-dim} shows that losses on the extensive margin $\theta^*  F_n(p^*_n)$ become negligible relative to the intensive margin $\theta^* - p^*_n$ at large $n$. Hence, the revenue gap relative to the first-best vanishes at the same rate as the intensive margin. Moreover, this rate takes a simple form, given by the standard deviation$\frac{\sigma}{\sqrt{n}}$ of the seller's posterior scaled by $\sqrt{\ln n}$.\footnote{To understand this rate, note that if the intensive margin is proportional to the standard deviation ($\theta^*-p_n=\kappa \frac{\sigma}{\sqrt{n}}$ for some $\kappa>0$), extensive-margin losses $F_n(p_n)=\Phi \left( \frac{ \left(p_n - \theta \right) \sqrt{n}}{\sigma} \right) =\Phi(-\kappa)$ do not vanish with $n$. For the extensive margin to vanish, $\theta^* - p_n$ must be scaled up, where the role of the scaling factor $\sqrt{\ln n}$ comes from the exponential form of the standard normal distribution.} While there is no closed-form solution for the optimal prices $p^*_n$, the result implies that the same optimal convergence rate is achieved by a simple pricing rule, 
\begin{equation}\label{eq:simple-price}
p_n=\theta^*-\sqrt{\ln n}\frac{\sigma}{\sqrt{n}},
\end{equation}
which shades the price by the amount $\sqrt{\ln n}\frac{\sigma}{\sqrt{n}}$ relative to the posterior mean $\theta^*$.

It may appear counterintuitive that the intensive and extensive margins are highly imbalanced at the optimum. Why can't the seller reduce the intensive margin a little by correspondingly increasing the extensive margin? The ``moreover'' part of Proposition~\ref{lem:1-dim} sheds light on this point: If the intensive margin is scaled down relative to the optimum $\theta^* - p^*_n$ by some factor $\delta < 1$, then the extensive margin explodes relative to the magnitude $\sqrt{\frac{\ln n}{n}}$ of the intensive margin, which is clearly suboptimal. Notably, this occurs even when $\delta$ is arbitrarily close to 1. This reflects the relatively thin tails of the Gaussian distribution, which make the extensive margin become infinitely more elastic than the intensive margin as $n$ grows large  (Appendix~\ref{app:non-gaussian} formalizes this point and extends Proposition~\ref{lem:1-dim} to more general distributions). 

\subsection{Bundling vs.\ Separate Sales: Role of Standard Deviation}\label{sec:bd-sep}

We return to the multi-good setting, $|G|\geq 2$, with seller belief ${\cal N} \left(\theta^*,  \frac{1}{n} I(\theta^*)^{-1} \right)$. Proposition~\ref{lem:1-dim} immediately yields the convergence rates of $R^{\rm bd}_{{\rm gauss}, n} (\theta^*) $ and $R^{\rm sep}_{{\rm gauss}, n} (\theta^*) $, as under both bundling and separate sales, the seller's problem reduces to single-good monopoly. Specifically, bundling is an instance of single-good monopoly, where the good is the grand bundle and the buyer's type is $\sum_{g \in G} \theta_g$. Since the standard deviation of $\sum_{g \in G} \theta_g$ is $\sqrt{\frac{{\mathbf 1}^G\cdot I(\theta^*)^{-1} {\mathbf 1}^G}{n}} = \frac{\lambda^G(\theta^*)}{\sqrt n}$,  (\ref{eq:1-dim R}) yields that the revenue gap relative to the first-best vanishes as fast as $\sqrt{\ln n}\frac{\lambda^G(\theta^*)}{\sqrt n}$. Under separate sales, the seller solves a collection of single-good monopoly problems, one for each good $g$, with corresponding type $\theta_g$. Since the standard deviation of $\theta_g$ is $\sqrt{\frac{{\mathbf 1}^g\cdot I(\theta^*)^{-1} {\mathbf 1}^g}{n}} =\frac{\lambda^g(\theta^*)}{\sqrt{n}}$,  (\ref{eq:1-dim R}) yields that the revenue gap relative to the first-best vanishes as fast as $\sqrt{\ln n} \frac{\sum_{g \in G} \lambda^{g}(\theta^*)}{\sqrt n}$.

Thus, the reason that convergence under bundling is faster than under separate sales boils down to the fact that the standard deviation of the sum of all valuations $\sum_{g \in G} \theta_g$ is less than the sum of the standard deviations of each individual valuation $\theta_g$. By Proposition~\ref{lem:1-dim}, this allows the seller to charge a bundling price $p^{\rm bd}_{n} (G)$ that at large $n$ exceeds the total price $\sum_{g \in G} p^{\rm sep}_{n} (g)$ under separate sales. By the dominance of the intensive margin, this ensures a higher revenue under bundling.

At a high level, this reflects the classic intuition that bundling reduces the seller's uncertainty about the buyer's type relative to separate sales, as overestimating the valuations of some goods but underestimating those of others can cancel out when estimating the valuation of the grand bundle. However, for this intuition to be valid in our setting, it is crucial that Proposition~\ref{lem:1-dim} shows that the relevant measure of uncertainty at large $n$ is \emph{standard deviation}, which is subadditive. The same logic would not have been valid under other natural measures of uncertainty: For example, the variance of $\sum_{g \in G} \theta_g$ can be greater or smaller than the sum of the variances of all $\theta_g$, depending on the correlation across dimensions.\footnote{To illustrate a setting where variance would have been the appropriate measure, consider a generalization of the monopolist's problem in Section~\ref{sec:1dim}, where $\theta$ is distributed $\mathcal{N} (\theta^*, \frac{\sigma^2}{n})$ and the price choice $p$ induces a payoff $u(p, \theta)$. If $u(p, \theta)$ is smooth, Taylor approximation arguments imply that the gap to the first-best payoff is proportional to the variance $\frac{\sigma^2}{n}$. However, in our setting, $u(p, \theta) = p \mathbf{1}_{\{ p \leq \theta \} }$ is discontinuous at $p = \theta$. This makes it more difficult to approximate the first-best payoff, leading to the slower convergence rate $\sqrt{\ln n} \frac{\sigma}{\sqrt{n}}$.}

\subsection{Bundling vs.\ Mixed Bundling}\label{sec:mixed bundling}

It remains to show that general mechanisms, which may involve menus of bundles and randomization, cannot achieve a faster convergence rate than pure bundling. As a first step, we focus on the effect of menus and ignore randomization. IC-IR deterministic mechanisms correspond to \textbf{\textit{mixed bundling}}: The seller sets prices $p (B)$ for all bundles $B$, and the buyer chooses which bundle to purchase.\footnote{We implicitly assume that a buyer cannot purchase multiple bundles at once.}
 Let $\left( p_n^{\rm mix} (B) \right)_{B \subseteq G}$ denote the optimal mixed bundling prices under the seller belief $\mathcal{N} \left(\theta^*,  \frac{1}{n} I(\theta^*)^{-1} \right)$.

Relative to pure bundling, mixed bundling allows the seller to extract revenue from buyers who are unwilling to purchase the grand bundle $G$, i.e., those types with $\sum_{g \in G} \theta_g<p^{\rm bd}_n (G)$. However, while this improves the seller's revenue at each $n$, we show that the benefit in terms of the convergence rate to the first-best is negligible.

The logic relies on the intensive vs.\ extensive margin analysis in Proposition~\ref{lem:1-dim}. Indeed, by (\ref{eq:1-dim R}), the extensive-margin losses under pure bundling from types $\sum_{g \in G} \theta_g<p^{\rm bd}_n (G)$ become negligible relative to the intensive-margin losses $\sum_{g \in G} \theta^*_g - p^{\rm bd}_n (G)$ at large $n$. Thus, the only way for mixed bundling to achieve a first-order improvement over pure bundling would be if mixed bundling allowed the seller to raise the price of the grand bundle in a significant enough way that the intensive-margin losses $\sum_{g \in G} \theta^*_g- p^{\rm mix}_n(G)$ vanish strictly faster than $\sum_{g \in G} \theta^*_g- p^{\rm bd}_n (G)$. However, this is impossible by the ``moreover'' part of Proposition~\ref{lem:1-dim}: Under any such price sequence $p^{\rm mix}_n (G)$, extensive-margin losses (i.e., the rejection probability of the grand bundle) would increase so substantially relative to pure bundling that this would more than offset the improvements on the intensive margin.

\subsection{Bounding the Optimal Revenue via a Relaxed Problem}\label{sec:vs optimal}  

Finally, we show that bundling achieves the same convergence rate as the second-best mechanism. Relative to mixed bundling, the second-best mechanism may involve randomization and, as noted, can be difficult to characterize. To get around this, we instead derive an upper bound on the second-best revenue $R^{\rm SB}_{{\rm gauss}, n} (\theta^*)$ by considering the seller's optimal revenue $\overline R^{\rm SB}_{{\rm gauss}, n} (\theta^*)$ in a more tractable relaxed problem. We then show that even $R^{\rm FB}(\theta^*)-\overline R^{\rm SB}_{{\rm gauss}, n} (\theta^*)$ does not vanish at a faster rate than $R^{\rm FB}(\theta^*)- R^{\rm bd}_{{\rm gauss}, n} (\theta^*)$.

\begin{figure}[t]
\begin{center}
\includegraphics[width=0.5 \textwidth]{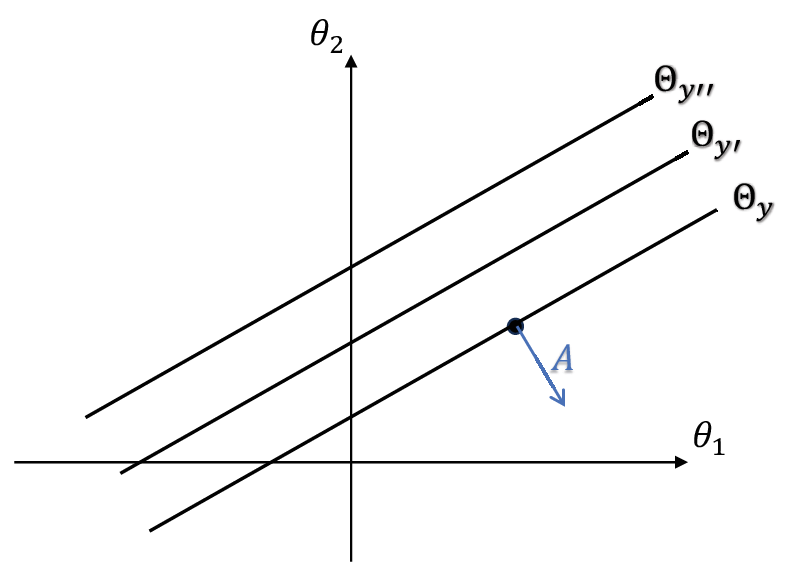}
\end{center}
\begin{caption}
{\label{fig:segmentation} \footnotesize Illustration of segmentation when $|G|=2$.  }
\end{caption}
\end{figure}

To define the relaxed problem, we split the type space into one-dimensional segments: Fix a full-rank matrix $A\in{\mathbb R}^{(|G|-1)\times |G|}$ and partition types into lines $(\Theta_y)_{y\in\mathbb R^{|G|-1}}$, where
\[
\Theta_y:=\{\theta\in \mathbb R^{|G|}: A \theta=y\}\quad \text{ for each } y \in \mathbb R^{|G|-1}.
\]
See Figure~\ref{fig:segmentation} for an illustration when $|G|=2$, in which case $A$ corresponds to a vector in $\mathbb{R}^2$ and each line segment is parametrized by a real number $y$. Given this segmentation, denote by $\overline R^{\rm SB}_{{\rm gauss}, n} (\theta^*)$ the second-best revenue when the IC constraint governing ${R}^{\rm SB}_{{\rm gauss}, n} (\theta^*)$ is relaxed to only hold across types in the same line segment, i.e., we only require that
\begin{equation*}
\sum_{B\subseteq G}q(B; \theta) \left(\mathbf{1}^B \cdot \theta \right) -t(\theta) \geq \sum_{B\subseteq G}q(B; \theta')\left(\mathbf{1}^B \cdot \theta \right) -t(\theta'),  \quad \forall y\in{\mathbb R}^{|G|-1}, \theta, \theta' \in \Theta_y.
\end{equation*}
Thus, it is as if the seller observes each type $\theta$'s segment (i.e., the vector $A\theta$) before choosing a mechanism. Clearly, $\overline{R}^{\rm SB}_{{\rm gauss}, n} (\theta^*) \geq {R}^{\rm SB}_{{\rm gauss}, n} (\theta^*)$.

An advantage of the relaxed problem is that solving for $\overline{R}^{\rm SB}_{{\rm gauss}, n} (\theta^*)$ reduces to a collection of problems with one-dimensional type spaces. That is, $\overline{R}^{\rm SB}_{{\rm gauss}, n} (\theta^*)$ can be written as the average
\[
\overline R^{\rm SB} _{{\rm gauss}, n} (\theta^*) ={\mathbb E} \left[ R^{\rm SB}_{{\rm gauss}, n} (\theta^* \mid y) \right],
\] 
where $R^{\rm SB}_{{\rm gauss}, n} (\theta^* \mid y)$ denotes the second-best revenue when the seller's belief over types is the conditional distribution on segment $\Theta_y$ induced by $\mathcal{N} \left(\theta^*,  \frac{1}{n} I(\theta^*)^{-1} \right)$, and the expectation on the right is with respect to $y$ being distributed ${\cal N}(A \theta^*, \frac{1}{n}A I(\theta^*)^{-1}A^\top)$.

Since each $\Theta_y$ is one-dimensional, analyzing $R^{\rm SB}_{{\rm gauss}, n} (\theta^* \mid y)$ is quite tractable. Section~\ref{sec:1-dim type} below characterizes optimal mechanisms in a more general environment with one-dimensional types (dropping the Gaussian assumption). The key observation is that optimal mechanisms at each $\Theta_y$ are ``almost deterministic:'' In addition to deterministic bundles, they offer (at most) a {\it single} lottery over bundles. Thus, ruling out more complicated mechanisms with infinitely many lotteries, the optimal mechanism at each $\Theta_y$ is a simple modification of mixed bundling. Based on this, a similar logic to the comparison between mixed and pure bundling in Section~\ref{sec:mixed bundling} implies that, at each $\Theta_y$, the seller's second-best revenue $R^{\rm SB}_{{\rm gauss}, n} (\theta^* \mid y)$ does not converge to the first-best faster than does her conditional pure bundling revenue $R^{\rm bd}_{{\rm gauss}, n} (\theta^* \mid y)$. That is, the seller's revenue gain from randomization becomes negligible at large $n$.

A crucial remaining issue is that the pure bundling price at each $\Theta_y$ may in general depend on $y$, so the bundling revenue ${\mathbb E} \left[ R^{\rm bd}_{{\rm gauss}, n} (\theta^* \mid y) \right]$ in the relaxed problem may be greater than the revenue $R^{\rm bd}_{{\rm gauss}, n} (\theta^*)$ in the original problem. However, this issue can be avoided by carefully choosing how to segment the type space in the relaxed problem. Specifically, we choose the matrix $A$ in such a way that the conditional distribution of $\sum_{g \in G} \theta_g$ at every segment $\Theta_y$ is the same as the unconditional distribution of $\sum_{g \in G} \theta_g$. That is, observing the buyer's line segment (i.e., the vector $A\theta$) provides the seller with no additional information about the buyer's valuation of the grand bundle. Picking an $A$ with this property is always possible, since in the current Gaussian setting this is equivalent to the requirement that $A \theta$ and $\mathbf{1}^G \cdot \theta$ are uncorrelated.\footnote{For example, if $|G|=2$ and ${\rm Var} (\theta_1) = {\rm Var} (\theta_2)$, then we can pick $A$ to be the vector $(1, -1)$.}
Under this choice of $A$, the optimal pure bundling price at each $\Theta_y$ is the same as the optimal pure bundling price in the original problem. Hence, the seller's pure bundling revenue in the original problem can be written as 
\[R^{\rm bd}_{{\rm gauss}, n} (\theta^*)= {\mathbb E} \left[ R^{\rm bd}_{{\rm gauss}, n} (\theta^* \mid y) \right]. \]
Combined with the previous paragraph, this yields the desired conclusion that $R^{\rm FB}(\theta^*)-\overline R^{\rm SB}_{{\rm gauss}, n} (\theta^*)$ does not vanish at a faster rate than  $R^{\rm FB}(\theta^*)-R^{\rm bd}_{{\rm gauss}, n} (\theta^*)$.

\section{Analysis of One-Dimensional Type Setting }\label{sec:1-dim type}

This section analyzes (a generalization of) the one-dimensional type environment at the center of the argument in Section~\ref{sec:vs optimal}. As claimed in Section~\ref{sec:vs optimal}, we show that optimal mechanisms take an almost deterministic form. This result may be of independent interest beyond its application in Section~\ref{sec:vs optimal}.

Let $\mathcal{B}$ denote a finite set of allocations, which we enumerate as $\{1,2, \ldots , m\}$. An agent's type is described by an $L^1$ random variable $\tau$ that admits a strictly positive density on some interval $T \subseteq \mathbb{R}$. We normalize types so that $0\in T$.  Each type $\tau\in T$ is associated with a vector $\alpha(\tau) \in \mathbb{R}^m$, where $\alpha_\ell(\tau)$ represents type $\tau$'s utility from allocation $\ell$.  Our key assumption is that $\alpha$ is linear, i.e., there is $\beta\in \mathbb{R}^m$ such that 
$\alpha(\tau) = \alpha(0) + \tau\beta$.

Let $\overline{\beta} := \max\{ \beta_1, \ldots , \beta_m\}$,  $\underline{\beta} := \min \{ \beta_1, \ldots, \beta_m\},$
 and assume there exists some allocation $\bar{\ell}$ for which $\alpha_{\bar{\ell}}(0) = \beta_{\bar{\ell}} = 0$. Section~\ref{sec:vs optimal} is a special case of this environment where $\mathcal{B}$ corresponds to the set of bundles (with $\bar \ell$ the empty bundle), but the current formulation does not require Gaussian type distributions or additive valuations.\footnote{More specifically, the current setting nests Section~\ref{sec:vs optimal} by letting $T = {\mathbb R}$ and choosing $\alpha $ with $\alpha_\ell(0)={\mathbb E}[\theta\cdot {\bf 1}^{B_\ell}| A\theta=y]$ and $\beta_\ell={\bf 1}^{B_\ell} \cdot z$ for each bundle $B_\ell$, where $z\in\mathbb R^G$ is such that $Az={\bf 0}$.}  The current setting also encompasses some other economic applications.\footnote{Indeed, the current linearity assumption can be viewed as an extension of analogous assumptions in one-dimensional screening models \`a la \cite{mussa1978}. \cite{boleslavsky2024} consider a setting where an agent purchases multi-dimensional allocations (influence bundles), whose payoff depends linearly on his one-dimensional type. \label{fn:almostdet}}

A direct mechanism consists of a pair of measurable functions $q:T\to \Delta(\cal B)$ and $t:T\to\mathbb R$. Let $\mathcal{M}$ denote the set of all direct mechanisms $(q, t)$ that satisfy both IC (i.e., $\alpha(\tau)\cdot q(\tau)-t(\tau) \geq \alpha(\tau)\cdot q(\tau')-t(\tau')$ for all $\tau,\tau'\in T$) and IR (i.e., $\alpha(\tau)\cdot q(\tau)-t(\tau) \geq 0$ for all $\tau\in T$). The designer's objective is to maximize the expected transfer ${\mathbb E}[t(\tau)]$ subject to IC and IR.

Since $\beta_{\bar{\ell}} = 0 \in [\underline\beta,\overline\beta]$, we can pick a lottery
\[
b_0\in \argmax_{q \in \Delta(\mathcal{B})}  \, \alpha(0) \cdot q \text{ such that } \beta \cdot q = 0;
\]
that is, $b_0$ yields maximal utility among all lotteries that give the same utility to all types. Let  ${\mathcal D} := \left\{\delta_k : k \in \mathcal{B} \right\}$ denote the set of deterministic allocations. Let 
$\mathcal{M}^s := \left\{(q, t) \in \mathcal{M} : q(T) \subseteq {\mathcal D} \cup \{b_0\} \right\}$ denote the subset of IC and IR mechanisms that involve only allocations that are either deterministic or $b_0$. 
The following result shows that it is without loss of optimality for the designer to restrict attention to such mechanisms.
Note that the choice of the lottery $b_0$ is independent of the type distribution.

\begin{prop}\label{prop:simplemech} We have $\sup_{(q, t) \in \mathcal{M} } \mathbb{E} \left[t(\tau) \right] = \sup_{(q, t) \in \mathcal{M}^s } \mathbb{E} \left[ t(\tau) \right]$.
\end{prop}

We prove Proposition~\ref{prop:simplemech} in Appendix~\ref{app:1-dim}. As in \cite{manelli2007} and \cite{daskalakis2017}, we reformulate the designer's problem as an optimization over agent value functions $V:T\to\mathbb R$ that are induced by IC-IR mechanisms. In contrast to these papers, a challenge is that $V$ does not uniquely pin down the corresponding mechanism (and hence the designer's payoff), as the allocation space is higher-dimensional than the type space.   Thus, we proceed in two steps: First, for each $V$, we optimize over the IC-IR mechanisms that induce $V$; second, we optimize over $V$.  While the objective in the second step is non-linear in $V$, we show that it has a piecewise linear structure. Given this, we verify that it suffices for optimality to consider $V$ functions induced by mechanisms in $\mathcal{M}^s$.

Relative to deterministic mechanisms,  the role of offering lottery $b_0$ is to extract more surplus from types for whom IR binds. To illustrate, consider the case where $ \alpha(0) \cdot b_0>0$ and $b_0$ is a non-degenerate lottery.  Suppose that under an optimal deterministic mechanism,  there is a non-trivial interval $\hat T\subseteq T$ of types who receive the degenerate lottery $\delta_{\overline\ell}$. Clearly, IR binds for these types and they pay nothing to the designer.  But then the designer can strictly improve her payoff by offering $b_0$ at price $ \alpha(0) \cdot b_0 >0$ to these types instead of $\delta_{\overline\ell}$. Under this modification, IC remains valid, because $\alpha(\tau)\cdot b_0=\alpha(0) \cdot b_0$ for all $\tau\in T$ by construction of $b_0$, i.e., all types are indifferent between receiving $b_0$ at price $b_0\cdot \alpha(0)$ or receiving $\delta_{\overline\ell}$ for free.

If instead $\alpha(0) \cdot b_0 = 0$, we can set $b_0 = \delta_{\overline\ell}$, so Proposition~\ref{prop:simplemech} implies the optimality of a deterministic mechanism. A sufficient condition for $\alpha(0) \cdot b_0 = 0$ is if $\beta_\ell \geq 0$ for all $\ell$, i.e., the utility to each allocation is nondecreasing in types.\footnote{This observation contrasts with \cite{ghili2023} and \cite{yang2023}, who consider settings with one-dimensional types but without our linear utility assumption.  In their settings, optimal mechanisms in general require randomization, even though all bundles' values are nondecreasing in types.}



\section{Discussion}\label{sec:dis}

\subsection{Extensions}\label{sec:extensions}

{\bf Production costs/negative valuations.} In our main model, the first-best $R^{\rm FB} = \mathbb{E} [\mathbf{1}^G \cdot \theta]$ involves supplying the grand bundle to all buyer types. This can fail if the seller faces a production cost or some types have negative valuations for some goods. Appendix~\ref{app:cost} extends the analysis to allow for both possibilities: First, to produce each bundle $B$, the seller incurs a cost $c(B)$, so that $R^{\rm FB} = \mathbb{E} \left[\mathbf{1}^{B_\theta} \cdot \theta - c(B_\theta) \right]$ involves supplying a possibly different bundle $B_\theta \in \argmax_{B \subseteq G} \left(\mathbf{1}^{B} \cdot \theta - c(B)\right)$ to each type $\theta$. Second, we drop the assumption that $\Theta\subseteq\mathbb R^{|G|}_{++}$ and instead assume $\mathbf{1}^{B_\theta} \cdot \theta - c(B_\theta)>0$ for all $\theta\in \Theta$, which ensures positive gains from trade.\footnote{If $c \equiv 0$, this simply requires each type to have a positive valuation for at least one good.} In this setting, a simple generalization of pure bundling---\textit{\textbf{single-bundle mechanisms}}---achieves the optimal convergence rate to the first-best: Let $R^{\text{1bd}}_n$ denote the expected revenue when, following each signal sequence $x^n$, the seller optimally chooses a \emph{single} bundle $B$ and fixed price $p (B)$ at which to offer this bundle. Theorem~\ref{thm:main-c} shows that $R^{\rm FB} - R^{\text{1bd}}_n$ vanishes at the same rate as $R^{\rm FB} - R^{\rm SB}_n$, namely as fast as $\mathbb{E} [ \lambda^{B_\theta} (\theta) ] \sqrt{\frac{\ln n}{n}}$.

{\bf Non-additive utilities.} While we focused on buyer utilities that are additive across goods, this is not essential for our main result. Appendix~\ref{sec:nonadd} assumes that the buyer's utility from receiving bundle $B$ and paying transfer $t$ takes the form $\omega_B - t$, where the only restriction on the valuations $\omega_B$ is that $\omega_G>\max\{\omega_{B}, 0\}$ for all $B \subsetneq G$. This allows for fairly general complementarity and substitutability patterns across goods, provided the buyer finds the grand bundle $G$ most attractive.\footnote{This can be further relaxed, similar to the negative type case in the previous paragraph.} Indeed, this abstract formulation can capture other multi-dimensional screening settings beyond multi-good monopoly; for example, bundles $B$ might be reinterpreted as discrete quality levels, where $G$ represents the highest quality. As in our main model, prior to choosing a mechanism, the seller observes $n$ i.i.d.\ signals about the buyer's type, which in this setting corresponds to the vector $\omega: = (\omega_B)_{B \subseteq G} \in \mathbb{R}^{2^G}$. Applying analogous arguments as in Theorem~\ref{thm:main}, Theorem~\ref{thm:main2} shows that the revenue gap to the first-best continues to vanish at the same rate under the second-best and optimal bundling mechanisms, viz., as fast as $\mathbb{E}[\lambda(\omega)] \sqrt{\frac{\ln n}{n}}$, where $\lambda (\omega) :=\sqrt{{\mathbf 1}^G\cdot I(\omega)^{-1}  {\mathbf 1}^G} $.


{\bf More general seller information.} To formalize precise seller information, we assumed that the seller observes $n$ i.i.d.\ signals about the buyer's type. This formulation allowed us to apply the Bernstein-von Mises theorem to reduce the analysis to a setting with a deterministic sequence of Gaussian seller beliefs. While we do not pursue this direction in the current paper, we conjecture that our results can be generalized to certain classes of correlated signals, by applying non-i.i.d.\ versions of the Bernstein-von Mises theorem to extend this Gaussian approximation approach.

As an alternative modeling approach, Appendix~\ref{app:non-gaussian} takes as a primitive a general (not necessarily Gaussian) deterministic sequence of seller beliefs indexed by $n$ that converge to a point-mass on the true type $\theta^*$. Such beliefs can be interpreted as the outcome of the seller observing increasingly precise information that need not take the form of $n$ (i.i.d.\ or correlated) signal draws. Under a tail condition on seller beliefs, we extend the key insight (Proposition~\ref{lem:1-dim}) that intensive-margin losses dominate the extensive margin at large $n$ under single-good monopoly. Based on this, Proposition~\ref{prop:non-Gaussian} shows that pure bundling continues to achieve the same convergence rate to the first-best as the optimal deterministic mechanism.\footnote{We do not analyze non-deterministic mechanisms, but conjecture that the approach via the relaxed problem in Section~\ref{sec:vs optimal} might be extended by exploiting the non-Gaussian analysis in Section~\ref{sec:1-dim type}.}

\subsection{Concluding Remarks}\label{sec:conclusion}

The primary message of this paper is methodological: Studying a multi-good monopolist with precise information about the buyer's valuations can yield sharp insights into an otherwise quite intractable problem. Whereas the comparison between pure bundling and separate sales is not clear-cut in general, we saw that bundling always outperforms separate sales when the seller's data is rich enough. What is more, pure bundling performs essentially as well as the second-best mechanism, because both mechanisms approximate the first-best revenue at the same rate as the amount of data grows rich. As discussed, our approach can be seen as a natural complement to a recent literature that analyzes the worst-case optimality of simple mechanisms when the seller has minimal information about buyer valuations.

As far as the practical relevance of our results is concerned, there is indeed evidence that firms in various settings employ pure bundling (or simple generalizations thereof) in conjunction with some degree of personalization.\footnote{For example, some job matching websites provide all subscribed employers with access to their entire database of job seekers (the grand bundle), but charge personalized subscription rates based on various employer characteristics (e.g., location, company and job type, benefits offered) that are correlated with willingness-to-pay \citep{dube2023}; and some SAT tutoring companies offer only a small number of different online tutoring bundles, but at prices that vary with customers' zip code and IP address \citep[e.g.,][]{sweeney2019}.} At the same time, it is important to keep in mind that sellers may face limitations (e.g., due to legal restrictions or fears of consumer backlash) on the extent to which they can use consumer data to personalize selling mechanisms. The numerical analysis in Section~\ref{sec:numerical} suggested that bundling can perform very close to the second-best even when the seller's ability to personalize is relatively moderate. To shed more light on this question, it would be valuable to derive analytical estimates on the second-order ($o( \sqrt{\frac{\ln n}{n} })$) terms that govern the convergence under pure bundling vs.\ second-best mechanisms. Relative to Theorem~\ref{thm:main}, such results may depend more finely on the details of the environment. However, this approach could also be helpful for understanding the extent to which pure bundling is outperformed by commonly observed generalizations such as nested bundling or two-part tariffs (see Remark~\ref{rem:thm}), whereas all these mechanisms perform equally well in terms of their convergence rates to the first-best.

\appendix

\section{Appendix: Main Proofs}\label{app:main-proofs}

Appendix~\ref{sec:Gaussian} proves Propositions~\ref{prop:gaussian}--\ref{lem:1-dim}. Appendix~\ref{app:main-pf} uses these to prove Theorem~\ref{thm:main}. Online Appendices~\ref{app:1-dim}--\ref{app:extensions} present proofs and omitted details for Sections~\ref{sec:1-dim type}--\ref{sec:dis}.


\subsection{Gaussian Environment}\label{sec:Gaussian}

We analyze the Gaussian environment from Section~\ref{sec:numerical}. Appendix~\ref{app:margin} proves Proposition~\ref{lem:1-dim}. Appendix~\ref{app:gaussian-pf} uses this to prove (a strengthening of) Proposition~\ref{prop:gaussian}.

Throughout Appendix~\ref{sec:Gaussian}, we denote by $R^{i} (\mu, \Sigma)$ ($i \in \{{\rm SB, bd, sep}\}$) the seller's optimal expected revenue (under general mechanisms, bundling, and separate sales) when her belief is distributed $\mathcal{N} (\mu, \Sigma)$; when $|G| = 1$, we drop the superscripts $i$. Let $\Phi$ and $\varphi$ be the cdf and pdf of the one-dimensional standard normal distribution. Let $\gamma: = \sup_{x \in \mathbb{R} } \varphi(x) x$. We will often use the following tail bound \citep{wainwright2019}:

\begin{equation}\label{eq:Gauss-tail}
\Phi(-z)\leq \exp[-\frac{z^2}{2}],\quad \text{ for all } z\geq 0.
\end{equation}
Moreover, the tail probability admits the approximation $\lim_{z\to\infty}\frac{\Phi(-z)}{\frac{1}{\sqrt{2\pi}|z|}\exp[-\frac{z^2}{2}]}=1$.

\subsubsection{Preliminaries for Single-Good Case}

Assume $|G| = 1$ and fix some $\mu, \sigma \in \mathbb{R}$. Then $R(\mu,\sigma^2)=\max_{p \in \mathbb{R}} p (1- \Phi\left(\frac{p-\mu}{\sigma}\right))$. Let $x^*(\mu, \sigma^2) := \argmax_{x\in\mathbb R} (\mu + x \sigma) (1 - \Phi(x))$ be the optimal normalized price.

\begin{lem}\label{lem:Gaussbd-prelim}
We have $|x^*(\mu, \sigma^2)| \geq  \sqrt{  2\left|\ln \frac{\mu}{\sigma \sqrt{2 \pi}(1 + \gamma)}\right|}$. If $\mu\geq 0$, then $x^*(\mu, \sigma^2) \leq 1$.
\end{lem}

\begin{proof}
Let $x^* := x^*(\mu, \sigma^2)$. To prove the first claim, assume $\mu\not=0$, as the claim is trivial if $\mu=0$. 
    The first-order condition for revenue maximization implies $\sigma \geq \sigma \left(1 - \Phi(x^*)\right) = ( \mu + \sigma x^*) \varphi(x^*)$, which yields $1-\varphi(x^*)x^*\geq \frac{\mu}{\sigma}\varphi(x^*)$. Thus, we have
    \begin{align*}
        \frac{1}{\sqrt{2 \pi}} e^{- {x^*}^2/2} = \varphi(x^* )   \leq \left(1 - \varphi(x^*)x^*\right) \frac{\sigma}{\mu} \leq \left(1 + \gamma\right) \frac{\sigma}{\mu } .
    \end{align*}
    After manipulation, we get $ e^{{x^*}^2} \geq \left(\frac{\mu}{\sigma}\right)^2 \frac{1}{2\pi (1 + \gamma)^2}$, which ensures
$    |x^*| \geq  \sqrt{  2\left|\ln \frac{\mu }{\sigma \sqrt{2 \pi}(1 + \gamma)}\right|}.$


If $\mu \geq 0$, the derivative of revenue with respect to $x  >  1$ satisfies
\[
\sigma (1 - \Phi(x)) - \varphi(x)(\mu + \sigma x)  \leq \sigma \left(\frac{\varphi(x)}{x} - \varphi(x) x \right) < 0.
\]
Therefore, $x^* \leq 1$ is needed to satisfy the first-order condition.
\end{proof}

\begin{lem}\label{lem:upperbound2} If $\mu\geq 0$, then $R(\mu, \sigma^2) \leq \max\left\{  \mu - \sigma \sqrt{  2\left|\ln \frac{\mu  }{\sigma \sqrt{2 \pi}(1 + \gamma)}\right|} , \frac{\mu + \sigma}{2}   \right\}$.
\end{lem}

\noindent {\it Proof.} As before, let $x^* = x^*(\mu, \sigma^2)$ denote the optimal normalized price. If $x^* \leq 0$, then by Lemma~\ref{lem:Gaussbd-prelim}, $x^* \leq - \sqrt{  2\left|\ln \frac{\mu }{\sigma \sqrt{2 \pi}(1 + \gamma)}\right|}$. Thus,
    \[
 R(\mu, \sigma^2) = (\mu + x^* \sigma) (1 - \Phi(x^*)) \leq \left( \mu + x^* \sigma\right) \leq \mu  - \sigma \sqrt{  2\left|\ln \frac{\mu  }{\sigma \sqrt{2 \pi}(1 + \gamma)}\right|}.
    \]
If $x^* > 0$, then by Lemma~\ref{lem:Gaussbd-prelim}, $x^* \leq 1$.  Thus,
    \[
R(\mu , \sigma^2) \leq (\mu  + \sigma x^*) (1 - \Phi(x^*))\leq \frac{\mu + \sigma}{2}. \qed
    \]

\begin{prop}\label{prop:boundmonopoly}
If $\mu\geq 0$, then for every $n\in\mathbb N$,
\begin{align*}
\max &\left\{  \mu - \sigma \sqrt{\frac{\ln n}{n}} + \frac{\sigma}{\sqrt{n}} \sqrt{  2\left|\ln \frac{\mu  }{\sigma \sqrt{2 \pi}(1 + \gamma)}\right|} , \frac{\mu + \sigma/ \sqrt{n}}{2}   \right\} \geq R(\mu , \sigma^2/n)\\
& \geq \mu - \sigma \sqrt{\frac{\ln n}{n}} - \frac{\mu}{\sqrt{ 2\pi  n \ln n}}.
\end{align*}
If  $\mu\leq 0$, then for every $n\in\mathbb N$, $R(\mu, \sigma^2/n) \leq  \sqrt{\frac{2 \pi}{n}} \sigma$.
\end{prop}

\begin{proof} First, consider $\mu \geq 0$. By Lemma~\ref{lem:upperbound2}, 
\begin{align*}
R(\mu, \sigma^2/n) \leq \max\left\{  \mu - \frac{\sigma}{\sqrt{n}} \sqrt{  2\left|\ln \frac{\mu  \sqrt{n}}{\sigma \sqrt{2 \pi}(1 + \gamma)}\right|} , \frac{\mu + \sigma/\sqrt{n}}{2}   \right\}. 
\end{align*}
Thus, the desired upper bound follows from the observation that
\[
\sqrt{ 2\left|\ln \frac{\mu  \sqrt{n}}{\sigma \sqrt{2 \pi}(1 + \gamma)}\right| }  = \sqrt{ \left| \ln n +  2\ln \frac{\mu  }{\sigma \sqrt{2 \pi}(1 + \gamma)}\right| } \geq \sqrt{\ln n} - \sqrt{2\left|\ln \frac{\mu  }{\sigma \sqrt{2 \pi}(1 + \gamma)}\right|}.
\]
To derive the lower bound, observe that the revenue from offering price 
$\mu - \sigma \sqrt{\frac{\ln n}{n}}$ is
\begin{align*}
\left(\mu - \sigma \sqrt{\frac{\ln n}{n}}  \right) \left(1 - \Phi\left( - \sqrt{\ln n} \right) \right) & \geq \mu - \sigma \sqrt{\frac{\ln n}{n}} - \mu \Phi\left( - \sqrt{\ln n} \right) \\
= \mu - \sigma \sqrt{\frac{\ln n}{n}} - \mu \left(1 - \Phi\left(  \sqrt{\ln n} \right)\right) & \geq  \mu - \sigma \sqrt{\frac{\ln n}{n}} - \mu \frac{\varphi(\sqrt{\ln n})}{\sqrt{ \ln n}} \\
&=  \mu - \sigma \sqrt{\frac{\ln n}{n}} - \frac{\mu}{\sqrt{ 2\pi  n \ln n}},
\end{align*}
where the second inequality uses the bound on the Mills ratio \citep{wainwright2019}.

Next, consider $\mu \leq 0$, and let $x^* := x^* (\mu, \sigma^2/n)$. Then 
\[
R(\mu, \sigma^2/n) \leq\mu + x^* \frac{\sigma}{\sqrt{n}} = \frac{\sigma}{\sqrt{n}} \frac{1 - \Phi(x^*)}{\varphi(x^*)} = \frac{\sigma}{\sqrt{n}} \frac{\Phi(- x^*)}{\varphi(x^*)} \leq \frac{\sigma}{\sqrt{n}} \frac{e^{- \frac{(x^*)^2}{2}}}{\varphi(x^*)} = \sqrt{2 \pi} \frac{\sigma}{\sqrt{n}},
\]
where the first equality uses the first-order condition $\frac{\sigma}{\sqrt n} (1 - \Phi(x^*)) = (\mu + x^* \frac{\sigma}{\sqrt n})\varphi(x^*)$, and the second inequality uses the tail bound (\ref{eq:Gauss-tail}) at $x^* \geq |\mu| / (\sigma/ \sqrt{n}) \geq 0$.
\end{proof}

\subsubsection{Proof of Proposition~\ref{lem:1-dim}}\label{app:margin}

Take any $\delta\geq 0$ and sequence of prices $(p_n)$ with $\theta^*-p_n\sim \delta   \sqrt{\ln n}\frac{\sigma}{\sqrt{n}}$.   Then
\begin{eqnarray*}
F_n(p_n)= \Phi \left(\frac{(p_n - \theta^*) \sqrt{n}}{\sigma} \right) = \exp\left[-\frac{\delta^2}{2}\ln n+ o\left(\ln n\right)\right]= n^{-\frac{\delta^2}{2}+o(1)},
\end{eqnarray*}
and thus
\begin{equation}\label{eq:discont}
\lim_{n\to\infty} \frac{\theta^* F_n (p_n)}{(\ln n)^{1/2}n^{-1/2}}=
\begin{cases}
0 \text{ if } \delta>1, \\
\infty \text{ if } \delta<1.
\end{cases}
\end{equation}
Additionally, if $\theta^*-p_n= \sqrt{\ln n}\frac{\sigma}{\sqrt{n}}$, then $\lim_n \frac{\theta^* F_n (p_n)}{(\ln n)^{1/2}n^{-1/2}}=0$, and hence $R^{\rm FB} (\theta^*) -R_n(p_n)\sim \sqrt{\ln n}\frac{\sigma}{\sqrt{n}}$, where $R_n(p_n):=p_n(1-F_n(p_n))$. This implies that, up to taking a convergent subsequence, any sequence of optimal prices $(p^*_n)$ must satisfy
\begin{equation}\label{eq:int-ineq}
1 \geq \lim_{k\to\infty} \frac{R^{\rm FB} (\theta^*) -R_{n_k}(p^*_{n_k})}{ \sqrt{\ln n_k}\frac{\sigma}{\sqrt{n_k}}} \geq \lim_{k\to\infty} \frac{\theta^*-p^*_{n_k}}{ \sqrt{\ln n_k}\frac{\sigma}{\sqrt{n_k}}}.
\end{equation}
If either inequality is strict, then $\theta^* - p^*_{n_k} \sim \delta \sqrt{\ln n_k}\frac{\sigma}{\sqrt{n_k}}$ for some $\delta \in (0, 1)$, so  ({\ref{eq:discont}) implies the contradiction that
\[
\infty = \lim_{k\to\infty} \frac{\theta^* F_{n_k} (p_{n_k})}{ \sqrt{\ln n_k}\frac{\sigma}{\sqrt{n_k}}} \leq \lim_{k\to\infty} \frac{R^{\rm FB} (\theta^*) -R^*_{n_k}(p^*_{n_k})}{ \sqrt{\ln n_k}\frac{\sigma}{\sqrt{n_k}}} .
\]
Thus, both inequalities in (\ref{eq:int-ineq}) hold with equality, which implies $R^{\rm FB} (\theta^*) -R_n(p^*_n)\sim  \sqrt{\ln n}\frac{\sigma}{\sqrt{n}}  \sim \theta^* - p^*_n$. This in turn implies that $\theta^* F_n (p^*_n) = o\left(\sqrt{\frac{\ln n}{n}}\right)$, as claimed.

\subsubsection{Proof of Proposition~\ref{prop:gaussian}}\label{app:gaussian-pf}

We prove the following stronger result than Proposition~\ref{prop:gaussian}, which we will use in the proof of Theorem~\ref{thm:main}. Assumption~\ref{asp} ensures that $I(\theta)$ is invertible for each $\theta\in\hat\Theta$, and that $\{I(\theta)^{-1}: \theta\in\hat\Theta\}$ is a compact set consisting of positive definite matrices. Let $\cal J$ be a compact neighborhood of $\{I(\theta)^{-1}: \theta\in\hat\Theta\}$. By choosing $\cal J$ to be sufficiently small, we can assume that every $J\in \cal J$ is positive definite.

\begin{prop}\label{prop:gaussian'}
There is $K>0$ such that for all $\theta^*\in\hat\Theta$, $J\in \cal J$, and $n\in\mathbb N$,
\[
\left|R^i\left(\theta^*, \frac{1}{n}J\right)-\theta^*\cdot {\bf 1}^G + \left(\frac{\ln n}{n} {\mathbf 1}^G\cdot J {\mathbf 1}^G\right)^{1/2}  \right| \leq K {n}^{-1/2} \quad \text{ for } i = {\rm bd, SB},
\]
\[
\text{ and } \quad \left|R^{\rm sep}\left(\theta^*, \frac{1}{n}J \right)-\theta^*\cdot {\bf 1}^G + \sum_{g\in G}\left(\frac{\ln n}{n} {\mathbf 1}^g\cdot J{\mathbf 1}^g \right)^{1/2}  \right| \leq K {n}^{-1/2} . 
\]
\end{prop}

\subsubsection{Proof of Proposition~\ref{prop:gaussian'}}\label{app:propgaussian-pf}

Since $R^{\rm bd}\left(\theta^*, \frac{1}{n}J\right)=R\left(\sum_{g \in G} \theta^*_g, \frac{{\mathbf 1}^G\cdot J {\mathbf 1}^G}{n}\right)$ and $R^{\rm sep}\left(\theta^*, \frac{1}{n}J\right)=\sum_{g\in G} R\left(\theta^*_g, \frac{{\mathbf 1}^g\cdot J {\mathbf 1}^g}{n}\right)$, the desired inequalities for bundling and separate sales follow from the single-good case (Proposition~\ref{prop:boundmonopoly}) combined with the compactness of $\hat\Theta$ and $\cal J$. Since $R^{\rm bd} \left(\theta^*, \frac{1}{n}J\right) \leq R^{\rm SB} \left(\theta^*, \frac{1}{n}J\right)$, the lower bound for $R^{\rm SB}\left(\theta^*, \frac{1}{n}J\right)$ is immediate from the bundling case, and it suffices to establish the desired upper bound for $R^{\rm SB}\left(\theta^*, \frac{1}{n}J\right)$.

To this end, suppose types are distributed ${\cal N}(\theta^*, \frac{1}{n}J)$ for some $\theta^*\in\hat\Theta$ and $J\in\cal J$.  Enumerate the set of bundles as ${\cal B}= \{B_\ell : \ell = 1, \ldots, m\}$, where $m = 2^{|G|}$ and $B_{m} := G$. Rewrite types as vectors $\omega\in \mathbb{R}^m$, where each coordinate $\omega_\ell$ represents the valuation of bundle $B_\ell$.\footnote{This more general formulation is useful for analyzing non-additive utilities (Appendix~\ref{sec:nonadd}).}  Note that $\omega$ is distributed ${\cal N}(y, \frac{1}{n}\Sigma)$, where $y\in\mathbb R^m$ satisfies $y_\ell=\sum_{g\in B_\ell}\theta^*_g$ for all $\ell$,  and $\Sigma\in\mathbb R^{m\times m}$ satisfies $\Sigma_{\ell\ell'}=\sum_{g\in B_\ell, g'\in B_{\ell'}}J_{gg'}$ for all $\ell, \ell'$.  Observe that $y$ is strictly positive and $y_m\geq y_\ell+\kappa$ for all $\ell\not=m$, where $\kappa:=\min_{g\in G, \theta\in \hat \Theta}\theta_g>0$.

Consider the following relaxed problem. Pick a full-rank matrix $A\in\mathbb R^{(m - 1) \times m}$ with $A \Sigma {\mathbf 1}^m = \mathbf{0}$.\footnote{In particular, if $\Sigma {\mathbf 1}^m= \mathbf{0}$, then we can choose any full-rank matrix $A$. If $\Sigma {\mathbf 1}^m\not= \mathbf{0}$, then $\{ a : a \cdot \Sigma {\mathbf 1}^m = 0\}$ is an  $(m - 1)$-dimensional hyperplane in $\mathbb{R}^m$, and hence includes $m -1$ linearly independent vectors $a_1, \ldots , a_{m - 1}$. Then the matrix $A$ with row vectors $a_1, \ldots, a_{m - 1}$ has row rank $m - 1$ and satisfies $A \Sigma {\mathbf 1}^m = \mathbf{0}$.} 
Conditional on $A\omega=z\in\mathbb R^{m-1}$, $\omega$ is distributed ${\cal N}(\hat y^z, \frac{1}{n}\hat\Sigma)$, where $\hat y^z:=y+B(z-Ay)$, $B:=\Sigma A^\top (A\Sigma A^\top)^{-1}$, $\hat\Sigma:=\Sigma \left({\mathbb I}-A^\top (A\Sigma A^\top)^{-1}A\Sigma\right)$, and ${\mathbb I}\in\mathbb R^{m\times m}$ is the identity matrix. Define 
\[
\overline R^{\rm SB}\left(y, \frac{1}{n}\Sigma\right):={\mathbb E}\left[R^{\rm SB}\left(\hat y^z, \frac{1}{n}\hat\Sigma\right)\right],
\]  
where the expectation is with respect to $z=A\omega$, which is distributed ${\cal N}(Ay, \frac{1}{n}A\Sigma A^\top)$. The value $\overline R^{\rm SB}\left(y, \frac{1}{n}\Sigma\right)$ represents the seller's revenue in the relaxed problem where each type $\omega$ is restricted to only report types in $\{\omega' \in \mathbb{R}^m: A\omega'=A\omega\}$. Clearly, $R^{\rm SB}\left(\theta^*, \frac{1}{n}J\right) \leq \overline R^{\rm SB}\left(y, \frac{1}{n}\Sigma\right)$.

Conditional on $ A \omega = z$, the buyer's type belongs to $\{ \omega' \in \mathbb{R}^m : A \omega' = z\}$, which is a line of the form $\{\hat y^z + \tau \Sigma {\mathbf 1}^m : \tau \in \mathbb{R} \}$. Thus, the current setting is a special case of the environment in Section~\ref{sec:1-dim type}, where $\alpha(0) :=\hat y^z$ and $\beta := \Sigma {\bf 1}^m$.  Moreover, conditional on $A\omega = z$, the valuation $\omega_\ell$ of each bundle $B_\ell$ is distributed ${\cal N}(\hat y^z_\ell,\frac{1}{n}\hat\sigma^2_{\ell})$, where $\hat\sigma^2_\ell:=\hat\Sigma_{\ell\ell}$.  Crucially, $\hat y^z_m=y_m$ and $\hat\sigma_m^2=\sigma_m^2:=\Sigma_{mm}$, i.e., the distribution of the valuation $\omega_m$ of the grand bundle  is unchanged by observing $z$. This is because $A \Sigma {\mathbf 1}^m = \mathbf{0}$ ensures that $\omega_m$ and $A\omega$ are uncorrelated.

To apply the analysis in Section~\ref{sec:1-dim type},  we define the lottery 
\[
b^z_0 \in \argmax_{q \in \Delta({\cal B})} \hat y^z \cdot q \text{ such that } \beta \cdot q = 0.
\]
Conditional on $A\omega = z$, the value of  $b^z_0$ is thus distributed ${\cal N}(\hat y_0, \frac{1}{n}\hat\sigma^2_0)$, where 
$
\hat y_0 := b_0^z\cdot \hat y^z$ and $\hat\sigma_0^2 := b^z_0\cdot \hat\Sigma b^z_0.
$
Let $R^s\left(\hat y^z, \frac{1}{n}\hat\Sigma\right)$ denote the optimal revenue when restricting attention to mechanisms that allocate to each type either a deterministic bundle $\{\delta_{B_\ell}:\ell=1,\ldots, m\}$ or $b_0^z$.  By Proposition~\ref{prop:simplemech}, this restriction is without loss.

The following result establishes an upper bound on $R^s\left(\hat y^z, \frac{1}{n}\hat\Sigma\right)$:

\begin{lem} \label{prop:upperbound} Suppose that $y_m  > \hat y^z_\ell + \kappa/2>0$ for all $\ell = 0,1,\ldots , m- 1$.    
    Then $R^s\left(\hat y^z, \frac{1}{n}\hat\Sigma\right)$ is bounded above by
     \[
       \max\left\{  y_m - \sigma_m \sqrt{\frac{\ln n}{n}} + \frac{\sigma_m}{\sqrt{n}} \sqrt{  2\left|\ln \frac{\kappa/2 }{\sigma_m \sqrt{2 \pi}(1 + \gamma)}\right|} , y_m  - \frac{\kappa}{4} + \frac{\sigma_m}{2\sqrt{n}}   \right\} + \sum_{\ell \neq m} \sqrt{\frac{2 \pi}{n}} \hat\sigma_\ell.
    \]
\end{lem}

\begin{proof}
Any IC-IR mechanism that allocates only deterministic bundles and $b_0^z$ can be represented as a collection of prices $p_\ell$ for each deterministic bundle $B_{\ell}$ ($\ell = 1, \ldots, m$) and price $p_0$ for lottery $b_0^z$. For the empty bundle $B_{\overline \ell}$, where $\omega_{\overline\ell}=0$ for all types, it is without loss to set $p_{\overline\ell}=0$. Denote by ${\mathbb P}_z$ the probability measure  conditional on $z=A\omega$.
The revenue $\sum_{\ell = 0}^{m} {\mathbb P}_z (\omega_\ell -p_\ell\geq \omega_{\ell'}-p_{\ell'} \forall \ell') p_\ell$ is bounded above by
\begin{eqnarray*}
&&y_m -\kappa/2+\sum_{\ell = 0}^{m} {\mathbb P}_z (\omega_\ell -p_\ell\geq \omega_{\ell'}-p_{\ell'} \forall \ell') \max\{p_\ell - (y_m - \kappa/2), 0\} \\
& \leq& y_m -\kappa/2+ \sum_{\ell = 0}^{m}{\mathbb P}_z (\omega_\ell \geq p_\ell)\max\{p_\ell - (y_m - \kappa/2), 0\}  \\
&\leq& y_m - \kappa/2 + \sum_{\ell = 0 }^{m} R(\hat y^z_\ell - y_m + \kappa/2,\hat \sigma_\ell^2/n) \\
& \leq & y_m - \kappa/2 + R(\kappa/2, \sigma_m^2/n) + \sum_{\ell \neq m}  \sqrt{\frac{2 \pi}{n}} \hat\sigma_\ell,
\end{eqnarray*}
where the last inequality uses part 2 of Proposition~\ref{prop:boundmonopoly}. Then part 1 of Proposition~\ref{prop:boundmonopoly} yields the desired bound.
\end{proof}

To complete the proof of Proposition~\ref{prop:gaussian'}, recall that $\hat y^z=y+B(z-Ay)$ and $\hat y^z_m=y_m>y_\ell+\kappa$ for all $z\in\mathbb R^{m-1}$ and $\ell=0,1,\ldots,m-1$. Pick $\varepsilon>0$ small enough that all $z$ with $\|z-Ay\|\leq\varepsilon$ satisfy $y_m  > \hat y^z_\ell + \kappa/2>0$ for all $\ell = 0,1,\ldots , m- 1$. 
Since $z$ is distributed  ${\cal N}(Ay, \frac{1}{n}A\Sigma A^\top)$, there is $K_1>0$ such that, for all $n$,
\[
\left |\overline R^{\rm SB}\left(y, \frac{1}{n}\Sigma\right) - {\mathbb E}\left[R^{\rm SB}\left(\hat y^z, \frac{1}{n}\hat\Sigma\right): \|z-Ay\|\leq\varepsilon \right]\right|\leq K_1n^{-1}.\] 
By Lemma~\ref{prop:upperbound}, there is $K_2>0$ such that, for all $n$, 
\[
{\mathbb E}\left[R^{\rm SB}\left(\hat y^z, \frac{1}{n}\hat\Sigma\right): \|z-Ay\|\leq\varepsilon \right] \leq y_m- \left(\frac{\ln n}{n}\right)^{1/2}  \sigma_m+K_2n^{-1/2}.
\]
Since $y_m=\sum_{g\in G}\theta^*_g$ and $\sigma_m=\left({\mathbf 1}^G\cdot J{\mathbf 1}^G\right)^{1/2}$, this implies the existence of $K>0$ such that $\overline R^{\rm SB}\left(y, \frac{1}{n}\Sigma\right)\leq \sum_{g\in G}\theta^*_g-\left(\frac{\ln n}{n}\right)^{1/2}  \sigma_m+Kn^{-1/2}$ for all $n\in\mathbb N$. Finally, by compactness of $\hat\Theta$ and $\cal J$, the coefficient  $K$ can be chosen uniformly in $y$ and $\Sigma$.

\subsection{Proof of Theorem~\ref{thm:main}}\label{app:main-pf}

\subsubsection{Preliminaries}\label{app:approximation}

Denote by $R^{i}(\lambda)$ ($i \in \{\rm SB, bd, sep \}$) the optimal expected revenue (under general mechanisms, bundling, and separate sales) when the buyer's type is drawn from $\lambda\in\Delta(\Theta)$. Denote by $\lambda_{x^n}\in\Delta(\Theta)$ the seller's posterior after observing signals $x^n$. 
For all $i \in \{\rm SB, bd, sep \}$, $R^i_n={\mathbb E}[R^i(\lambda_{x^n})]$. For all signals $x^n$, the compactness of $\hat \Theta$ and continuity of $\ln f$ implies the existence of a maximum likelihood estimate (MLE)
$\hat\theta_{x^n}\in\argmax_{\theta\in\hat\Theta}\sum_{i=1}^n\ln f(x_i, \theta)$. (Note that the maximum is taken over $\hat\Theta$ instead of $\Theta$). The \textit{\textbf{empirical Fisher information}} at $\theta$ is the matrix
$I_{x^n}(\theta):=\left(-\frac{1}{n}\sum_{i=1}^n\frac{\partial^2}{\partial \theta_g \partial \theta_{g'}}\ln  f(x_i, \theta)\right)_{g,g' \in G}$. For each $x^n$, denote by $\hat\lambda_{x^n}$ the Gaussian distribution with mean $\hat\theta_{x^n}$ and covariance matrix $\frac{1}{n} I_{x^n}(\hat\theta_{x^n})^{-1}$ if $I_{x^n}(\hat\theta_{x^n})$ is invertible; if $I_{x^n}(\hat\theta_{x^n})$ is not invertible, we set the covariance matrix to be the identity matrix. Denote by ${\rm TV}$ the total-variation distance on $\Delta(\mathbb R^{|G|})$, i.e., ${\rm TV}(\lambda,\lambda'):=\sup_{\Theta'} \left( \lambda(\Theta')-\lambda'(\Theta') \right)$, where the supremum is over all measurable subsets $\Theta' \subseteq \mathbb R^{|G|}$.

By the Bernstein-von Mises theorem, the seller's posterior $\lambda_{x^n}$ can be approximated (with respect to the total-variation metric) by the Gaussian belief $\hat\lambda_{x^n}$. Moreover, \cite{hipp1976} bound the approximation errors. Denote by  $\mathbb{P}_\theta$ and $\mathbb{E}_\theta$ probabilities and expectations conditional on the true type being $\theta$:

\begin{lem}[\cite{hipp1976} ]\label{lem:BvM}
 For each $\theta\in{\rm int} \, \Theta$, there is $c_\theta\geq 0$ with 
${\mathbb P}_\theta({\rm TV}(\lambda_{x^n}, \hat\lambda_{x^n})>c_\theta n^{-1/2} )=o(n^{-1})$.
\end{lem}

It is also known that the MLE $\hat\theta_{x^n}$ is asymptotically normal conditional on each $\theta$. Specifically, the distribution of $n^{1/2} (\hat\theta_{x^n}-\theta)$ is approximated by the Gaussian distribution  with mean $\mathbf{0}$ and covariance matrix $I(\theta)^{-1}$, denoted by $N_{I(\theta)^{-1}}$. Moreover, \cite{pfanzagl1973} bounds the approximation errors, uniformly in $\theta$:

\begin{lem}[\cite{pfanzagl1973}]
\label{lem:MLE} 
There is $c'> 0$ such that for every $\theta\in\Theta$, convex set $E\subseteq\mathbb R^G$, and $n\in\mathbb N$,  we have $\left|{\mathbb P}_\theta \left( n^{1/2} (\hat\theta_{x^n}-\theta) \in E \right)-N_{I(\theta)^{-1}}(E) \right|\leq c'n^{-1/2}$.
\end{lem}

Let $T:=\sup_{\theta\in\Theta, B\subseteq G} \left({{\bf 1}}^B \cdot I(\theta)^{-1}{{\bf 1}}^B \right)^{1/2}$, which is finite by compactness of $\Theta$ and Assumption~\ref{asp}. For every $B\subseteq G$ and $n\in\mathbb N$, let $E^B_n :=\left\{z\in\mathbb R^{|G|}: z\cdot {\bf 1}^B > (\ln n)^{1/2}T\right\}$. Then for all $n \in \mathbb{N}$,
\begin{equation}\label{eq:theta-d pbound}
\begin{split}
{\mathbb P}_\theta\left[ |(\hat\theta_{x^n}-\theta) \cdot {\bf 1}^B|> \left(\frac{\ln n}{n}\right)^{1/2} T\right]\leq N_{I(\theta)^{-1}}(E^B_n) +c'n^{-1/2} \\
\leq 2n^{-\frac{T}{2\left({{\bf 1}}^B \cdot I(\theta)^{-1}{{\bf 1}}^B \right)^{1/2}}} +c'n^{-1/2} \leq (2+c')n^{-1/2},
\end{split}
\end{equation}
where the first inequality uses Lemma~\ref{lem:MLE} (with corresponding coefficient $c'$). The second inequality uses the Gaussian tail bound (\ref{eq:Gauss-tail}).

We first prove some preliminary lemmas. Endow $|G|\times |G|$-matrices with the norm $\|A\|:=\sup_{y\not={\bf 0}}\frac{\|Ay\|}{\|y\|}=\sum_{g\in G}\max_{g'\in G}|A_{gg'}|$ induced by the $L^1$-norm on $\mathbb{R}^{|G|}$:

\begin{lem}\label{lem:regular}  There is $C>0$ such that for all $\theta\in\Theta$ and $n\in\mathbb N$, we have
\begin{equation}\label{eq:Fisherpbound}
{\mathbb P}_\theta\left[A_{\theta,n}  \right]  \geq 1-Cn^{-1/2},
\end{equation}
where $A_{\theta, n}$ is the event that $\|I_{x^n}(\hat\theta_{x^n})-I(\theta)\|\leq \left(\frac{\ln n}{n}\right)^{1/2}C$. Moreover, there is $\overline n$ such that for all $\theta\in\Theta$ and $n\geq\overline n$, $I_{x^n}(\hat\theta_{x^n})$ is invertible with $\|I_{x^n}(\hat\theta_{x^n})^{-1}-I(\theta)^{-1}\|\leq C \|I_{x^n}(\hat\theta_{x^n})-I(\theta)\|$ for any $x^n$ in $A_{\theta, n}$. 
\end{lem}

\begin{proof} \noindent{\bf Inequality (\ref{eq:Fisherpbound}):}  By Assumption~\ref{asp}.2, there is a Lipschitz coefficient $L$ of $\frac{\partial^2 \ln f(x, \cdot)}{\partial\theta_g\partial\theta_{g'}}$ that is uniform in $g,g'\in G$ and $x\in X$. Then for every $n\in\mathbb N$ and $\theta\in\Theta$,
\begin{eqnarray*}
{\mathbb P}_\theta\left[ \|I_{x^n}(\theta)-I_{x^n}(\hat\theta_{x^n})\|> \left(\frac{\ln n}{n}\right)^{1/2}  LT \right] &\leq&  {\mathbb P}_\theta\left[ \| \theta-\hat\theta_{x^n}\|> \left(\frac{\ln n}{n}\right)^{1/2} T  \right] \\
\leq \sum_{g\in G}{\mathbb P}_\theta\left[ |(\theta-\hat\theta_{x^n})\cdot {{\bf 1}}^g|> \left(\frac{\ln n}{n}\right)^{1/2} T  \right]
&\leq & |G|(2+c')n^{-1/2}
\end{eqnarray*}
where the last inequality uses (\ref{eq:theta-d pbound}). By Assumption~\ref{asp}.2, $L\geq \sup_{x\in X, \theta\in\Theta, g,g'\in G}\left|\frac{\partial^2 \ln f(x,\theta)}{\partial\theta_g\partial\theta_{g'}}\right|$. Then for every $n\in\mathbb N$ and $\theta\in\Theta$, Hoeffding's inequality implies that
\begin{eqnarray*}
&&{\mathbb P}_\theta\left[ \|I(\theta)-I_{x^n}(\theta)\|> \left(\frac{\ln n}{n}\right)^{1/2}\frac{L}{2}\right]  \\
&\leq& \sum_{g,g'\in G}{\mathbb P}_\theta\left[ \left| \int \frac{\partial^2\ln f(x,\theta)}{\partial \theta_g\partial\theta_{g'}} dP_\theta(x)- \frac{1}{n}\sum_{i=1}^n\frac{\partial^2\ln f(x_i,\theta)}{\partial \theta_g\partial\theta_{g'}} \right|> \left(\frac{\ln n}{n}\right)^{1/2}\frac{L}{2}\right] \leq |G|^22n^{-1/2}.
\end{eqnarray*}

Combining the above observations and the triangle inequality, (\ref{eq:Fisherpbound}) holds for all $C\geq \max\{LT, |G|(2+c'), \frac{L}{2}, 2|G|^2\}$, $n\in\mathbb N$, and $\theta\in\Theta$.

{\bf Invertibility:}  By Assumption~\ref{asp}.4,  $\|I(\theta)^{-1}\|>0$ for each $\theta$. Thus, there is $\varepsilon>0$ such that for any $\theta\in\Theta$, $n\in\mathbb N$, and $x^n$ with $\|I(\theta)-I_{x^n}(\hat\theta_{x^n}) \|\leq\varepsilon$, the matrix $I_{x^n}(\hat\theta_{x^n})$ is  invertible with $\|I_{x^n}(\hat\theta_{x^n})^{-1}\|\leq 2 \|I(\theta)^{-1}\|$. Since $\Theta$ is compact and $I(\theta)$ is continuous in $\theta$ by Assumption~\ref{asp}, $\varepsilon$ can be chosen uniformly in $\theta$. Under such realizations of $x^n$,
\begin{eqnarray*}
\|I(\theta)^{-1}-I_{x^n}(\hat\theta_{x^n})^{-1}\| &= & \|I(\theta)^{-1}(I_{x^n}(\hat\theta_{x^n})-I(\theta))I_{x^n}(\hat\theta_{x^n})^{-1}\| \\
 \leq \|I(\theta)^{-1}\| \|I_{x^n}(\hat\theta_{x^n})-I(\theta)\| \|I_{x^n}(\hat\theta_{x^n})^{-1}\|  &\leq& 2\sup_{\theta\in\Theta}\|I(\theta)^{-1}\|^2\|I_{x^n}(\hat\theta_{x^n})-I(\theta)\|.
\end{eqnarray*}

By the proof of (\ref{eq:Fisherpbound}), the conclusion follows by taking $C\geq \max\{LT, |G|(2+c'), \frac{L}{2}, 2|G|^2, 2\sup_{\theta\in\Theta}\|I(\theta)^{-1}\|^2\}$ and  $\overline n$ large enough that $\left(\frac{\ln \overline n}{\overline n}\right)^{1/2}C\leq \varepsilon$. \end{proof}

Henceforth, we will often use the constant $\overline n$ and $C$ and events $(A_{\theta, n})$ obtained above without explicitly referencing Lemma~\ref{lem:regular}.

\begin{lem}\label{lem:bounds} There is $C'>0$ such that for all $\theta\in\Theta$, $n\geq\overline n$, and $B\subseteq G$, 
\begin{equation}\label{eq:MLE error}
{\mathbb E}_\theta\left[\|\hat\theta_{x^n}-\theta\| \mid A_{\theta,n}\right]  \leq C'n^{-1/2},
\end{equation}
\begin{equation}\label{eq:Fisher error}
\text{ and } \quad \left|{\mathbb E}_\theta\left[ \left( {\bf 1}^B\cdot I_{x^n}(\hat\theta_{x^n})^{-1}{\bf 1}^B\right)^{1/2} \mid A_{\theta,n} \right]-\left({\bf 1}^B \cdot I(\theta)^{-1} {\bf 1}^B\right)^{1/2}  \right| \leq C'n^{-1/2}.
\end{equation}
\end{lem}

\begin{proof} {\bf Inequality (\ref{eq:MLE error}):}  Denote by $\hat F^g_{\theta,n}$ the cdf of $(\hat\theta_{x^n}-\theta)\cdot {\bf 1}^g$ at $\theta$, and by $F^g_{\theta,n}$ the cdf of the Gaussian distribution with mean $0$ and variance $\frac{1}{n}{\bf 1}^g\cdot I(\theta)^{-1}{\bf 1}^g$. Then for every $n\in\mathbb N$, $g\in G$, and $\theta\in \Theta$,
\begin{eqnarray*}
&& \left|\int_{-\kappa_n}^{\kappa_n} zd\hat F_{\theta,n}^g(z)\right| =   \left|\int_{-\kappa_n}^{\kappa_n} zd\hat F_{\theta,n}^g(z) -\int_{-\kappa_n}^{\kappa_n} zd F_{\theta,n}^g(z)\right|\\
 &= & \left|\kappa_n \left(\hat F_{\theta,n}^g\left(\kappa_n\right)-F_{\theta,n}^g\left(\kappa_n\right)+\hat F_{\theta,n}^g\left(-\kappa_n\right)-F_{\theta,n}^g\left(-\kappa_n\right)\right)-  \int_{-\kappa_n}^{\kappa_n} \left(\hat F_{\theta,n}^g(z)-F_{\theta,n}^g\left(\kappa_n\right)\right)dz \right|   \\
&\leq&  4\kappa_nc'n^{-1/2} \leq 4 Tc' n^{-1/2},
\end{eqnarray*}
where $\kappa_n:=\left(\frac{\ln n}{n}\right)^{1/2}T$,  the second equality uses integration by parts, and the first inequality uses Lemma~\ref{lem:MLE}. Let $M:=\sup_{\theta\in\hat\Theta}\|\theta\|$. For all $n\in\mathbb N$, $\theta\in\Theta$, the above inequality and (\ref{eq:theta-d pbound}) yield
\begin{eqnarray}\label{eq:MLE-E}
{\mathbb E}_\theta\left[\|\hat\theta_{x^n}-\theta\|\right]  = \sum_{g\in G} \left|\int zd\hat F_{\theta,n}^g(z) \right| \leq  |G| (2Tc' + M(2+c'))n^{-1/2}.
\end{eqnarray}
Then by Lemma~\ref{lem:regular} and as $\|\hat\theta_{x^n}-\theta\|\leq M$ for all $x^n$, (\ref{eq:MLE error}) holds for large enough $C'$. 


{\bf Inequality (\ref{eq:Fisher error}):} By part 2 of Lemma~\ref{lem:regular}, it suffices to show that  $\sup_{\theta\in\Theta}{\mathbb E}_\theta[\|I_{x^n}(\hat\theta_{x^n})-I(\theta)\| \mid A_{\theta, n}]$ is bounded by a constant times $n^{-1/2}$ for all $n\geq \overline n$. By the part 1 of Lemma~\ref{lem:regular} and since $\sup_{\theta\in\Theta, n\in\mathbb N, x^n\in X^n}\|I_{x^n}(\hat\theta_{x^n})-I(\theta)\|<\infty$ (by Assumption~\ref{asp}), it then suffices to show that ${\mathbb E}_\theta[\|I_{x^n}(\hat\theta_{x^n})-I(\theta)\| ]$  is bounded by a constant times $n^{-1/2}$ for all $n\geq \overline n$.

By Assumption~\ref{asp}.2, there is a Lipschitz coefficient $L$ of $\frac{\partial^2 \ln f(x, \cdot)}{\partial\theta_g\partial\theta_{g'}}$ that is uniform in $g,g'\in G$ and $x\in X$. For all $n\in\mathbb N$ and $\theta\in\Theta$, 
\[
{\mathbb E}_\theta[\|I_{x^n}(\hat\theta_{x^n})-I_{x^n}(\theta)\|] \leq |G|L{\mathbb E}_\theta[\|\hat\theta_{x^n}-\theta\|] \leq L|G|^2 (2Tc'+ M(2+c')n^{-1/2},
\]
where the second inequality uses (\ref{eq:MLE-E}). Let $\ell:=\sup_{x\in X, \theta\in\Theta, g,g'\in G}\left|\frac{\partial^2 \ln f(x,\theta)}{\partial\theta_g\partial\theta_{g'}}\right|$, which is finite by Assumption~\ref{asp}.2. For all $n\in\mathbb N$ and $\theta\in\Theta$, 
\begin{eqnarray*}
{\mathbb E}_\theta[\|I_{x^n}(\theta)-I(\theta)\| ] &\leq & |G|\max_{g,g'\in G} {\mathbb E}_\theta\left[\left|\sum_{i=1}^n\frac{1}{n}\frac{\partial^2}{\partial \theta_g\theta_{g'}}\ln f(x_i,\theta) -{\mathbb E}_\theta\left[\frac{\partial^2}{\partial \theta_g\theta_{g'}}\ln f(x,\theta)\right] \right|\right] \\
&\leq&  |G|\max_{g,g'\in G} \left({\rm Var}_\theta\left[\sum_{i=1}^n\frac{1}{n}\frac{\partial^2}{\partial \theta_g\theta_{g'}}\ln f(x_i,\theta) \right]\right)^{1/2} 
\leq |G| \ell n^{-1/2},
\end{eqnarray*}
where the second (resp.\ third) inequality uses H\"older's (resp.\ Popoviciu's) inequality. Combining these observations with the triangle inequality yields the desired claim.
\end{proof}

\subsubsection{Conditional Payoffs}

Next, we prove a conditional analog of Theorem~\ref{thm:main}:

\begin{prop}\label{prop:conditional} For all $\theta\in{\rm int} \, \Theta$, there exists $\kappa_\theta>0$ such that
\[
\left|\theta\cdot{\bf 1}^G- \left(\frac{\ln n}{n} {\mathbf 1}^G\cdot I(\theta)^{-1}  {\mathbf 1}^G\right)^{1/2}-{\mathbb E}_\theta[R^{i}(\lambda_{x^n})]\right|\leq \kappa_\theta n^{-1/2}+o(n^{-1/2}), \, \text{ for } i\in\{{\rm SB}, {\rm bd}\},
\]
\[
\text{ and } \left|\theta\cdot{\bf 1}^G- \sum_{g\in G}\left(\frac{\ln n}{n} {\mathbf 1}^g\cdot I(\theta)^{-1}  {\mathbf 1}^g\right)^{1/2}-{\mathbb E}_\theta[R^{\rm sep}(\lambda_{x^n})]\right|\leq \kappa_\theta n^{-1/2}+o(n^{-1/2}).
\]
\end{prop}

The proof is immediate from the following two lemmas:

\begin{lem}\label{lem:Gapproximation-value}  For all $\theta\in{\rm int} \, \Theta$, there is $K_\theta> 0$ such that, for all $i \in \{\rm SB, bd, sep \}$,
\[
\left|{\mathbb E}_\theta\left[R^i(\lambda_{x^n})\right]-{\mathbb E}_\theta\left[R^i\left(\hat\lambda_{x^n}\right) \mid A_{\theta, n} \right] \right|
\leq K_\theta n^{-1/2}+o\left(n^{-1/2}\right).
\]
\end{lem}

\begin{proof}  Let $c_\theta> 0$ denote the coefficients obtained in Lemma~\ref{lem:BvM}. Since $R^i(\lambda_{x^n})$ and $R^i(\hat\lambda_{x^n})$ are uniformly bounded under realizations in $(A_{\theta, n})$, it suffices by Lemmas~\ref{lem:BvM} and  \ref{lem:regular} to find $K_\theta$ such that  
\begin{equation}\label{eq:Gapp1}
{\mathbb E}_\theta\left[\left|R^i(\lambda_{x^n})-R^i\left(\hat\lambda_{x^n}\right)\right| \mid A_{\theta, n}\cap\{ {\rm TV}(\lambda_{x^n}, \hat\lambda_{x^n})\leq c_\theta n^{-1/2}\} \right] \leq K_\theta n^{-1/2}+o\left(n^{-1/2}\right). 
\end{equation}

Let $\hat\Theta'\subseteq\mathbb R^{|G|}$ be a compact neighborhood of $\hat\Theta$.  Let $\hat\lambda_{x^n}'$ denote the conditional probability measure on $\hat\Theta'$ induced by $\hat\lambda_{x^n}$. By Lemma~\ref{lem:regular} and compactness of $\{I(\theta): \theta\in\Theta\}$, there is $M>0$ such that $\|\frac{1}{n}I_{x^n}(\hat\theta_{x^n})^{-1}\|\leq n^{-1}M$ for every $n\geq \overline n$ and any realization $x^n$ in $A_{\theta, n}$.
 Thus, since $\hat\theta_{x^n}$ is bounded away from $\mathbb R^{|G|}\setminus \hat\Theta'$, there is $M'>0$ such that  $\hat\lambda_{x^n}(\mathbb R^{|G|}\setminus \hat\Theta')\leq Mn^{-1}$ and $\int_{\mathbb R^{|G|}\setminus \hat\Theta'} \sum_{g\in G}\max\{\theta_g, 0\} \, d\hat\lambda_{x^n}(\theta)\leq M'n^{-1}$  for any sequence of such realizations.

The seller's expected revenue under belief $\hat\lambda_{x^n}$ can be bounded as follows:
\[
\hat\lambda_{x^n}(\hat\Theta') R^i\left(\hat\lambda'_{x^n} \right)+ \int_{\mathbb R^{|G|}\setminus \hat\Theta'} \sum_{g\in G}\max\{\theta_g, 0\} d\hat\lambda_{x^n}(\theta) \geq R^i\left(\hat\lambda_{x^n}\right)\geq \hat\lambda_{x^n}(\hat\Theta') R^i\left(\hat\lambda_{x^n}' \right).
\]
The left-hand side is the seller's expected payoff when all types outside $\hat\Theta'$ are perfectly observable. The right-hand side is the seller's expected payoff when all types outside $\hat\Theta'$ do not purchase anything. 
Thus, there is $M''>0$ such that $\left|R^i(\hat\lambda'_{x^n})-R^i\left(\hat\lambda_{x^n}\right)\right|\leq M'' n^{-1}$ for all $x^n$ in $(A_{\theta,n})$.

Observe that, for all $\theta\in\Theta$, $n\geq 2M$, and $x^n$ in $A_{\theta,n}$,
\begin{eqnarray*}
{\rm TV}(\lambda_{x^n},\hat\lambda'_{x^n})&=&\sup_{E\subseteq \hat\Theta'}|\lambda_{x^n}(E)-\hat\lambda'_{x^n}(E)| =\sup_{E\subseteq \hat\Theta'}|\lambda_{x^n}(E)-\hat\lambda_{x^n}(E)|+\hat\lambda_{x^n}(E)\frac{1-\hat\lambda_{x^n}(\hat\Theta')}{\hat\lambda_{x^n}(\hat\Theta')} \\
&\leq &{\rm TV}(\lambda_{x^n},\hat\lambda_{x^n})+\frac{Mn^{-1}}{1-Mn^{-1}}\leq {\rm TV}(\lambda_{x^n},\hat\lambda_{x^n})+2Mn^{-1}.
\end{eqnarray*}

Since $\hat\Theta'$ is bounded, there is $K>0$ such that for all $x^n$ in $A_{\theta, n}$, 
\begin{eqnarray*}
\left|R^i(\lambda_{x^n})-R^i\left(\hat\lambda_{x^n}\right)\right| &\leq&\left|R^i(\lambda_{x^n})-R^i\left(\hat\lambda'_{x^n}\right)\right|+\left|R^i(\hat\lambda'_{x^n})-R^i\left(\hat\lambda_{x^n}\right)\right| \\
&\leq& K {\rm TV}(\lambda_{x^n}, \hat\lambda'_{x^n}) + M''n^{-1}.
\end{eqnarray*}
The above observations yield (\ref{eq:Gapp1}) with $K_\theta=Kc_\theta$. 
\end{proof}

\begin{lem}\label{lem:Dapproximation-value} There exist $K'>0$ and $\overline n'\in\mathbb N$ such that for all $\theta\in\Theta$ and $n\geq\overline n'$, 
\[
\left|\theta\cdot {\mathbf 1}^G - \left( \frac{\ln n}{n} {\mathbf 1}^G \cdot I(\theta)^{-1}  {\mathbf 1}^G\right)^{1/2}-{\mathbb E}_\theta[R^i(\hat\lambda_{x^n}) \mid A_{\theta,n}] \right|\leq K' n^{-1/2} \quad \text{ for  } i \in \{\rm SB, bd \},
\]
\[
\text{ and } \quad \left|\theta\cdot {\mathbf 1}^G - \sum_{g\in G}\left({\frac{\ln n}{n}\mathbf 1^g}\cdot I(\theta)^{-1}  {\mathbf 1}^g\right)^{1/2}-{\mathbb E}_\theta[R^{\rm sep}(\hat\lambda_{x^n}) \mid A_{\theta,n}] \right|\leq K' n^{-1/2}.
\]

\end{lem}

\begin{proof}

By Proposition~\ref{prop:gaussian'}, there is $K>0$ such that for all $n\in\mathbb N$ and $x^n$ with $I_{x^n}(\hat\theta_{x^n})^{-1} \in \cal J$, we have
\[
\left|R^i\left(\hat\lambda_{x^n}\right)-\hat\theta_{x^n}\cdot {\bf 1}^G + \left(\frac{\ln n}{n} {\mathbf 1}^G \cdot I_{x^n}(\hat\theta_{x^n})^{-1}  {\mathbf 1}^G \right)^{1/2}  \right| \leq K \left(\frac{\ln n}{n}\right)^{1/2} \quad \text{ for  } i \in \{\rm SB, bd \},
\]
\[
\text{ and } \quad \left|R^{\rm sep}\left(\hat\lambda_{x^n}\right)-\hat\theta_{x^n}\cdot {\bf 1}^G + \sum_{g\in G}\left(\frac{\ln n}{n} {\mathbf 1}^g\cdot I_{x^n}(\hat\theta_{x^n})^{-1}{\mathbf 1}^g \right)^{1/2}  \right| \leq K \left(\frac{\ln n}{n}\right)^{1/2}. 
\]
Lemma~\ref{lem:regular} yields $\overline n'$ such that for all $n\geq\overline n'$, $\theta\in\Theta$,  and $x^n$ in $A_{\theta, n}$, we have $I_{x^n}(\hat\theta_{x^n})^{-1} \in \cal J$. 
Thus, Lemma~\ref{lem:bounds} implies the claim by setting $K'=K+C'$.
\end{proof}

\subsubsection{Completing the Proof of Theorem~\ref{thm:main}}\label{app:MLEprice}

\noindent{\bf Lower bound of revenue gap:} 
We first show that $R^{\rm FB}-R^{\rm SB}_n$ and $R^{\rm FB}-R^{\rm bd}_n$ (resp.\ $R^{\rm FB}-R^{\rm sep}_n$) vanish no faster than at rate $\left(\frac{\ln n}{n}\right)^{1/2} {\mathbb E}\left[({\mathbf 1}\cdot I(\theta)^{-1}  {\mathbf 1})^{1/2}\right]$ (resp.\ $\left(\frac{\ln n}{n}\right)^{1/2} \sum_{g\in G}{\mathbb E}\left[(\mathbf{1}^g\cdot I(\theta)^{-1}  \mathbf{1}^g)^{1/2}\right]$).  For $i \in \{\rm SB, bd \}$, 
\begin{eqnarray*}
\liminf_{n\to\infty}(R^{\rm FB}-R^i_n) \left(\frac{\ln n}{n}\right)^{-1/2}&=&\liminf_{n\to\infty}\int (\theta\cdot {\mathbf 1}^G-{\mathbb E}_\theta[R^i(\lambda_{x^n})]) \left(\frac{\ln n}{n}\right)^{-1/2}g(\theta)d\theta \\
&\geq& \int  ({\mathbf 1}^G\cdot I(\theta)^{-1}  {\mathbf 1}^G)^{1/2}g(\theta)d\theta={\mathbb E}\left[({\mathbf 1}^G \cdot I(\theta)^{-1}  {\mathbf 1}^G)^{1/2}\right],
\end{eqnarray*}
where the inequality holds by Fatou's Lemma, as $\liminf_{n\to\infty}(\theta\cdot {\mathbf 1}^G-{\mathbb E}_\theta[R^i(\lambda_{x^n})]) \left(\frac{\ln n}{n}\right)^{-1/2}=({\mathbf 1}^G\cdot I(\theta)^{-1}  {\mathbf 1}^G)^{1/2}$ for almost all $\theta\in\Theta$ by Proposition~\ref{prop:conditional}. The same argument implies 
\[
\liminf_{n\to\infty}(R^{\rm FB}-R^{\rm sep}_n) \left(\frac{\ln n}{n}\right)^{-1/2}\geq  \sum_{g\in G}{\mathbb E}\left[(\mathbf{1}^g\cdot I(\theta)^{-1}  \mathbf{1}^g)^{1/2}\right].
\]

\noindent{\bf Upper bound of revenue gap:} Next, we show that $R^{\rm FB}-R^{\rm SB}_n$ and $R^{\rm FB}-R^{\rm bd}_n$ (resp.\ $R^{\rm FB}-R^{\rm sep}_n$) vanish no slower than at rate  $\left(\frac{\ln n}{n}\right)^{1/2} {\mathbb E}\left[({\mathbf 1}^G\cdot I(\theta)^{-1}  {\mathbf 1}^G)^{1/2}\right]$ (resp. $\left(\frac{\ln n}{n}\right)^{1/2} \sum_{g\in G}{\mathbb E}\left[(\mathbf{1}^g\cdot I(\theta)^{-1}  \mathbf{1}^g)^{1/2}\right]$). For each $x^n$, consider the mixed bundling mechanism that prices each $B \subseteq G$ at $0$ if $ I_{x^n}(\hat\theta_{x^n})$ is not invertible, and otherwise at
\[
p_{x^n} (B):=\hat\theta_{x^n}\cdot \mathbf{1}^B-  \left(\frac{\ln n}{n} {\mathbf 1}^B\cdot I_{x^n}(\hat\theta_{x^n})^{-1}  {\mathbf 1}^B\right)^{1/2}.
\]

We first lower-bound the probability that the buyer's willingness-to-pay for $B$ exceeds $p_{x^n} (B)$:

\begin{lem}\label{lem:MLrev1} There exists $M>0$ such that for all $B \subseteq G$, $n\geq \overline n$, and $\theta\in\Theta$, 
\[
{\mathbb P}_\theta[p_{x^n} (B)\leq \theta\cdot {\bf 1}^B] \geq  1- M n^{-1/2}.
\]
\end{lem}

\begin{proof}
Let $C$ denote the coefficient obtained in Lemma~\ref{lem:regular}. By Lemma~\ref{lem:regular}, there is $M_1>0$ such that $\left|\left({\bf 1}^B\cdot I_{x^n}(\hat\theta_{x^n})^{-1}{\bf 1}^B\right)^{1/2}-\left({\bf 1}^B \cdot I(\theta)^{-1}{\bf 1}^B\right)^{1/2}\right| \leq \left(\frac{\ln n}{n}\right)^{1/2}M_1$  for all $B\subseteq G$, $\theta\in \Theta$, $n\geq\overline n$ and $x^n$ with $ \|I(\theta)-I_{x^n}(\hat\theta_{x^n})\|\leq \left(\frac{\ln n}{n}\right)^{1/2}C$.

Letting $E^B_n:=\left\{z\in\mathbb R^{|G|}: z\cdot {\bf 1}^B > (\ln n)^{1/2}\left( \left({\bf 1}^B\cdot I(\theta)^{-1}{\bf 1}^B\right)^{1/2}+M_1\left(\frac{\ln n}{n}\right)^{1/2} \right) \right\}$ for all $B\subseteq G$, $n\geq \overline n$, and $\theta\in\Theta$, we have 
\begin{eqnarray*}
&&{\mathbb P}_\theta\left[(\hat\theta_{x^n}-\theta)\cdot{\bf 1}^B>\left(\frac{\ln n}{n}\right)^{1/2}\left( \left({\bf 1}^B\cdot I(\theta)^{-1}{\bf 1}^B\right)^{1/2}+M_1\left(\frac{\ln n}{n}\right)^{1/2} \right) \right] \\
&\leq & N_{I(\theta)^{-1}}(E^B_{n}) + c'n^{-1/2} \\
&\leq & \exp\left[-\frac{(\ln n )\left( \left({\bf 1}^B\cdot I(\theta)^{-1}{\bf 1}^B\right)+2M_1\left({\bf 1}^B\cdot I(\theta)^{-1}{\bf 1}^B\frac{\ln n}{n}\right)^{1/2}+M_1^2\frac{\ln n}{n} \right)}{2{\bf 1}^B\cdot I(\theta)^{-1}{\bf 1}^B}\right] +c'n^{-1/2} \\
&=& n^{-1/2}n^{\frac{2M_1\left({\bf 1}^B\cdot I(\theta)^{-1}{\bf 1}^B\frac{\ln n}{n}\right)^{1/2}+M_1^2\frac{\ln n}{n}}{2{\bf 1}^B\cdot I(\theta)^{-1}{\bf 1}^B}}+c'n^{-1/2} \\
&=& n^{-1/2}\exp\left[(\ln n){\frac{2M_1\left({\bf 1}^B\cdot I(\theta)^{-1}{\bf 1}^B\frac{\ln n}{n}\right)^{1/2}+M_1^2\frac{\ln n}{n}}{2{\bf 1}^B\cdot I(\theta)^{-1}{\bf 1}^B}} \right]+c'n^{-1/2} \leq M_2 n^{-1/2},
\end{eqnarray*}
where $M_2>0$ is chosen uniformly in $\theta$. The first inequality uses Lemma~\ref{lem:MLE}, the second inequality uses the Gaussian tail bound (\ref{eq:Gauss-tail}), and the last inequality uses the inequality $\ln n\leq \frac{n^\alpha}{\alpha}$, which holds for all $n>0$ and $\alpha>0$.

For every $B\subseteq G$, $n\geq\overline n$ and $\theta\in\Theta$, by  definition of $p_{x^n} (B)$ and choice of $M_1$, ${\mathbb P}_\theta[p_{x^n} (B)\leq\theta\cdot {\bf 1}^B] $ is weakly greater than the probability
\begin{equation*}
\begin{split}
{\mathbb P}_\theta\Bigg[ & \|I(\theta)-I_{x^n}(\hat\theta_{x^n})\|> \left(\frac{\ln n}{n}\right)^{1/2}C   \text{ and } \\
 &(\hat\theta_{x^n}-\theta)\cdot{\bf 1}^B\leq\left(\frac{\ln n}{n}\right)^{1/2}\left( \left({\bf 1}^B\cdot I(\theta)^{-1}{\bf 1}^B\right)^{1/2}+M_1\left(\frac{\ln n}{n}\right)^{1/2} \right)\Bigg],
 \end{split}
\end{equation*}
which in turn is no less than
\begin{eqnarray*}
&& 1- {\mathbb P}_\theta\left[ \|I(\theta)-I_{x^n}(\hat\theta_{x^n})\|> \left(\frac{\ln n}{n}\right)^{1/2}C   \right] \\
&& -{\mathbb P}_\theta\left[(\hat\theta_{x^n}-\theta)\cdot{\bf 1}^B>\left(\frac{\ln n}{n}\right)^{1/2}\left( \left({\bf 1}^B\cdot I(\theta)^{-1}{\bf 1}^B\right)^{1/2}+M_1\left(\frac{\ln n}{n}\right)^{1/2} \right) \right]\\
&&\geq 1-(C+M_2)n^{-1/2},
\end{eqnarray*}
where the inequality uses (\ref{eq:Fisherpbound}). This establishes the claim with $M:=C+M_2$. 
\end{proof}

Next, we approximate the revenue from offering each bundle $B$ at price $p_{x^n}(B)$: 
\begin{lem}\label{lem:MLrev2} There exists $M'>0$ such that for all $B\subseteq G$,  $n\geq \overline n$, and $\theta\in\Theta$,
\[
\left|{\mathbb E}_\theta[p_{x^n} (B) \mathbf{1}_{\{p_{x^n} (B)\leq \theta\cdot {\bf 1}^B\}}] -  \theta\cdot {\bf 1}^B +\left(\frac{\ln n}{n} {\mathbf 1}^B\cdot I(\theta)^{-1}  {\mathbf 1}^B\right)^{1/2}  \right|\leq M'n^{-1/2}.
\]
\end{lem}
\begin{proof}
Let $C'$ be the coefficient obtained in Lemma~\ref{lem:bounds}.  For all $B\subseteq G$, $n\geq \overline n$ and $\theta\in\Theta$, (\ref{eq:MLE error})--(\ref{eq:Fisher error}) from Lemma~\ref{lem:bounds} imply
\begin{eqnarray*}
&&\left|{\mathbb E}_\theta[p_{x^n} (B) \mid A_{\theta, n}]- \theta\cdot {\bf 1}^B +\left(\frac{\ln n}{n} {\mathbf 1}^B\cdot I(\theta)^{-1}  {\mathbf 1}^B\right)^{1/2} \right| \\
&\leq& \left|{\mathbb E}_\theta[\hat\theta_{x^n}\cdot {\mathbf 1}^B \mid A_{\theta, n}]-\theta\cdot {\bf 1}^B\right| \\
&& +\left(\frac{\ln n}{n} \right)^{1/2} \left|{\mathbb E}_\theta\left[ \left( {\bf 1}^B\cdot I_{x^n}(\hat\theta_{x^n})^{-1}{\bf 1}^B\right)^{1/2}| A_{\theta, n}\right]-\left({\mathbf 1}^B\cdot I(\theta)^{-1}  {\mathbf 1}^B\right)^{1/2}\right| 
\leq 2 C'n^{-1/2}.
\end{eqnarray*}

Since $\sup_{x\in X,\theta\in {\rm int} \, \hat\Theta, g, g' \in G}|\frac{\partial^2 \ln f(x,\theta)}{\partial\theta_g\partial\theta_{g'}}|<\infty$ by Assumption~\ref{asp}.2 and $\hat\Theta$ is bounded, $\sup_{n\in\mathbb N, x^n\in X^n}|p_{x^n} (B)|<\infty$.
Given this and Lemmas~\ref{lem:regular} and \ref{lem:MLrev1}, $|{\mathbb E}_\theta[p_{x^n} (B) \mid A_{\theta, n}]-{\mathbb E}_\theta[p_{x^n} (B)\mathbf{1}_{\{p_{x^n} (B)\leq \theta\cdot {\bf 1^B}\}}]|$ is bounded by $n^{-1/2}$ times a constant that is uniform in $\theta$. Thus, the desired bound follows from the above inequality. 
\end{proof}

To complete the proof of Theorem~\ref{thm:main}, denote by $\hat R^{\rm bd}_n := {\mathbb E}[p_{x^n} (G) \mathbf{1}_{\{p_{x^n} (G)\leq \theta\cdot {\bf 1}^G\}}]$ the expected revenue from the pure bundling mechanism that prices the grand bundle at $p_{x^n} (G)$ at each $x_n$ (as in (\ref{eq:price-bd})). By definition, $\hat R^{\rm bd}_n\leq R^{\rm bd}_n$ for each $n$. By Lemma~\ref{lem:MLrev2} applied to $B=G$, the difference between $\hat R^{\rm bd}_n$ and ${\mathbb E}\left[\theta\cdot{\bf 1}^G-\left(\frac{\ln n}{n} {\mathbf 1}^G\cdot I(\theta)^{-1}  {\mathbf 1}^G\right)^{1/2} \right]$ is bounded by $n^{-1/2}$ times a constant. 
This ensures that, for $i \in \{\rm SB, bd \}$,
\[
{\mathbb E}\left[({\mathbf 1}^G\cdot I(\theta)^{-1}  {\mathbf 1}^G)^{1/2}\right] \geq \limsup_{n\to\infty}(R^{\rm FB}-\hat R^{\rm bd}_n) \sqrt{\frac{n}{\ln n}}\geq \limsup_{n\to\infty}(R^{\rm FB}-R^{i}_n)  \sqrt{\frac{n}{\ln n}}.
\]

Analogously, denote by $\hat R^{\rm sep}_n$ the expected revenue from the separate sales mechanism that prices each good $g$ at $p_{x^n} (g)$ at each $x^n$.
Then, by the same arguments,
\[
\sum_{g\in G}{\mathbb E}\left[({\mathbf 1}^g\cdot I(\theta)^{-1} {\mathbf 1}^g)^{1/2}\right] \geq \limsup_{n\to\infty}(R^{\rm FB}-\hat R^{\rm sep}_n) \sqrt{\frac{n}{\ln n}} \\
\geq   \limsup_{n\to\infty}(R^{\rm FB}-R^{\rm sep}_n) \sqrt{\frac{n}{\ln n}}. \]

\singlespacing
\scriptsize
\begin{spacing}{0.6}
\bibliographystyle{econometrica}
\bibliography{screening}
\end{spacing}


\newpage
\setcounter{page}{1}

\singlespacing

\begin{center}
 {\Large\textbf{Online Appendix to ``Multidimensional Screening with Precise Seller Information''\\[0.7cm]}}
 {\large Mira Frick, Ryota Iijima, and Yuhta Ishii\\[1cm]}
\end{center}

\normalsize

\section{Proof of Proposition~\ref{prop:simplemech}}\label{app:1-dim}

\subsection{Preliminaries}

For each $x\in [\underline\beta,\overline\beta]$, define
\begin{align*}
H(x) &:= \max_{q \in \Delta(\mathcal{B})} \alpha(0) \cdot q \text{ such that } \beta \cdot q = x,\\
\mathcal{H}(x) &:= \argmax_{q \in \Delta(\mathcal{B})} \alpha(0) \cdot q \text{ such that } \beta \cdot q = x.
\end{align*}
Note that the constraint set $\{ q \in \Delta(\mathcal{B}) : \beta \cdot q = x\}$ is nonempty for all $x \in [ \underline{\beta}, \overline{\beta}]$.  Furthermore, $0 \in [ \underline{\beta}, \overline{\beta}]$ because $\beta_{\bar{\ell}} = 0$.  Henceforth, we fix some arbitrary $b_0 \in \mathcal{H}(0)$, and we order the allocations such that $\underline{\beta} = \beta_1 \leq \ldots \leq \beta_m = \overline{\beta}$.

Each mechanism $(q,t)$ induces the agent indirect utility function given by $V(\tau)=\alpha(\tau)\cdot q(\tau)-t(\tau)$ for each $\tau$.  For any convex function $V: T \rightarrow \mathbb{R}$, the set of subgradients at $\tau$ is 
\[
\partial V(\tau) :=  \left\{\lambda \in \mathbb{R}: V(\tau') \geq V(\tau) + \lambda (\tau' - \tau) \ \forall \tau' \right\}.
\]
It is well-known that (i) $\partial V(\tau) \neq \emptyset$; (ii) $\partial V$ is a monotone correspondence, i.e., for all $\tau' > \tau$, $\inf \partial V(\tau') \geq \sup \partial V(\tau)$; and (iii) $\partial V(\tau)$ is a singleton for almost all $\tau \in T$.
The last property implies that for any two functions $h_1$ and $h_2$ with $h_1(\tau), h_2(\tau) \in \partial V(\tau)$ for all $\tau$ and any measurable function $\varphi$, we have $\mathbb{E} \left[ \varphi(h_1(\tau))   \right] = \mathbb{E} \left[\varphi(h_2(\tau)) \right].$ Thus, we write $\mathbb{E} \left[ \varphi(\partial V(\tau)) \right]$ for this expectation.

Let $\mathcal{V}$ denote the set of all convex functions defined on $T$ such that $V(\tau) \geq 0$ and $\partial V(\tau) \subseteq [\underline{\beta}, \overline{\beta}]$ for all $\tau$. By standard arguments, any IC-IR mechanism $(q,t)$ induces an indirect utility function $V\in{\mathcal V}$. In particular,  its subgradient takes the form 
\begin{equation}\label{eq:IC slope}
\beta\cdot q(\tau) \in  \partial V(\tau)
\end{equation}
for all $\tau$.  
However, in contrast to the single-good monopoly problem, in our setting there are typically many mechanisms that induce the same indirect utility function but may lead to different payoffs for the designer. Let $\mathcal{M} (V)$ be the set of all IC-IR mechanisms $(q, t) \in \mathcal{M} $ that induce the indirect utility function $V$, and let $\mathcal{M}^s (V)$ be the subset of such mechanisms that are in $\mathcal{M}^s $.

The following four preliminary lemmas are proved in Appendix~\ref{app:1-dim lemmas}. First, we derive the designer's optimal payoff among all mechanisms that induce the same indirect utility function.  Moreover, we explicitly construct an optimal mechanism that induces the indirect utility function $V$:

\begin{lem}\label{lem:mechconstruct}
Consider any $V \in \mathcal{V}$.  Then $\mathcal{M} (V)\not=\emptyset$ and 
\[
\max_{(q, t) \in \mathcal{M} (V)} \mathbb{E} \left[t(\tau) \right] =  \mathbb{E} \left[H(\partial V(\tau)) + \tau \partial V(\tau) - V(\tau) \right].
\]
Moreover, for any mechanism $(q, t)$  such that for all $\tau$,
\begin{enumerate}
\item $q(\tau) \in \mathcal{H}(\partial V(\tau)):= \bigcup_{x \in \partial V(\tau)} \mathcal{H}(x)$,
\item $t(\tau) = \alpha(\tau) \cdot q(\tau) - V(\tau)$,
\end{enumerate}
we have $(q,t) \in \arg \max_{(q', t') \in \mathcal{M} (V)} \mathbb{E} \left[t'(\tau) \right]$. \end{lem}

Lemma~\ref{lem:mechconstruct} immediately yields the following reformulation of the designer's problem, in terms of optimization with respect to $V$:

\begin{cor}\label{prop:profitutility} We have
    \[
    \sup_{(q, t) \in \mathcal{M} } \mathbb{E} \left[t(\tau) \right] = \sup_{V \in \mathcal{V}} \mathbb{E} \left[ H(\partial V(\tau)) + \tau \partial V(\tau) - V(\tau) \right].
    \]
\end{cor}

The next lemma establishes a structural property of $H$ that is useful for simplifying the optimization problem with respect to $V$:

\begin{lem}\label{prop:pwlinear}
There exists a subsequence $(\beta_{\ell_i})_{i=1}^{k+1}$ of $(\beta_\ell)_{\ell=1}^m$ with $\beta_{\ell_1}=\underline\beta$ and $\beta_{\ell_{k+1}}=\overline\beta$ such that 
\begin{enumerate}
\item  $H$ is linear on $[\beta_{\ell_i}, \beta_{\ell_{i + 1}}]$ for every $i=1,\ldots, k$, 
\item $\mathcal{H}(\beta_{\ell_i}) \cap ( {\cal D} \cup \{b_0\}) \neq \emptyset$ for every $i=1,\ldots, k+1$,
\item $\beta_{\ell_i}=0$ for some $i=1,\ldots, k+1$.
\end{enumerate}
\end{lem}

For the remainder of this section, we fix the sequence $(\beta_{\ell_i})_{i=1,\ldots, k+1}$ derived in Lemma~\ref{prop:pwlinear}. Let ${\cal V}^s$ be the set of all $V\in {\cal V}$ such that $\partial V(\tau)\in \{\beta_{\ell_{1}}, \ldots, \beta_{\ell_{k+1}}\}$ for almost all $\tau$. 
The following lemma shows that the designer can restrict attention to indirect utility functions in ${\cal V}^s$ (as long as the type space is compact):

\begin{lem}\label{lem:convexmech}
Assume $T$ is compact.  For every $V \in \mathcal{V}$, there exists $\tilde V\in {\cal V}^s$ such that
\[
\mathbb{E} \left[ \left(H(\partial \tilde V(\tau)) + \tau \partial \tilde V(\tau) - \tilde V(\tau)\right) \right]  \geq  \mathbb{E} \left[ H(\partial V(\tau)) + \tau \partial V(\tau) - V(\tau) \right].
\]
\end{lem}

The final lemma guarantees that any indirect utility function in ${\cal V}^s$ can be induced by some mechanism in ${\cal M}^s$:

\begin{lem}\label{lem:simplemech}
For every $\tilde V\in {\cal V}^s$, there exists $(q, t) \in \mathcal{M}^s (\tilde V)$ such that 
\[
 \mathbb{E} \left[ t(\tau) \right] = \mathbb{E} \left[ H(\partial \tilde V(\tau)) + \tau \partial \tilde V(\tau) - \tilde V(\tau) \right].
\]
\end{lem}

\subsection{Completing the Proof of Proposition~\ref{prop:simplemech}}

First, assume that $T$ is compact. 
Take any $(q,t)\in {\cal M}$ and let $V$ denote the corresponding indirect utility. By Lemma~\ref{lem:convexmech}, there exists $\tilde V\in {\cal V}^s$ such that  
\[
 \mathbb{E} \left[ H(\partial \tilde V(\tau)) + \tau \partial \tilde V(\tau) - \tilde V(\tau) \right] \geq \mathbb{E} \left[ \left(H(\partial V(\tau)) + \tau \partial V(\tau) - V(\tau)\right) \right] \geq {\mathbb E}[t(\tau)]
\]
where the second inequality uses Lemma~\ref{lem:mechconstruct}. By Lemma~\ref{lem:simplemech}, there exists some $(\tilde q, \tilde t)\in \mathcal{M}^s$ with ${\mathbb E}[\tilde t(\tau)] =\mathbb{E} \left[ H(\partial \tilde V(\tau)) + \tau \partial \tilde V(\tau) - \tilde V(\tau) \right]$ that delivers a weakly higher payoff than $(q,t)$. This proves the result when $T$ is compact. 

To prove the result for a general interval $T \subseteq \mathbb{R}$, we want to show that, for any $(q,t)\in{\cal M}$ and $\varepsilon>0$, there is $(q^*,t^*)\in {\cal M}^s$ such that ${\mathbb E}[t(\tau)]\leq{\mathbb E}[t^*(\tau)]+\varepsilon$. To this end, note that for any $\varepsilon>0$, there is a compact interval $\tilde T\subseteq \mathbb R$} of types with ${\mathbb E}[{\mathbf 1}_{\tau\in \tilde T} \max_{\ell}\alpha_\ell(\tau)]\leq \varepsilon/2$ because $\tau$ is $L^1$. For any $(q,t)\in{\cal M}$, IR implies $t(\tau)\leq \max_{\ell}\alpha_\ell(\tau)$ for every $\tau$, and hence ${\mathbb E}[t(\tau)]\leq \sup_{(\tilde q,\tilde t)\in {\cal M}}{\mathbb E}[{\mathbf 1}_{\tau \in \tilde T}\tilde t(\tau)]+\varepsilon/2$.  By the result for the compact case,  there is $(q^*, t^*)\in {\cal M}^s$ such that ${\mathbb E}[{\bf 1}_{\tau \in \tilde T} t^*(\tau)]+\varepsilon/2 \geq \sup_{(\tilde q,\tilde t)\in {\cal M}}{\mathbb E}[{\bf 1}_{\tau \in \tilde T}\tilde t(\tau)]$, where $t^*(\cdot)\geq 0$. Thus, 
\[
{\mathbb E}[t(\tau)]\leq {\mathbb E}[{\bf 1}_{\tau \in \tilde T}t^*(\tau)]+\varepsilon\leq {\mathbb E}[t^*(\tau)]+\varepsilon. 
\]

\subsection{Proofs of Lemmas~\ref{lem:mechconstruct}--\ref{lem:simplemech}}\label{app:1-dim lemmas}

\begin{proof}[{\bf Proof of Lemma~\ref{lem:mechconstruct}}] 
We first show that 
\[
\sup_{(q, t) \in \mathcal{M} (V)} \mathbb{E} \left[t(\tau) \right] \leq  \mathbb{E} \left[H(\partial V(\tau)) + \tau  \partial V(\tau) - V(\tau) \right].
\]
To see this, consider any $(q, t) \in \mathcal{M} (V)$.  By (\ref{eq:IC slope}), $\beta \cdot q(\tau) \in \partial V(\tau)$ for all $\tau$.  Thus, 
\begin{eqnarray*}
\mathbb{E} \left[ t(\tau) \right] =  \mathbb{E} \left[\alpha(0) \cdot q(\tau) + (\beta \cdot q(\tau)) \tau - V(\tau) \right]  &\leq&  \mathbb{E} \left[H(\beta \cdot q(\tau)) + (\beta \cdot q(\tau)) \tau - V(\tau) \right] \\
&=& \mathbb{E} \left[H(\partial V(\tau)) + \tau \partial V(\tau) - V(\tau) \right].
\end{eqnarray*}

Now, consider any mechanism $(q, t)$ such that for all $\tau$,
\begin{enumerate}
\item $q(\tau) \in  \mathcal{H}(\partial V(\tau))$,
\item $t(\tau) = \alpha(\tau) \cdot q(\tau) - V(\tau)$.
\end{enumerate}
To see that $(q, t)$ is IC, note that type $\tau$'s utility from misreporting to be type $\hat{\tau}$ is
\begin{align*}
\alpha(\tau) \cdot q(\hat{\tau}) - t(\hat{\tau}) &= (\tau - \hat{\tau})( \beta \cdot q(\hat{\tau})) + V(\hat{\tau}) \leq V(\tau) = \alpha(\tau) \cdot q(\tau) - t(\tau),
\end{align*}
where the inequality uses $\beta\cdot q(\hat\tau)=\partial V(\hat\tau)$, which follows from $q(\hat{\tau})  \in \mathcal{H}(\partial V(\hat{\tau}))$.  Thus, $(q, t)$ is IC.  Moreover, by construction it induces $V\geq 0$ as the indirect utility function.  As a result, $(q, t) \in \mathcal{M} (V)$.  

Finally, because again $q(\tau) \in \mathcal{H}(\partial V(\tau))$ for all $\tau$, 
\[
\mathbb{E} \left[t(\tau) \right]  = \mathbb{E} \left[\alpha(0) \cdot q(\tau) + p (\beta \cdot q(\tau)) - V(\tau) \right] = \mathbb{E} \left[H(\partial V(\tau)) + \tau \partial V(\tau) - V(\tau) \right],
\]
as claimed. \end{proof}

\begin{proof}[{\bf Proof of Lemma~\ref{prop:pwlinear}}] The following lemma implies Lemma~\ref{prop:pwlinear}. This is because one can add $0$ to the sequence $(\beta_{\ell_i})_{i=1}^{k+1}$ constructed in Lemma~\ref{lem:wlinear-lem}, since we have $[\underline\beta, \overline\beta]\ni 0$ and ${\cal H}(0)\ni b_0$ by construction.

\begin{lem}\label{lem:wlinear-lem} There exists a subsequence $(\beta_{\ell_i})_{i=1}^{k+1}$ of $(\beta_\ell)_{\ell=1}^m$ with $\beta_{\ell_1}=\underline\beta$ and $\beta_{\ell_{k+1}}=\overline\beta$ such that 
\begin{enumerate}
\item  $H$ is linear on $[\beta_{\ell_i}, \beta_{\ell_{i + 1}}]$ for every $i=1,\ldots, k$; and
\item $\mathcal{H}(\beta_{\ell_i}) \cap {\cal D}  \neq \emptyset$ for every $i=1,\ldots, k+1$.
\end{enumerate}
\end{lem}

\noindent{\it Proof of Lemma~\ref{lem:wlinear-lem}}.
Consider the linear program associated with the value $H(x)$:
\[
H(x) = \max_{y \in \Delta(\mathcal{B})} \alpha(0) \cdot y \text{ such that } \beta \cdot y = x.
\]
By strong duality, the above is equal to the value of the dual program:
\begin{align*}
H(x) &= \min_{(z_1,z_2) \in \mathbb{R}^2}  z_1 x + z_2 \text{ such that } z_1 \beta + z_2 \mathbf{1} \geq \alpha(0)\\
&= \min_{z \in \mathbb{R}} z  x + \left(\max_{\ell} \alpha_\ell(0) - z \beta_\ell\right)\\
&= \min_{z \in \mathbb{R}} z x + G(z),
\end{align*}
where $G(z) := \max_\ell \alpha_\ell(0) - z \beta_\ell.$

Note that $G(z)$ is a piecewise linear function. That is, there exist $z_1>  \ldots > z_k$ and 
$S =  \{\ell_1, \ldots , \ell_{k + 1}\} \subseteq \{1,2, \ldots , m\}$ with $1= \ell_1 < \ell_2 < \cdots < \ell_{k + 1} = m$ such that 
\begin{align}
G(z) = \begin{cases}
\alpha_{\ell_1}(0) - z\beta_{\ell_1}  & \text{if } z \in [z_1, + \infty), \\
\alpha_{\ell_i}(0) - z \beta_{\ell_i} &\text{if } z \in [z_i , z_{i - 1}]  \text{ for } i =2, \ldots , k, \\
\alpha_{\ell_{k + 1}}(0) - z \beta_{\ell_{k + 1}} &\text{if } z \in (- \infty, z_k].
\end{cases}
\label{eqn:1}
\end{align}
Observe that $\partial G(z_i) = [- \beta_{\ell_{i+ 1}}, - \beta_{\ell_i}]$ for all $i$. Thus, whenever $-x \in \partial G(z_i)$, which occurs when $x \in [\beta_{\ell_{i}}, \beta_{\ell_{i+1}}]$, we have
\begin{align}
H(x) = z_i x+ G(z_i). \label{eqn:2}
\end{align}
This proves the first part of the lemma.

To prove the second part, suppose that $i = 1,\ldots , k$.  Then by (\ref{eqn:1}) and (\ref{eqn:2}),
\[
 H(\beta_{\ell_i}) =  z_{i} \beta_{\ell_{i }} + G(z_i) = \alpha_{\ell_i}(0) .
\]
Thus, $\delta_{\ell_i} \in \mathcal{H}(\beta_{\ell_i})$, whence ${\cal D} \cap \mathcal{H}(\beta_{\ell_i}) \neq \emptyset$.
If instead $i = k + 1$, then by (\ref{eqn:1})--(\ref{eqn:2}),
\[
H(\beta_{\ell_{k + 1}}) = z_k \beta_{\ell_{k + 1}} + G(z_k) = \alpha_{\ell_{k + 1}}(0).
\]
Again, $\delta_{\ell_{k + 1}} \in \mathcal{H}(\beta_{\ell_{k + 1}})$, and hence ${\cal D} \cap \mathcal{H}(\beta_{\ell_{k + 1}}) \neq \emptyset$. 
\end{proof}

\begin{proof}[{\bf Proof of Lemma~\ref{lem:convexmech}}] Write $T=[\underline\tau,\overline\tau]$, and let $(I_1, \ldots , I_{k})$ be any interval partition of $T$.\footnote{We allow $I_j = \emptyset$ for some $j$.}  Let 
$M(I_1, \ldots , I_{k})$ be the set of non-decreasing functions $h: T
\rightarrow [ \underline{\beta}, \overline{\beta}]$ for which $h(I_i) \subseteq [\beta_{\ell_i}, \beta_{\ell_{i+1}}] $ for each $i=1,\ldots, k$. We endow $M(I_1, \ldots , I_k)$ with the $L^1$ norm, making it a metric space. Let $S(I_1, \ldots , I_k)\subseteq M(I_1, \ldots , I_k)$ denote the set of all functions in $M(I_1, \ldots , I_k)$ that are step functions taking values in $\{ \beta_{1}, \ldots , \beta_{{k + 1}}\}$.

\begin{claim}\label{lem:extreme}
The set $M(I_1, \ldots , I_k)$ is compact and convex, with ${\rm ext}(M(I_1, \ldots , I_k)) = S(I_1, \ldots , I_k)$.
\end{claim}

\noindent{\it Proof of Claim~\ref{lem:extreme}.} Clearly, $M(I_1, \ldots , I_k)$ is convex.  To prove compactness, it suffices to show that $M(I_1, \ldots , I_k)$ is sequentially compact, since $M(I_1, \ldots , I_k)$ is a metric space. To this end, consider a sequence $h_n \in M(I_1, \ldots , I_k)$.  By Helly's selection theorem, there exist a subsequence $(n_k)$ and a function $h$ such that $\lim_{j \rightarrow \infty} h_{n_j}(\tau) = h(\tau)$  for all $\tau$.  Hence, we also have $h \in M(I_1, \ldots , I_k)$.  Moreover, by the Lebesgue dominated convergence theorem,  $\intop_{\underline{\tau}}^{\overline{\tau}} |h_{n_j}(\tau) - h(\tau)| d \tau = 0.$
Thus, $M(I_1, \ldots , I_k)$ is sequentially compact. The proof that ${\rm ext}(M(I_1, \ldots , I_k)) = S(I_1, \ldots , I_k)$ follows from standard arguments \citep[cf.\ Lemma 2.7 in][]{borgers2015}. \qed

\medskip

To complete the proof of Lemma~\ref{lem:convexmech}, consider any function $V \in \mathcal{V}$.  Let $p^* = \min \left(\arg \min_{\tau \in T}V(\tau)\right)$.  
Consider any monotone function $\varphi : T \rightarrow \mathbb{R}$ such that $\varphi(\tau) \in \partial V(\tau)$ for all $\tau\in T$.  Define the interval partition $I_1, I_2, \ldots , I_k$ by
\[
I_1 =  \varphi^{-1}( [ \beta_{\ell_1}, \beta_{\ell_2})), \quad \ldots , \quad I_k = \varphi^{-1}([ \beta_{\ell_k}, \beta_{\ell_{k + 1}}]).
\]
Note the following two properties of $I_1, \ldots , I_k$:
\begin{enumerate}
\item  $\varphi \in M(I_1, \ldots , I_k)$,
\item for all $h \in M(I_1, \ldots , I_k)$, $h(\tau) (\tau - p^*) \geq 0$ for all $\tau$.
\end{enumerate}
To see that the second property holds, note that $\beta_i=0$ for some $i$. Thus, $p^*$ is the lower bound of interval $I_i$. Then any $h\in M(I_1, \ldots , I_k)$ is decreasing at $\tau<p^*$ and weakly increasing at $\tau\geq p^*$.

Next, define the functional $T: M(I_1, \ldots , I_k) \rightarrow \mathbb{R}$ by 
\[
T(h) = \mathbb{E} \left[\left( H(h(\tau)) + \tau h(\tau) - \intop_{p^*}^\tau h(\tau') d\tau' \right) \right].
\]
By the dominated convergence theorem, $T$ is continuous.  Moreover, $T$ is linear, since $H$ is linear on $[\beta_i, \beta_{i + 1}]$ for every $i=1,\ldots, k$ by Proposition~\ref{prop:pwlinear}. Thus,
\begin{align*}
\mathbb{E} \left[\mathbf{1}_{\{\tau \in I\}} \left(H (\partial V(\tau)) + \tau \partial V(\tau) - V(\tau) \right) \right] &= \mathbb{E} \left[\mathbf{1}_{\{\tau \in I\}} \left(H (\varphi(\tau)) + \tau \varphi(\tau) - V(\tau) \right) \right] \\
= T(\varphi) - V(p^*)  
 \leq \sup_{h \in M(I_1, \ldots , I_k)} T(h) - V(p^*)  &= \max_{h \in S(I_1, \ldots , I_k)} T(h) - V(p^*),
\end{align*}
where the last equality follows from Claim~\ref{lem:extreme} and Bauer's maximum principle.

Thus, there exists some $\tilde  h \in S(I_1, \ldots , I_k)$ for which 
\[
T(\tilde h) - V(p^*)\geq \mathbb{E} \left[ \left(H (\partial V(\tau)) + \tau \partial V(\tau) - V(\tau) \right) \right].
\]
Define $\tilde V$ by $\tilde V(\tau) = \intop_{p^*}^\tau \tilde h(\tau') d\tau'$. Since $ h \in S(I_1, \ldots , I_k)$, $\partial \tilde V(\tau) \in \{\beta_{\ell_1}, \ldots , \beta_{\ell_{k + 1}}\}$ for almost all $\tau \in T$.  Moreover, because $\tilde h \in M(I_1, \ldots , I_k)$, we have $\tilde h(\tau)(\tau - p^*) \geq 0$ for all $\tau \in T$.  Thus, 
\[
\min_{\tau \in T} \tilde V(\tau) = \tilde V(p^*) =  0.
\]
Together, this implies $\tilde V\in{\cal V}^s$. Hence, 
\begin{align*}
 \mathbb{E} \left[ (H(\partial \tilde V(\tau)) + \tau \partial \tilde V(\tau) - \tilde V(\tau) )\right] &= \mathbb{E} \left[  (H(\tilde h(\tau)) + \tau \tilde h(\tau) - \intop_{p^*}^{\tau} \tilde h(\tau') d\tau' )\right] \\
 \geq T(\tilde h) - V(p^*)
 &\geq \mathbb{E} \left[\left(H (\partial V(\tau)) + \tau \partial V(\tau) - V(\tau) \right) \right],
\end{align*}
 as claimed. \end{proof}

\begin{proof}[{\bf Proof of Lemma~\ref{lem:simplemech}}]
Let $q$ be an allocation rule with $q(\tau) \in \mathcal{H}(\partial \tilde V(\tau))$
for all $\tau$.  Since by Lemma~\ref{prop:pwlinear} $\partial \tilde V (\tau) \cap \{ \beta_{\ell_1}, \ldots , \beta_{\ell_{k + 1}}\} \neq \emptyset$ for all $\tau$, we can choose $q$ such that $q(\tau) \in \{b_0\} \cup {\cal D}$ for all $\tau$.  Define transfers $t(\tau) := \alpha(\tau) \cdot q(\tau) - \tilde V(\tau).$
By Lemma~\ref{lem:mechconstruct}, $(q,t) \in \arg\max_{(q', t') \in \mathcal{M} (\tilde V)} \mathbb{E} \left[t'(\tau) \right]$ and $\mathbb{E} \left[ t(\tau) \right] = \mathbb{E} \left[ H(\partial \tilde V(\tau)) + \tau \partial \tilde V(\tau) - \tilde V(\tau) \right]$.  Moreover, by construction, $q(\tau) \subseteq {\cal D} \cup \{b_0\}$.  Hence, $(q, t) \in \mathcal{M}^s (\tilde V)$.
\end{proof}

\section{Details for Section~\ref{sec:extensions}}\label{app:extensions}

\subsection{Production Costs and Negative Types}\label{app:cost}

In our main model, the first-best $R^{\rm FB} = \mathbb{E} [\mathbf{1}^G \cdot \theta]$ involves supplying the grand bundle to all buyer types. This can fail if the seller faces a production cost or some types have negative valuations for some goods. We now extend the analysis to allow for both possibilities: First, to produce each bundle $B$, the seller incurs a cost $c(B)$, so that $R^{\rm FB} = \mathbb{E} \left[\mathbf{1}^{B_\theta} \cdot \theta - c(B_\theta) \right]$ involves supplying a possibly different bundle $B_\theta \in \argmax_{B \subseteq G} \left(\mathbf{1}^{B} \cdot \theta - c(B)\right)$ to each type $\theta$. Second, we drop the assumption that $\Theta\subseteq\mathbb R^{|G|}_{++}$ and instead assume $\mathbf{1}^{B_\theta} \cdot \theta - c(B_\theta)>0$ for all $\theta\in \Theta$, which ensures positive gains from trade.

 In this setting, a simple generalization of pure bundling---\textit{\textbf{single-bundle mechanisms}}---achieves the optimal convergence rate to the first-best: For each signal sequence $x^n$, let $R^{\text{1bd}}_{x^n}$ denote the expected revenue when, conditional on $x^n$, the seller optimally chooses a single bundle $B$ (possibly a strict subset of $G$) and price $p (B)$ at which to offer this bundle. Let $R_n^{\text{1bd}} := \mathbb{E} [R_{x^n}^{\text{1bd}}]$.

\begin{thm}\label{thm:main-c} Under the second-best mechanism and optimal single-bundle mechanism, the revenue gap to the first-best vanishes equally fast: Letting $\lambda:={\mathbb E}[\lambda^{B(\theta)}(\theta)]$, 
\begin{equation*}\label{eq:SB-conv-c}
R^{\rm FB}-R^{\rm SB}_n\sim R^{\rm FB}-R_n^{\text{\rm 1bd}}\sim \lambda\sqrt{\frac{\ln n}{n}}.  
\end{equation*}
\end{thm}

\subsubsection{Proof of Theorem~\ref{thm:main-c}}

The proof proceeds analogously to Theorem~\ref{thm:main}. 
Denote by $\Theta^*$ the set of types $\theta\in {\rm int} \, \Theta$ such that $\argmax_{B\subseteq G} \theta\cdot {\mathbf 1}^{B}-c(B)$ is a singleton. Types are in $\Theta^*$ with probability 1. The analysis of the one-dimensional environment in Appendix~\ref{app:1-dim} extends unchanged; that is, optimal mechanisms only use deterministic bundles and one specific random bundle. 

We extend the analysis of the Gaussian environment in Appendix~\ref{sec:Gaussian}. Denote by $R^i\left(\theta^*, \frac{1}{n}J\right)$ ($i \in \{ {\rm SB, 1bd} \}$) the seller's (second-best and single-bundling) payoffs under the belief ${\cal N}(\theta^*, \frac{1}{n}J)$.  An analogous argument as in Proposition~\ref{prop:gaussian'} yields a $K>0$ such that for all $\theta^*\in\Theta^*$, $J\in \cal J$, $n\in\mathbb N$, and $i \in \{ {\rm SB, 1bd} \}$,
\[
\left|R^i\left(\theta^*, \frac{1}{n}J\right)-\theta^*\cdot {\bf 1}^{B(\theta^*)} +c({B(\theta^*)}) + \left(\frac{\ln n}{n} {\mathbf 1}^{B(\theta^*)}\cdot J {\mathbf 1}^{B(\theta^*)}\right)^{1/2}  \right| \leq K {n}^{-1/2}.  \]

Combined with the Gaussian approximation in Appendix~\ref{app:approximation}, we obtain the following extension of Proposition~\ref{prop:conditional} that characterizes the convergence rate of the seller's payoff conditional on each type: For each $\theta\in\Theta^*$ and $i \in \{ {\rm SB, 1bd} \}$,
\[
\sum_{g\in B(\theta)}\theta_g-c(B(\theta))-{\mathbb E}_\theta[R^{\rm i}(\lambda_{x^n})]= \left(\frac{\ln n}{n}\right)^{1/2}\lambda^{B(\theta)}(\theta)+ o\left(\left(\frac{\ln n}{n}\right)^{1/2} \right). 
\]
This approximation, together with Fatou's Lemma, implies that both $R^{\rm FB}-R^{\rm SB}_n$ and $R^{\rm FB}-R_n^{\text{\rm 1bd}}$ vanish at least as slow as $ \lambda\sqrt{\frac{\ln n}{n}}$.

To show that $R^{\rm FB}-R^{\rm SB}_n$ and $R^{\rm FB}-R_n^{\text{\rm 1bd}}$ vanish at least as fast as $ \lambda\sqrt{\frac{\ln n}{n}}$, we extend the argument in Appendix~\ref{app:MLEprice}. Denote by $\hat R^{\rm 1bd}_n$ the expected revenue when, following each realization $x^n$, the seller uses a single-bundle mechanism that offers the bundle $B_{x^n}$ 
at price $p_{x^n} (B_{x^n})$, where 
\[
B_{x^n}\in\argmax_{B\subseteq G} p_{x^n} (B)-c(B)
\]
and $p_{x^n} (B)$ for each $B\subseteq G$ is as defined in Appendix~\ref{app:MLEprice}. Clearly, $R^{\rm 1bd}_n\geq \hat R_n^{\rm 1bd}$.

By the same argument as in Lemma~\ref{lem:MLrev2}, there exists $M'>0$ such that for all $n\geq \overline n$, $\theta\in\Theta$, and $B \subseteq G$,
\[
\left|{\mathbb E}_\theta[(p_{x^n} (B)-c(B))\mathbf{1}_{\{p_{x^n} (B)\leq \theta\cdot {\bf 1}^{B}\}}] -  \theta\cdot {\bf 1}^B+c(B) +\left(\frac{\ln n}{n} {\mathbf 1}^B\cdot I(\theta)^{-1}  {\mathbf 1}^B\right)^{1/2}\right|  \leq M'n^{-1/2}.
\]
For every $n\geq\overline n$ and $\theta\in\Theta$, we have
\begin{eqnarray*}
&& {\mathbb E}_\theta[(p_{x^n} (B_{x^n}) -c(B_{x^n}))\mathbf{1}_{\{p_{x^n} (B_{x^n})\leq \theta\cdot {\bf 1}^{B_{x^n}}\}}] \geq  {\mathbb E}_\theta[p_{x^n} (B_{x^n})-c(B_{x^n})] -M''n^{-1/2} \\
&\geq & \max_{B\subseteq G}{\mathbb E}_\theta[p_{x^n} (B)-c(B)] -M''n^{-1/2} \\
&\geq&  \max_{B\subseteq G} {\mathbb E}_\theta[(p_{x^n} (B)-c(B))\mathbf{1}_{\{p_{x^n} (B)\leq \theta\cdot {\bf 1}^{B}\}}] -M'''n^{-1/2} \\
&\geq&  \max_{B\subseteq G} \theta\cdot {\bf 1}^B-c(B) -\left(\frac{\ln n}{n} {\mathbf 1}^B\cdot I(\theta)^{-1}  {\mathbf 1}^B\right)^{1/2}  -(M'+M''')n^{-1/2} 
\end{eqnarray*}
for some $M'', M'''$>0 that are chosen uniformly in $\theta$,  where the first and third inequalities use Lemma~\ref{lem:MLrev1}, the second uses the definition of $B_{x^n}$, 
and the fourth uses the above extension of Lemma~\ref{lem:MLrev2}.

Since $\hat R^{\rm 1bd}_n={\mathbb E}[(p_{x^n} (B_{x^n}) -c(B_{x^n}))\mathbf{1}_{\{p_{x^n} (B_{x^n})\leq \theta\cdot {\bf 1}^{B_{x^n}}\}}]$, we can lower-bound $\hat R^{\rm 1bd}_n-{\mathbb E}\left[\theta\cdot{\bf 1}^{B(\theta)}-c(B(\theta))-\left(\frac{\ln n}{n} {\mathbf 1}^{B(\theta)}\cdot I(\theta)^{-1}  {\mathbf 1}^{B(\theta)}\right)^{1/2} \right]$ by $n^{-1/2}$ times a constant. 
Then, as claimed, we have that for $i \in \{\rm SB, 1bd \}$,
\begin{eqnarray*}
{\mathbb E}\left[({\mathbf 1}^{B(\theta)}\cdot I(\theta)^{-1}  {\mathbf 1}^{B(\theta)})^{1/2}\right] &\geq& \limsup_{n\to\infty}(R^{\rm FB}-\hat R^{\rm 1bd}_n) \left(\frac{\ln n}{n}\right)^{-1/2} \\
&\geq& \limsup_{n\to\infty}(R^{\rm FB}-R^{i}_n) \left(\frac{\ln n}{n}\right)^{-1/2}.
\end{eqnarray*}

\subsection{Non-Additive Utilities}\label{sec:nonadd}

We extend the main result to a setting where the buyer's payoffs are not additive across goods.  Let ${\cal B}=2^G$ be the space of all bundles with $m=|{\cal B}|$. The buyer's type is represented by a vector $\omega\in\mathbb R^{\cal B}$, where entry $\omega_B$ represents his valuation of bundle $B$. His utility from receiving bundle $B\in{\cal B}$ and paying  transfer $t$ is $\omega_B-t$. Type $\omega$ is drawn from a density $g$ whose support is some compact set $\Omega\subseteq \mathbb R^{\cal B}$ with non-empty interior. We assume that $\omega_G>\max\{\omega_{B}, 0\}$ for all $B\subsetneq G$ and $\omega\in \Omega$. Thus, every buyer type finds the grand bundle $G$ most attractive. We note that the analysis in this appendix does not make use of the fact that bundles $B$ are sets of goods that are partially ordered by set inclusion. Thus, in addition to multi-good monopoly, our analysis applies more generally to any setting where $\cal B$ represents a finite set of allocations (e.g., discrete quality levels) and $G$ represents an allocation (e.g., highest quality level) that is valued the most by all types.\footnote{Analogously to Appendix~\ref{app:cost}, it is straightforward to further extend the analysis to settings where different types $\omega$ have different most preferred allocations $B_\omega$.}

As in our main model, conditional on buyer type $\omega$, the seller observes a sequence of $n$ signals, $x^n = (x_1, \ldots, x_n)$, that are drawn i.i.d.\ from a distribution $P_\omega \in \Delta(X)$ with density $f(\cdot, \omega)$. The seller then chooses a direct mechanism, described by mappings $q:\Omega\to \Delta(2^G)$ and $t:\Omega\to \mathbb R$. Conditional on $x^n$, the second-best revenue is $R^{\rm SB}_{x^n}:=\sup_{q, t}\mathbb E[t(\omega)|x^n]$
subject to
\begin{equation}\tag{IC}\label{eq:IC''}
\sum_{B\in{\cal B}}q(B; \omega) \omega_B -t(\omega) \geq\sum_{B\in{\cal B}}q(B; \omega') \omega_B -t(\omega'), \;\; \forall\omega, \omega' \in \Omega,
\end{equation}
\begin{equation}\tag{IR}\label{eq:IR''}
\sum_{B\in{\cal B}}q(B; \omega) \omega_B -t(\omega) \geq 0, \;\; \forall \omega\in \Omega.
\end{equation}
Let $R_n^{\rm SB}:={\mathbb E}[R^{\rm SB}_{x^n}]$ and $R^{\rm FB}:=\mathbb{E}\left[ \omega_G \right]$. Let $R_{x^n}^{\text{bd}}$ denote the seller's optimal expected revenue at $x^n$ when she is restricted to using (pure) bundling (i.e., posting a single price $p_{x^n} (G)$ for bundle/allocation $G$). Let $R_n^{\text{bd}} := \mathbb{E} [R_{x^n}^{\text{bd}}]$.\footnote{We do not analyze separate sales (posting prices $p_B$ for each bundle $B$ that are additive, i.e., $p_B = \sum_{g \in B} p_{\{g\}}$): Under non-additive utilities, this need not reduce to single-good monopoly, as the optimal price of each good $g$ may depend on the prices of other goods. However, Theorem~\ref{thm:main2} implies that convergence under separate sales is no faster than under pure bundling (and by Theorem~\ref{thm:main}, convergence under separate sales can be strictly slower).}

 We impose an analog of Assumption~\ref{asp}. As before, we extend the signal distribution $P_\omega$ and density $f(\cdot,\omega)$ to all types in some compact neighborhood $\hat\Omega \supseteq \Omega$. Here, we pick $\hat\Omega$ sufficiently close to $\Omega$ so that $\omega_G>\max\{\omega_B, 0\}$ holds for all $\omega\in\hat\Omega$ and $B \subsetneq G$.

\begin{assume}\label{asp'} \
\begin{enumerate}
\item  The prior density $g$ is strictly positive and locally Lipschitz continuous for all $\omega \in \Omega$.

\item The signal densities $f(x,\omega)$ are strictly positive and bounded in $(x, \omega)$, and $C^2$ in $\omega\in {\rm int} \, \hat\Omega$. Moreover, for all $B,B'\in{\cal B}$,  $\frac{\partial^2 \ln f(x, \omega)}{\partial\omega_B\partial\omega_{B'}}$ is bounded in $(x, \omega)$ and Lipschitz continuous in $\omega$ uniformly across $x$.

\item We have $\sup_{\omega\in\Omega} \int \left(\inf_{\omega'\in\Omega}\ln f(x,\omega')\right)^2 \, dP_\omega(x)<\infty$. 

\item The Fisher information matrix $I(\omega) \in \mathbb{R}^{\cal B \times \cal B}$, given by 
 $$I(\omega) := \left(\int-\frac{\partial^2}{\partial \omega_B\omega_{B'}}\ln f(x,\omega)dP_\omega(x)\right)_{B,B'\in{\cal B}},$$
 is well-defined and positive definite for each $\omega\in\hat\omega$. 
 \end{enumerate}
 \end{assume}

The following result extends Theorem~\ref{thm:main}. Analogously to Section~\ref{sec:convergence}, conditional on each buyer type $\omega^*$, the seller's posterior standard deviation about $\omega_G$ is approximated by $\frac{\lambda_{\omega^*}}{ \sqrt{n}}$ at large $n$, where $\lambda_{\omega^*} := \sqrt{{\mathbf 1}^G\cdot I(\omega^*)^{-1}  {\mathbf 1}^G}$. Let $\lambda := \mathbb{E} [\lambda(\omega)]$.

\begin{thm}\label{thm:main2} Under the second-best mechanism and optimal bundling mechanism, the revenue gap to the first-best revenue vanishes equally fast: 
\begin{equation*}\label{eq:SB-conv2}
R^{\rm FB}-R^{\rm SB}_n\sim R^{\rm FB}-R_n^{\text{\rm bd}}\sim  \lambda\sqrt{\frac{\ln n}{n}}.  
\end{equation*}
\end{thm}

We omit the proof, which is analogous to that of Theorem~\ref{thm:main}. 
In particular, as in Appendix~\ref{app:main-pf}, the seller's posterior at large $n$ can be approximated by a Gaussian distribution whose mean is $
\hat\omega_{x^n}\in\argmax_{\omega\in\hat\Omega}\sum_{i=1}^n\ln f(x_i, \omega)$
and whose covariance matrix is the inverse of 
\[
I_{x^n}(\omega):=\left(-\frac{1}{n}\sum_{i=1}^n\frac{\partial^2}{\partial \omega_B \partial \omega_{B'}}\ln  f(x_i, \omega)\right)_{B,B'\in{\cal B}}
\]
evaluated at $\omega=\hat\omega_{x^n}$.  
To upper-bound $R^{\rm SB}_n$, we make use of a one-dimensional relaxed problem as in Appendix~\ref{app:propgaussian-pf}; note that the arguments there are presented for the type space $\mathbb{R}^{{\cal B}}$ and do not assume additive values.  


\subsection{Deterministic Setting with Non-Gaussian Beliefs}\label{app:non-gaussian}

Extending the deterministic Gaussian environment studied in the main text, we consider a setting where the seller has a general deterministic sequence of beliefs indexed by $n$ that converge to a point-mass on the true type $\theta^*$. Such beliefs can be interpreted as the outcome of the seller observing increasingly precise information that need not take the form of $n$ (i.i.d.\ or correlated) signal draws. 

Formally, conditional on true type $\theta^*\in\mathbb R^{|G|}_{++}$, the seller's belief at $n$ is that the buyer's type is $\theta=\theta^*+n^{-1/2}z$, where $z\in\mathbb R^{|G|}$ is drawn from a cdf $F$ that is continuously differentiable and positive. For each $B\subseteq G$, let $F^B$ denote the induced cdf of ${\bf 1}^B\cdot z$. We impose the following tail conditions: There exist $\alpha^+(B),\alpha^-(B), \beta^+(B), \beta^-(B)>0$ such that, as $z \to \infty$,
\begin{eqnarray*}
1-F^B(z)&=&\exp[-\alpha^+(B)z^{\beta^+(B)}+o(z^{\beta^+(B)})], \\
\text{ and } \quad F^B(-z)&=&\exp[-\alpha^-(B)z^{\beta^-(B)}+o(z^{\beta^-(B)})].
\end{eqnarray*} 
These conditions generalize the Gaussian setting in the main text, which corresponds to the case $\alpha^+(B)=\alpha^-(B)=\frac{1}{2(\lambda^B(\theta^*))^2}$ and $\beta^+(B)=\beta^-(B)=2$ (based on the Gaussian tail bound in (\ref{eq:Gauss-tail})).


We restrict attention to deterministic mechanisms. 
Denote by $R^{\rm mix}_{n} (\theta^*)$ the seller's optimal expected revenue under IC-IR deterministic mechanisms (which can be represented as mixed bundling). Denote by $R^{\rm bd}_{n} (\theta^*)$ the optimal revenue under (pure) bundling. Denote by $R^{\rm FB} (\theta^*):=\sum_{g\in G}\theta^*_g$ the first-best revenue.

\begin{prop}\label{prop:non-Gaussian} 
Under the optimal deterministic mechanism and optimal bundling mechanism, the revenue gap to the first-best vanishes equally fast:
\[
R^{\rm FB} (\theta^*) -R^{\rm mix}_n(\theta^*)\sim R^{\rm FB} (\theta^*) -R^{\rm bd}_n(\theta^*)\sim \left(\frac{\ln n}{2\alpha^-(G)}\right)^{1/\beta^-(G)}n^{-1/2}.
\] 
\end{prop}

Thus, optimizing over pure bundling achieves the same convergence rate to the first-best as optimizing over all possible deterministic mechanisms. 
This generalizes our main insight that pure bundling is an effective mechanism under precise seller beliefs, as using menus of bundles only allows for a negligible improvement relative to pure bundling. The main qualification is that, unlike Theorem~\ref{thm:main}, Proposition~\ref{prop:non-Gaussian} does not allow for stochastic mechanisms, whose analysis we leave as an open question.

Appendix~\ref{app:1-dim non-Gaussian} first analyzes the single-good setting, extending the intensive vs.\ extensive margin analysis from the Gaussian environment in Section~\ref{sec:1dim} to the current setting. Appendix~\ref{app:non-Gaussian proof} then proves Proposition~\ref{prop:non-Gaussian}. Finally, Appendix~\ref{app:elasticity} relates the dominance of intensive vs.\ extensive margins in the single-good setting to their elasticities.

\subsubsection{Single-Good Case}\label{app:1-dim non-Gaussian}

Assume $|G| = 1$, so we can omit the dependence of parameters $\alpha^+, \alpha^-, \beta^+, \beta^-$ on $B$.  Denote by $R_n^*$ the optimal revenue, which is achieved by a posted price $p_n^*$. Let $F_n$ denote the cdf of the seller's belief, i.e., $F_n(\theta)=F((\theta-\theta^*)\sqrt{n})$.

Generalizing Proposition~\ref{lem:1-dim}, the following result shows that at large $n$, the seller optimally sets prices in such a way that revenue losses relative to the first-best are driven by the intensive margin, while the extensive margin becomes negligible. Write $\alpha = \alpha^-$, $\beta = \beta^-$.

\begin{prop}\label{lem:1-dim'} 
Any optimal price sequence $p^*_n$ satisfies
\[
R^{\rm FB} (\theta^*) -R^*_n\sim \theta^*- p^*_n\sim \left(\frac{\ln n}{2\alpha}\right)^{1/\beta}n^{-1/2}  \quad  \text{ and } \quad \theta^* F_n(p^*_n)=o\left((\ln n)^{1/\beta}n^{-1/2}\right).
\]
Moreover, under any price sequence $p_n$ with $\theta^*-p_n\sim \delta  \left(\frac{\ln n}{2\alpha}\right)^{1/\beta}n^{-1/2}$ for some $\delta\in [0,1)$, we have $\lim_n \frac{\theta^* F_n (p_n)}{(\ln n)^{1/\beta}n^{-1/2}}=\infty$. 
\end{prop}

\begin{proof}
Take any $\delta\geq 0$ and sequence of prices $(p_n)$ with $\theta^*-p_n\sim \delta  \left(\frac{\ln n}{2\alpha}\right)^{1/\beta}n^{-1/2}$. Then by the tail conditions on $F$,
\begin{eqnarray*}
F_n(p_n)= \exp\left[-\frac{\delta^\beta}{2}\ln n+ o\left(\ln n\right)\right]= n^{-\frac{\delta^\beta}{2}+o(1)},
\end{eqnarray*}
and thus
\[
\lim_{n\to\infty} \frac{\theta^* F_n (p_n)}{(\ln n)^{1/\beta}n^{-1/2}}=
\begin{cases}
0 \text{ if } \delta>1, \\
\infty \text{ if } \delta<1.
\end{cases}
\]
Additionally, if $\theta^*-p_n= \left(\frac{\ln n}{2\alpha}\right)^{1/\beta}n^{-1/2}$ for all $n$, then $\lim_n \frac{\theta^* F_n (p_n)}{(\ln n)^{1/\beta}n^{-1/2}}=0$, and hence $R^{\rm FB} (\theta^*) -R_n(p_n)\sim  \left(\frac{\ln n}{2\alpha}\right)^{1/\beta}n^{-1/2}$, where $R_n(p_n):=p_n(1-F_n(p_n))$. 

Given these observations, the remainder of the proof is analogous to that of Proposition~\ref{lem:1-dim}. \end{proof}

\subsubsection{Proof of Proposition~\ref{prop:non-Gaussian}}\label{app:non-Gaussian proof}

Let $(p_n(B))_{B\subseteq G}$ denote optimal mixed-bundling prices that yield $R^{\rm mix}_n(\theta^*)$, where $p_n(\emptyset)=0.$ 
Observe first that for any $B\subseteq G$ with $\lim_{n\to\infty}p_n(B)>{\bf 1}^B\cdot\theta^*$,
\begin{equation}
\begin{split}\label{eq:B prob}
&{\rm Prob}[{\bf 1}^B\cdot \theta - p_n(B)\geq  {\bf 1}^{B'}\cdot \theta - p_n(B'), \forall B'\subseteq G]\leq {\rm Prob}[{\bf 1}^B\cdot \theta - p_n(B)\geq 0] \\
=&1-F^B( n^{1/2}(p_n(B)-{\bf 1}^B\cdot\theta^*))\\
=&\exp[-\alpha^+(B) n^{\frac{\beta^+(B)}{2}}+o(n^{\frac{\beta^+(B)}{2}})]=o(n^{-1}),
\end{split}
\end{equation}
where the second equality uses the assumption on right-tail probabilities.

Write $\alpha=\alpha^-(G)$ and $\beta=\beta^-(G)$. To prove the proposition, note that 
$R^{\rm FB} (\theta^*)-R^{\rm bd}_n(\theta^*)\sim \left(\frac{\ln n}{2\alpha}\right)^{1/\beta}n^{-1/2}$ follows from Proposition~\ref{lem:1-dim'}.  Suppose toward a contradiction that $R^{\rm FB} (\theta^*) -R^{\rm mix}_n(\theta^*)\sim \left(\frac{\ln n}{2\alpha}\right)^{1/\beta}n^{-1/2}$ fails.  Since $R^{\rm FB} (\theta^*) -R^{\rm mix}_n(\theta^*)\leq R^{\rm FB} (\theta^*) -R^{\rm bd}_n(\theta^*)$ holds for each $n$ by definition,  we have 
\begin{equation}\label{eq:mix-cont}
\liminf_{n\to\infty}\frac{R^{\rm FB} (\theta^*) -R^{\rm mix}_n(\theta^*)}{\left(\frac{\ln n}{2\alpha}\right)^{1/\beta}n^{-1/2}}<1. 
\end{equation}

Observe by definition of $R^{\rm mix}_n(\theta^*)$ that 
\begin{equation}\label{eq:mix-gap}
R^{\rm FB} (\theta^*) -R^{\rm mix}_n(\theta^*)=\sum_{B\subseteq G} {\rm Prob}[{\bf 1}^B\cdot \theta - p_n(B)\geq  {\bf 1}^{B'}\cdot \theta - p_n(B'), \forall B'\subseteq G]\left({\bf 1}^G \cdot\theta^*-p_n(B)\right).
\end{equation}

Consider the case $\liminf_{n\to\infty}\frac{{\bf 1}^G\cdot \theta^*-p_n(G)}{\left(\frac{\ln n}{2\alpha}\right)^{1/\beta}n^{-1/2}}\geq 1$. Given (\ref{eq:mix-cont})-(\ref{eq:mix-gap}), there exists $B\subsetneq G$ such that $\liminf_{n\to\infty}{\rm Prob}[{\bf 1}^B\cdot \theta - p_n(B)\geq  {\bf 1}^{B'}\cdot \theta - p_n(B'), \forall B'\subseteq G]>0$ and $\liminf_{n\to\infty}\frac{{\bf 1}^G\cdot \theta^*-p_n(B)}{\left(\frac{\ln n}{2\alpha}\right)^{1/\beta}n^{-1/2}}< 1$. This contradicts (\ref{eq:B prob}).

Consider the case $\liminf_{n\to\infty}\frac{{\bf 1}^G \cdot \theta^*-p_n(G)}{\left(\frac{\ln n}{2\alpha}\right)^{1/\beta}n^{-1/2}}< 1$. Then
\begin{eqnarray*}
&&\liminf_{n\to\infty}\frac{\sum_{B\subsetneq G} {\rm Prob}[{\bf 1}^B\cdot \theta - p_n(B)\geq  {\bf 1}^{B'}\cdot \theta - p_n(B'), \forall B'\subseteq G]}{\left(\frac{\ln n}{2\alpha}\right)^{1/\beta}n^{-1/2}} \\
&\geq& \liminf_{n\to\infty}\frac{{\rm Prob}[{\bf 1}^G \cdot\theta-p_n(G)<0]}{\left(\frac{\ln n}{2\alpha}\right)^{1/\beta}n^{-1/2}}=\infty,
\end{eqnarray*}
where the equality uses the second part of Proposition~\ref{lem:1-dim'}. But given (\ref{eq:mix-cont})-(\ref{eq:mix-gap}), there exists $B\subsetneq G$ such that $\liminf_{n\to\infty}\frac{{\rm Prob}[{\bf 1}^B\cdot \theta - p_n(B)\geq  {\bf 1}^{B'}\cdot \theta - p_n(B'), \forall B'\subseteq G]}{\left(\frac{\ln n}{2\alpha}\right)^{1/\beta}n^{-1/2}}> 0$ and $\liminf_{n\to\infty}\frac{{\bf 1}^G \cdot \theta^*-p_n(B)}{\left(\frac{\ln n}{2\alpha}\right)^{1/\beta}n^{-1/2}}< 1$. This contradicts  (\ref{eq:B prob}).

\subsubsection{Elasticity of Extensive vs.\ Intensive Margins}\label{app:elasticity}

We revisit the single-good environment from Appendix~\ref{app:1-dim non-Gaussian} and relate the dominance of the intensive vs.\ extensive margins to their elasticities. We drop the tail conditions on $F$ and assume only that $F$ admits a positive density $f$. For any price sequence $p_n$, denote by $\gamma_n:=(\theta^*-p_n)n^{1/2}$ the normalized price discount. The corresponding extensive and intensive margins are $\theta^* F_n (p_n) = \theta^* F(- \gamma_n)$ and $\theta^* - p_n = \gamma_nn^{-1/2}$, respectively. The elasticities of these margins with respect to $\gamma_n$ are $\odv{\theta^* F(- \gamma_n)}{\gamma_n}\frac{\gamma_n}{\theta^* F(- \gamma_n)} = \frac{-\gamma_n f(-\gamma_n)}{F(-\gamma_n)}$ and $\odv{\gamma_nn^{-1/2}}{\gamma_n}\frac{\gamma_n}{\gamma_nn^{-1/2}} = 1$, respectively. 
 
The following lemma shows that, along the optimal price sequence, the relative size of the intensive and extensive margins is determined by the relative magnitudes of these elasticities at large $\gamma_n$:

\begin{lem}\label{lem:elasticity} Assume that along the optimal price sequence $p^*_n$ (with $\gamma^*_n := (\theta^*-p^*_n)n^{1/2}$) both the extensive and intensive margins vanish as $n \to \infty$. Assume $\lim_{\gamma\to\infty}\frac{-\gamma f(-\gamma)}{F(-\gamma)}$ is well-defined. Then 
\[
\lim_{n \to \infty} \frac{ \theta^* F(- \gamma^*_n)}{\gamma^*_n n^{-1/2}} = \lim_{\gamma\to\infty}\frac{F(-\gamma)}{\gamma f(-\gamma)}.
\]
\end{lem}

\begin{proof} By the first-order condition for revenue maximization, 
\begin{equation}\label{eq:elasticity-FOC}
(\theta^* - \gamma^*_n n^{-1/2}) f(- \gamma^*_n) = n^{-1/2} (1 - F(- \gamma^*_n)).
\end{equation}
By assumption, $1 - F(- \gamma^*_n) \rightarrow 1$ and $\gamma^*_n n^{-1/2} \rightarrow 0$. Thus, (\ref{eq:elasticity-FOC}) yields
\[
\lim_{n \to \infty} \frac{\theta^* f(- \gamma^*_n)}{n^{-1/2}} = 1.
\]
Therefore, 
\[
\lim_{n \to \infty} \frac{ \theta^* F(- \gamma^*_n)}{\gamma^*_nn^{-1/2}} = \lim_{n \to \infty} \frac{ \theta^* F(- \gamma^*_n)}{\gamma^*_n\theta^* f(- \gamma^*_n)} = 
\lim_{n \to \infty} \frac{F(- \gamma^*_n)}{\gamma^*_n f(- \gamma^*_n)} = \lim_{\gamma\to\infty}\frac{F(-\gamma)}{\gamma f(-\gamma)},
\]
where the last equality uses the fact that $\gamma^*_n \to \infty$ as $n \to \infty$ (by the assumption that the extensive margin vanishes).
\end{proof}

Lemma~\ref{lem:elasticity} implies that the intensive margin dominates, i.e., $\lim_{n \to \infty} \frac{\gamma^*_n n^{-1/2}}{ \theta^* F(- \gamma^*_n)} =\infty$,  whenever the elasticity $\frac{-\gamma f(-\gamma)}{F(-\gamma)}$ of the extensive margin diverges as $\gamma \to \infty$, which happens as long as the reverse hazard rate $\frac{f(-\gamma)}{F(-\gamma)}$ does not vanish too quickly as $\gamma$ grows large. This is the case under many distributions (in particular, Gaussian distributions as in Section~\ref{sec:1dim}), where $\frac{f(-\gamma)}{F(-\gamma)}$ is in fact increasing in $\gamma$. More generally, $\lim_{\gamma\to\infty}\frac{\gamma f(-\gamma)}{F(-\gamma)}=\infty$ also holds under the tail conditions in Appendix~\ref{app:1-dim non-Gaussian} (i.e., when there exist $\alpha, \beta > 0$ such that $F(-\gamma)=\exp[-\alpha \gamma^\beta+o(\gamma^\beta)]$ as $\gamma\to\infty$).

\end{document}